\documentclass[reprint, superscriptaddress, amsmath,amssymb, aps, longbibliography]{revtex4-2}

\usepackage[version=3]{mhchem} 
\usepackage{graphicx}
\usepackage{dcolumn}
\usepackage{bm}
\usepackage{chemformula}
\usepackage{amsmath,amsthm,amssymb}
\usepackage{gensymb}
\usepackage{xr}
\usepackage{hyperref}
\usepackage{soul}
\usepackage[mathlines]{lineno}

\DeclareMathOperator\erfc{erfc}

\begin{document}

\author{Alessandra Milloch}
\email{alessandra.milloch@unicatt.it}
\affiliation{Department of Mathematics and Physics, Università Cattolica del Sacro Cuore, Brescia I-25133, Italy}
\affiliation{ILAMP (Interdisciplinary Laboratories for Advanced
Materials Physics), Università Cattolica del Sacro Cuore, Brescia I-25133, Italy}
\affiliation{Department of Physics and Astronomy, KU Leuven, B-3001 Leuven, Belgium}

\author{Umberto Filippi}
\affiliation{Italian Institute of Technology (IIT), Genova 16163, Italy}

\author{Paolo Franceschini}
\affiliation{CNR-INO (National Institute of Optics), via Branze 45, 25123 Brescia, Italy}
\affiliation{Department of Information Engineering, University of Brescia, Brescia I-25123, Italy}

\author{Selene Mor}
\affiliation{Department of Mathematics and Physics, Università Cattolica del Sacro Cuore, Brescia I-25133, Italy}
\affiliation{ILAMP (Interdisciplinary Laboratories for Advanced
Materials Physics), Università Cattolica del Sacro Cuore, Brescia I-25133, Italy}

\author{Stefania Pagliara}
\affiliation{Department of Mathematics and Physics, Università Cattolica del Sacro Cuore, Brescia I-25133, Italy}
\affiliation{ILAMP (Interdisciplinary Laboratories for Advanced
Materials Physics), Università Cattolica del Sacro Cuore, Brescia I-25133, Italy}

\author{Gabriele Ferrini}
\affiliation{Department of Mathematics and Physics, Università Cattolica del Sacro Cuore, Brescia I-25133, Italy}
\affiliation{ILAMP (Interdisciplinary Laboratories for Advanced
Materials Physics), Università Cattolica del Sacro Cuore, Brescia I-25133, Italy}

\author{Franco V. A. Camargo}
\affiliation{IFN-CNR, Piazza Leonardo da Vinci 32, I-20133, Milano, Italy}

\author{Giulio Cerullo}
\affiliation{IFN-CNR, Piazza Leonardo da Vinci 32, I-20133, Milano, Italy}
\affiliation{Department of Physics, Politecnico di Milano, Piazza Leonardo da Vinci 32, I-20133 Milano, Italy}

\author{Dmitry Baranov}
\affiliation{Italian Institute of Technology (IIT), Genova 16163, Italy}
\affiliation{Division of Chemical Physics, Department of Chemistry, Lund University, P.O. Box 124, SE-221 00 Lund, Sweden}
 
\author{Liberato Manna}
\affiliation{Italian Institute of Technology (IIT), Genova 16163, Italy}

\author{Claudio Giannetti}
\email{claudio.giannetti@unicatt.it}
\affiliation{Department of Mathematics and Physics, Università Cattolica del Sacro Cuore, Brescia I-25133, Italy}
\affiliation{ILAMP (Interdisciplinary Laboratories for Advanced
Materials Physics), Università Cattolica del Sacro Cuore, Brescia I-25133, Italy}
\affiliation{CNR-INO (National Institute of Optics), via Branze 45, 25123 Brescia, Italy}

\title{The fate of optical excitons in \ch{FAPbI_3} nanocube superlattices}

\begin{abstract}
Understanding the nature of the photoexcitation and ultrafast charge dynamics pathways in organic halide perovskite nanocubes and their aggregation into superlattices is key for the potential applications as tunable light emitters, photon harvesting materials and light-amplification systems.
In this work, we apply two-dimensional coherent electronic spectroscopy (2DES) to track in real time the formation of near-infrared optical excitons and their ultrafast relaxation in \ch{CH(NH_2)_2PbI_3} nanocube superlattices. Our results unveil that the coherent ultrafast dynamics is limited by the combination of the inherent short exciton decay time ($\simeq 40$ fs) and the dephasing due to the coupling with selective optical phonon modes at higher temperatures. On the picosecond timescale, we observe the progressive formation of long-lived localized trap states. The analysis of the temperature dependence of the excitonic intrinsic linewidth, as extracted by the anti-diagonal components of the 2D spectra, unveils a dramatic change of the excitonic coherence time across the cubic to tetragonal structural transition. Our results offer a new way to control and enhance the ultrafast coherent dynamics of photocarrier generation in hybrid halide perovskite synthetic solids. 

\end{abstract}

\maketitle

\section{Introduction}

\begin{figure}[]
\includegraphics[width=8cm]{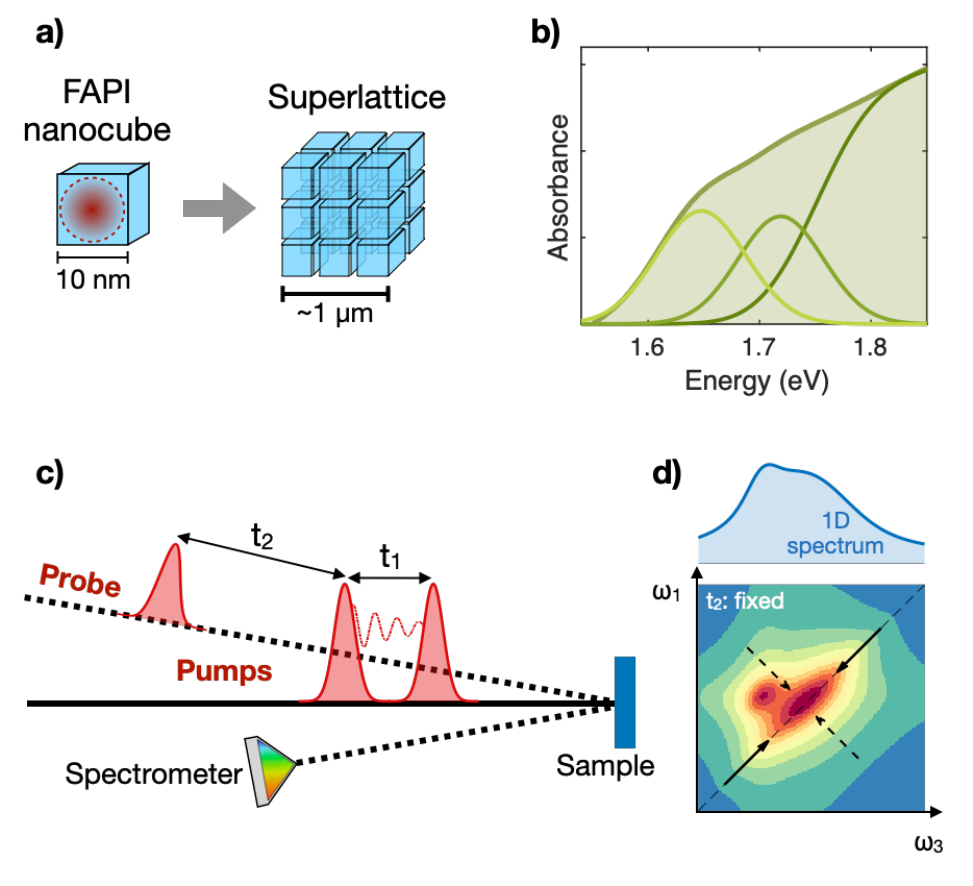}
\caption{a) Cartoon of FAPI nanocubes, hosting quantum confined excitons, and FAPI superlattices. b) Absorption spectrum (green area) of the FAPI superlattices sample measured in this work. The green solid lines represent a free carriers edge and two peaks, as obtained from Elliott analysis of the room temperature absorbance (Supporting Information Section S1-D). c) Scheme of 2DES experiment. d) Sketch of a typical 2D spectrum (vertical axis: excitation, horizontal axis: detection) where, as opposed to 1D experiments (top panel), homogeneous and inhomogeneous linewidths and correlations between spectral features can be resolved.}
\label{fig1: sample}
\end{figure}

The outstanding optical and electronic properties \cite{herz2016charge,kanemitsu2017luminescence, quan2019perovskites} make lead halide perovskites ideal materials for a wide variety of applications ranging from optoelectronics to photovoltaics \cite{stranks2015metal,quan2019perovskites}.
The recent advances in the synthesis of hybrid halide perovskite nanocubes \cite{shamsi2019metal} have enabled unprecedented control of the emission properties by combining the tunability of the chemical composition with that of the nanocube shape and size. The further advance given by the synthesis of nanocube superlattices (Figure \ref{fig1: sample}a) \cite{jurow2017tunable,tong2018spontaneous,baranov2019investigation} not only provides additional control parameters - such as the superlattice size, periodicity and structure \cite{Cherniukh2021} - but also opens the way to the development of novel light-emitting devices based on cooperative superradiant effects \cite{raino2018superfluorescence, zhou2020cooperative,krieg2020monodisperse,pashaei2023superradiance} as well as to the implementation of quantum simulators of solid state problems \cite{Milloch_NanoLett}. Collective superradiant phenomena in nanocube superlattices are driven by the long-range electromagnetic interactions among excitons in different nanocubes \cite{cong2016dicke, mattiotti2020thermal,ghonge2023enhanced} and are therefore crucially affected by the exciton dynamics following light excitation of the individual nanocubes \cite{blach2022superradiance}. Light emission in lead halide perovskites can also be affected by strong coupling of excitons with the lattice, which results in broad spectra originating from self-trapped exciton (STE) states \cite{smith2018white,li2019self}. 

An important case is given by hybrid organic-inorganic halide perovskite superlattices, whose excitonic energy extends down to the near infrared (1.5 eV), thus significantly broadening the potential exploitation of these synthetic materials \cite{manser2016intriguing,chen2015under,shamsi2019metal,Fang2017}. Formamidinium lead iodide \ch{FAPbI_3} (FAPI, \ch{FA^+} = \ch{CH(NH_2)_2^+}) is especially promising because of its smaller band gap and increased chemical and thermal stability as compared to the archetypal compound MAPI (\ch{MA^+} = \ch{CH_3NH_3^+}) \cite{levchuk2017brightly,tan2020temperature,fang2016photoexcitation}. In these systems, the presence of collective superradiance effects, analogous to those reported in fully inorganic superlattices, remains unclear; the possibility of extending superradiant phenomena to hybrid organic-inorganic perovskite superlattices requires an understanding of the ultrafast exciton dynamics of these systems. The photophysics of these materials is in fact characterized by a complex interplay of free carrier excitations, bound excitons, multi-excitons and trapped states which generally give rise to multiple optical resonances in the visible/near-infrared range. These resonances overlap in energy and cannot be disentangled by linear or pump-probe spectroscopies.
In Figure \ref{fig1: sample}b, for example, we report a typical absorption spectrum of FAPI superlattices which shows multiple features compatible with the existence of multiple excitonic-like resonances in the 1.6-1.8 eV range (see Supporting Information Section S1-D for details of the analysis and fitting of the linear spectra). Due to the overlap of different structures, however, a linear fit can hardly offer a decisive answer regarding the nature of the absorption. Therefore, the nature of the photocarrier generation and recombination processes in hybrid halide perovskite superlattices is still a matter of debate \cite{yamada2014photocarrier,phuong2016free,tahara2016experimental,li2021exciton}. Since addressing the early-stage dynamics of these processes is of central importance for controlling and enhancing macroscopic coherent effects, advanced optical techniques that go beyond linear spectroscopies and combine high temporal and spectral resolution should be employed \cite{sutherland2016perovskite,zhou2020cooperative}.

In this work, we use two-dimensional coherent electronic spectroscopy (2DES) to investigate the exciton dynamics in FAPI nanocube superlattices. 2DES allows us to unveil the hierarchy of the photoexcitation and relaxation processes. We report evidence of early-stage excitons which rapidly decay and form novel bound states compatible with bi-excitons, both at room temperature and at cryogenic temperatures. Within a few hundred femtoseconds, the excitonic resonance is lost in favour of the generation of lower energy trap-states ascribable to self-trapping of excitons or defect-trapping \cite{smith2018white}. Analysis of the excitonic decoherence times suggests that coherent ultrafast diffusion is limited by the short exciton lifetime at cryogenic temperatures and that the interaction with optical phonon modes starts playing a role at higher temperatures. The exciton decoherence time is also directly affected by the tetragonal-to-cubic phase transition occurring upon raising the temperature: the symmetry increase suppresses the number of optical phonon modes, thus offering a new parameter for controlling the photo-conversion process and the onset of coherent phenomena. Overall these results provide a snapshot of the ultrafast exciton dynamics in FAPI nanocubes deposited on a solid substrate, which is key for future modelling and control of possible collective properties emerging in more complex structures, such as superlattice with different symmetries or twisted 2D-perovskites configurations \cite{Cherniukh2021, zhang2024moire}.

\begin{figure*}[t]
\includegraphics[width=16cm]{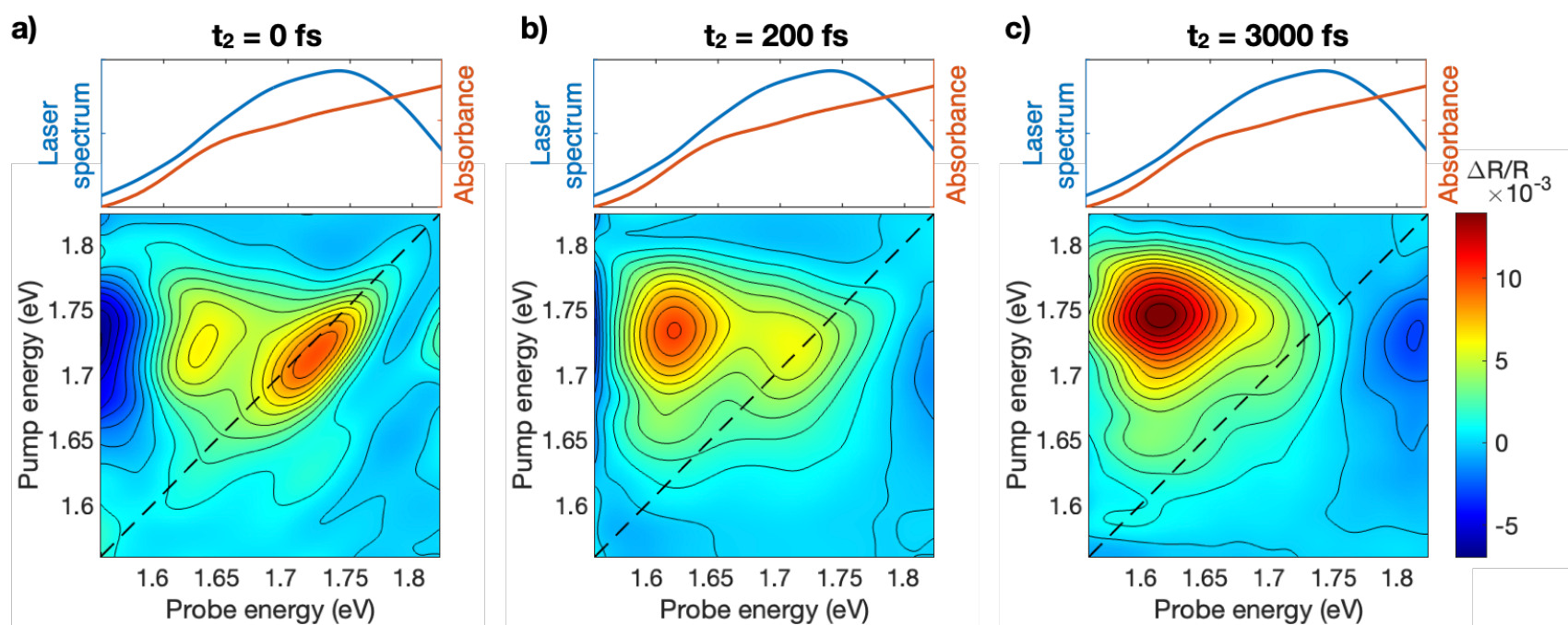}
\caption{2D spectra of FAPI superlattice measured at 200 K with 175 \textmu J/cm$^2$ excitation fluence for three different $t_2$ delays:  a) $t_2$ = 0 fs, b) 200 fs, and c) 3 ps. The top panels show the sample absorption spectrum (red) and the laser pulse spectrum (blue).}
\label{fig2: 2D}
\end{figure*}

\section{Experimental section}
2DES \cite{jonas2003two,cho2008coherent,mukamel2000multidimensional} is a multi-dimensional approach based on the use of a pair of phase-coherent broadband excitation pulses (see Fig. \ref{fig1: sample}c), separated by a variable time delay $t_1$ (coherence time), and a probe pulse delayed by $t_2$ (waiting time) to measure the transient reflectivity/transmissivity of the sample. Spectral resolution along the excitation axis is achieved by Fourier transforming the signal with respect to $t_1$, while resolution along the detection axis is achieved by spectrally resolving the detection of the signal \cite{fuller2015experimental,tollerud2017coherent}.
The 2DES setup adopted here is based on the Translating-Wedge-Based Identical Pulses eNcoding System (TWINS) scheme \cite{oriana2016scanning}. As detailed in Materials and Methods, two common-path birefringent interferometers are employed: the first interferometer (GEMINI 2D by NIREOS) generates the pair of collinear phase-coherent pump pulses \cite{brida2012phase}, the second one (GEMINI by NIREOS) is used in detection to spectrally resolve the transient reflectivity measured by the probe \cite{preda2016broadband}. The partially collinear geometry employed here (see Materials and Methods and Supporting Information Sec. S2 for experimental detail) measures the purely absorptive spectrum which corresponds to the sum of the real part of rephasing and non-rephasing signals \cite{fuller2015experimental}. This technique allows to detect, for each $t_2$ delay, a two-dimensional map which connects excitation at a given frequency $\omega_1$ to the response of the system at another frequency $\omega_3$, while retaining the femtosecond temporal resolution given by the pulse duration. The 2D spectra allow to disentangle the intrinsic decoherence dynamics of optical excitations, which manifest as a  broadening along the anti-diagonal direction (see Fig. \ref{fig1: sample}d), from disorder-induced inhomogeneous broadening along the diagonal of the $\omega_1$-$\omega_3$ map. By monitoring the temporal evolution of different peaks in the map, it is also possible to infer a cause-effect relation between different spectral features. This capability is key for monitoring the real-time transformation of the initial photoexcited carriers into the long-lived states responsible for the slow fluorescence emission.

The FAPI superlattices investigated in the present work are assembled from a colloidal dispersion of uniform $\sim$10 nm nanocubes. The studied samples comprise $\sim 0.3$ - 1 \textmu m superlattices of close-packed nanocubes deposited on a silicon substrate (see Materials and Methods for sample preparation details). The 2DES spectra are collected in the reflection geometry, which guarantees a better signal-to-noise ratio compared to the transmission configuration (see Supporting Information Section S6).

\section{Results and discussion}

In Figure \ref{fig2: 2D} we report the spectra measured on FAPI superlattices at $T$ = 200 K as a function of the excitation ($\hbar\omega_1$, vertical axis) and detection ($\hbar\omega_3$, horizontal axis) photon energies and for different $t_2$ delays. They display two distinct spectral features: i) a component located along the diagonal of the spectrum around 1.72 eV (Fig. \ref{fig2: 2D}a); ii) a broad structure centred off-diagonal, at $\hbar\omega_1 \simeq$ 1.72 eV pump photon energy and $\hbar\omega_3 \simeq$ 1.61 eV probe photon energy (Fig. \ref{fig2: 2D}b and \ref{fig2: 2D}c), which can be ascribed to transient photobleaching and stimulated emission (see also the positive transmissivity variation reported in the Supporting Information Section S6). As observed from 2D spectra at different $t_2$ time delays, these two spectral features are characterized by different dynamics, with feature i) decaying on a sub-picosecond timescale and feature ii) building up on a longer timescale. The well-defined and well-localized diagonal structure reveals the existence in FAPI superlattices of a short-lived bound exciton at 1.72 eV, whereas the delayed off-diagonal feature suggests the formation of long-lived trap-states. In the following, we analyse in detail the characteristics of these two components. 

 \begin{figure*}[]
\includegraphics[width=16cm]{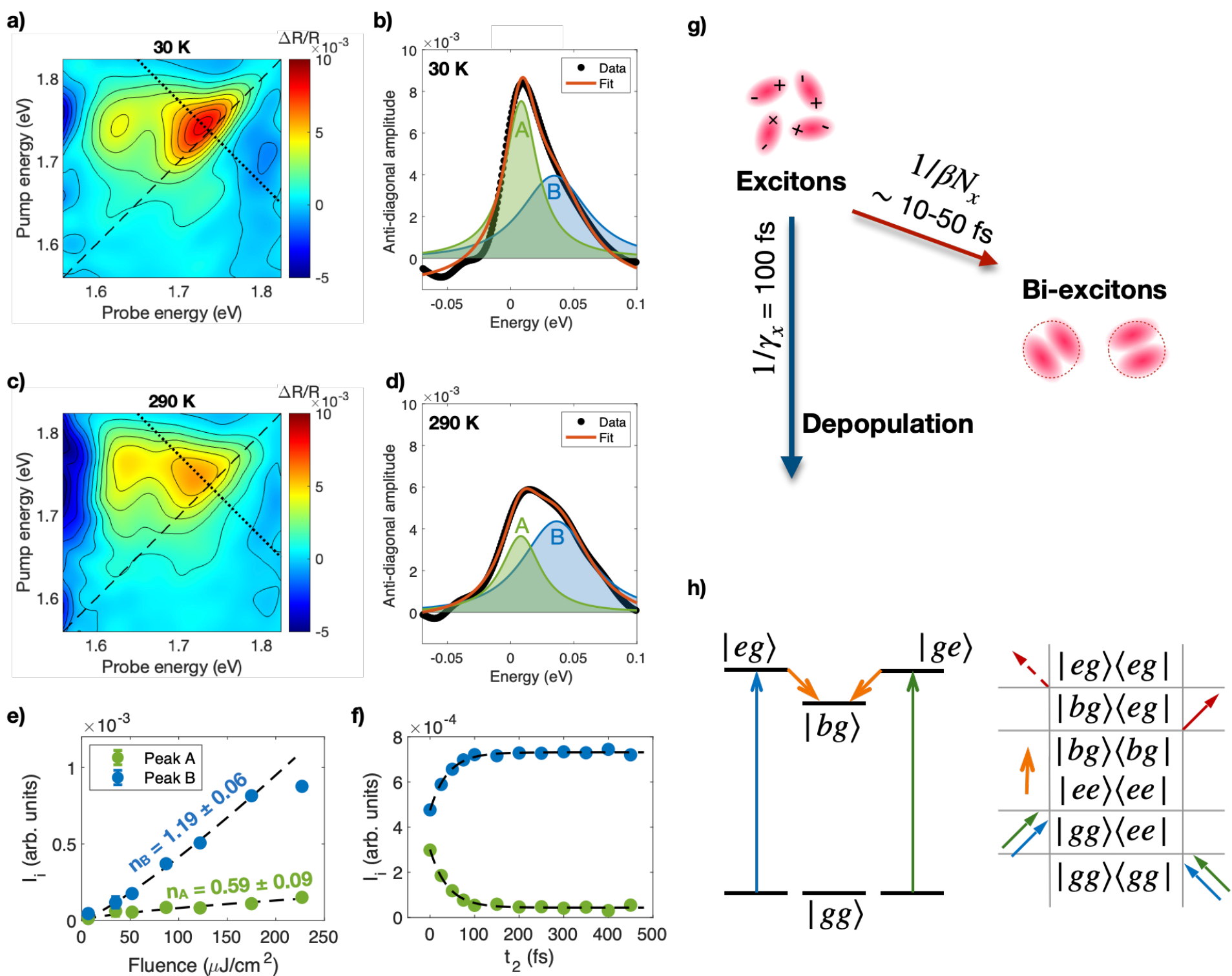}
\caption{a) 2D spectrum of FAPI superlattices measured at 30 K and $t_2$ = 0 fs, with 175 \textmu J/cm$^2$ excitation fluence. b) Anti-diagonal profile of 2D spectrum in a) taken along the dotted line in a). The black dots are the experimental data points, the red line is the profile fit, the green and blue filled areas are peaks A and B associated with, respectively, exciton and bi-exciton resonances. c) and d) display 2D spectrum and anti-diagonal profile analogous to a) and b) but collected at a higher sample temperature (290 K). e) Fluence dependence of the integrated amplitude $I_i$ of peaks A and B, extracted from fitting the 2D spectra anti-diagonal profiles at 200 K and $t_2$ = 0 fs. The black dashed lines represent the power-law fit to the data. f) Dynamics of the exciton (peak A) and bi-exciton (peak B) peak amplitudes $I_i$ at 200K for 175 \textmu J/cm$^2$ excitation fluence. g) Sketch of the relaxation channels for excitons, which can decay at a $\gamma_x$ rate or bind into bi-excitons at a rate $\beta N_x$, $N_x$ being the instantaneous number of excitons. h) Representation of the energy levels scheme (left) and fifth-order rephasing double-sided Feynman diagrams (right) for the bi-exciton formation process.}
\label{fig3: Exciton}
\end{figure*}

We first examine the short time delay ($t_2 = 0$ fs) 2D spectra and investigate the temperature dependence of the excitonic peak. The elongation along the diagonal of the 2D spectrum allows estimating the inhomogeneous Full-Width-Half-Maximum (FWHM) that results $\simeq$ 83 $\pm$ 4 meV when projected along the probe axis. This value is compatible with the width of the resonance at 1.72 eV in the steady state absorbance (see Supporting Information Section S1-D). The large inhomogeneous broadening originates from the size and shape dispersion of FAPI nanocubes and superlattices. 
The intrinsic homogeneous linewidth can instead be obtained from slicing the 2D spectrum measured at $t_2$ = 0 fs along the anti-diagonal direction crossing the exciton peak (see dotted lines in Figures \ref{fig3: Exciton}a and \ref{fig3: Exciton}c). The resulting anti-diagonal cuts are plotted in Fig. \ref{fig3: Exciton}b and \ref{fig3: Exciton}d for experiments conducted at two different temperatures; the horizontal axis represents the energy axis along the anti-diagonal direction, with positive values referring to the region above the diagonal in the 2D spectra (i.e. the region corresponding to pump photon energy larger than probe energy). The anti-diagonal profiles so obtained show a clear asymmetry that reveals the existence of two structures, one centred on the 2D spectrum diagonal (peak A) and one located above the diagonal (peak B). We stress that the signal acquired in a conventional pump-probe experiment, which corresponds to integrating along the $\hbar\omega_1$ axis (see Fig. \ref{fig1: sample}c), would be dominated by the large inhomogeneous broadening of the exciton, thus preventing the possibility of inferring information about the decoherence and the emergence of secondary structures.

A multi-peak fitting, described in detail in Supporting Information Section S4, is performed in order to extract the peaks amplitudes $I_i$, linewidths $\Gamma_i$ and positions $x_{0,i}$, where $i$=A,B (see Section S4 for the detailed list of values).
In order to assess the origin of these two components, we perform fluence-dependent and time-resolved ($t_2$ dependence) 2DES at $T = 200$ K. In Figure \ref{fig3: Exciton}e, we report the peak amplitudes (integrated areas along the anti-diagonal line-cut) as a function of the incident light fluence $F$. The power law $I_i$=$cF^{n_i}$ is fit to the data to determine possible non-linearities, $c$ being a free coefficient. We observe that the peak amplitudes clearly display two different behaviours: $I_{\mathrm{A}}$ scales sub-linearly with the fluence ($n_{\mathrm{A}}$ = 0.59 $\pm$ 0.09) whereas $I_{\mathrm{B}}$ follows a superlinear power law ($n_{\mathrm{B}}$ = 1.19 $\pm$ 0.06), before saturating at large excitation intensity. From the $t_2$ scan between 0 and 450 fs (Figure \ref{fig3: Exciton}f), we observe a fast decay of $I_{\mathrm{A}}(t_2)$, counterbalanced by an increase of $I_{\mathrm{B}}(t_2)$. 
This behaviour indicates that spectral weight transfers from peak A to B within tens of femtoseconds. To quantitatively analyze the $I_i$ dynamics, we fit the data using the following exponential laws: $I_{\mathrm{A}}(t_2) = I_{\mathrm{A}0} \exp(-t_2/\tau_{\mathrm{A}}) + I_{\mathrm{A \infty}}$ and $I_{\mathrm{B}}(t_2)=I_{\mathrm{B}0} + K [1-\exp(-t_2/\tau_{\mathrm{B}})]$. The best fit to the data returns $\tau_{\mathrm{A}}$ = $\tau_{\mathrm{B}}$ = 39 $\pm$ 6 fs and $K\simeq I_{\mathrm{A}0}$ thus indicating the same time constant for the two processes and the conservation of the total spectral weight ($I_{\mathrm{A}}(t_2)$+$I_{\mathrm{B}}(t_2)$=$constant$ for any delay $t_2$). The constant $I_{\mathrm{A \infty}} $ in $I_{\mathrm{A}}(t_2)$ accounts for a slow exciton decay occurring over a timescale that is longer than the $t_2$ range scanned here. Investigation of the dynamics of peak B on timescales longer than 500 fs is hindered by the appearance of the strong and broad off-diagonal structure. These results indicate that peak B corresponds to a new bound state that is formed following the photoinjection of excitons. One possible scenario is that peak B arises from the formation of bi-excitons that can live on a longer timescale \cite{aneesh2017ultrafast}. Occupation of the bi-exciton state, i.e. a bound state originating from the interaction between two excitons and resulting in a lower energy state, gives rise to a transient signal at a probe photon energy smaller than the excitation energy, which corresponds to an off-diagonal signal at a distance from the diagonal equal to the exciton-exciton binding energy. The energy separation between peak A and B, i.e.  $x_{0,\mathrm{A}}$-$x_{0,\mathrm{B}}$, therefore provides the exciton-exciton interaction strength, $\delta E_{bi-exc}$. The estimated value, $\delta E_{bi-exc}= 29 \pm 4 $ meV, is compatible with the bi-exciton binding energy reported in the literature for similar systems \cite{cho2021luminescence,huang2020inhomogeneous,shulenberger2019setting}. In order to further support this picture, the fluence dependence in Fig. \ref{fig3: Exciton}e and the dynamics in Fig.  \ref{fig3: Exciton}f are compared with numerical solutions of exciton and bi-exciton rate equations, discussed in Supporting Information Section S3. The rate equation model shows that both the experimental $t_2$-dynamics and the power-law exponents of peak A and peak B are compatible with the behaviour expected for exciton decay and bi-exciton formation originating from exciton-exciton interaction. More specifically, the rate equations model that reproduces the experimental trends, provides an out-of-equilibrium scenario where photo-injected excitons are subject to multiple decay channels, as sketched in Figure \ref{fig3: Exciton}g. The excitonic population decays at a rate $\gamma_x$, which effectively includes all the direct depopulation channels, including relaxation to the ground state, formation of trap states and any mechanism that changes the exciton density; at the same time, exciton-exciton interaction leads to the formation of the bi-exciton bound states, which takes place at a rate $\beta N_x$, where $N_x$ is the instantaneous number of excitons and $\beta$ is a fit parameter. The parameters that better match the experimental findings suggest that direct excitonic decay takes place on a timescale of the order of $\gamma_x^{-1}=$ 100 fs, whereas the bi-exciton formation rate, which depends on the product $\beta N_x$ ($N_x$ being the instantaneous number of excitons) strongly depends on the time considered. In the early stage dynamics (10-50 fs), the bi-excitonic production characteristic time ranges from $1/\beta N_x =$ 10 to 70 fs. This indicates that the generation of bi-excitonic bound states occurs already within the 30 fs duration of both the pump and probe pulses. The formation of bi-excitons necessarily corresponds to fifth-order or higher nonlinear signals, since the generation of each exciton requires two field-matter interactions. In order to represent such a process, we describe the system with a density matrix that can describe two independent exciton states, $|eg\rangle\langle eg|$ and $|ge\rangle\langle ge|$ (the labels $e$ and $g$ indicate exciton and ground state, respectively), which account for the third order response, as well as two interacting exciton states, $|ee\rangle\langle ee|$, which account for the fifth-order response \cite{dostal2018direct,kriete2019interplay}. The interacting excitons can then form the bi-exciton state, which is represented by $|bg\rangle\langle bg|$ (the label $b$ refers to bi-exciton) and lead to a red-shifted signal due to the bi-exciton binding energy. Figure 3h shows a possible rephasing Feynman diagram representing this process that is phase-matched in the direction of the probe pulses. The pathway displayed in Fig. 3h corresponds to a stimulated emission signal, in agreement with 2DES experiments in transmission geometry (Supporting Information Section S6), which, despite being noisier, show a transient increase of transmission in the energy region corresponding to peak B.

\begin{figure*}[]
\includegraphics[width=11cm]{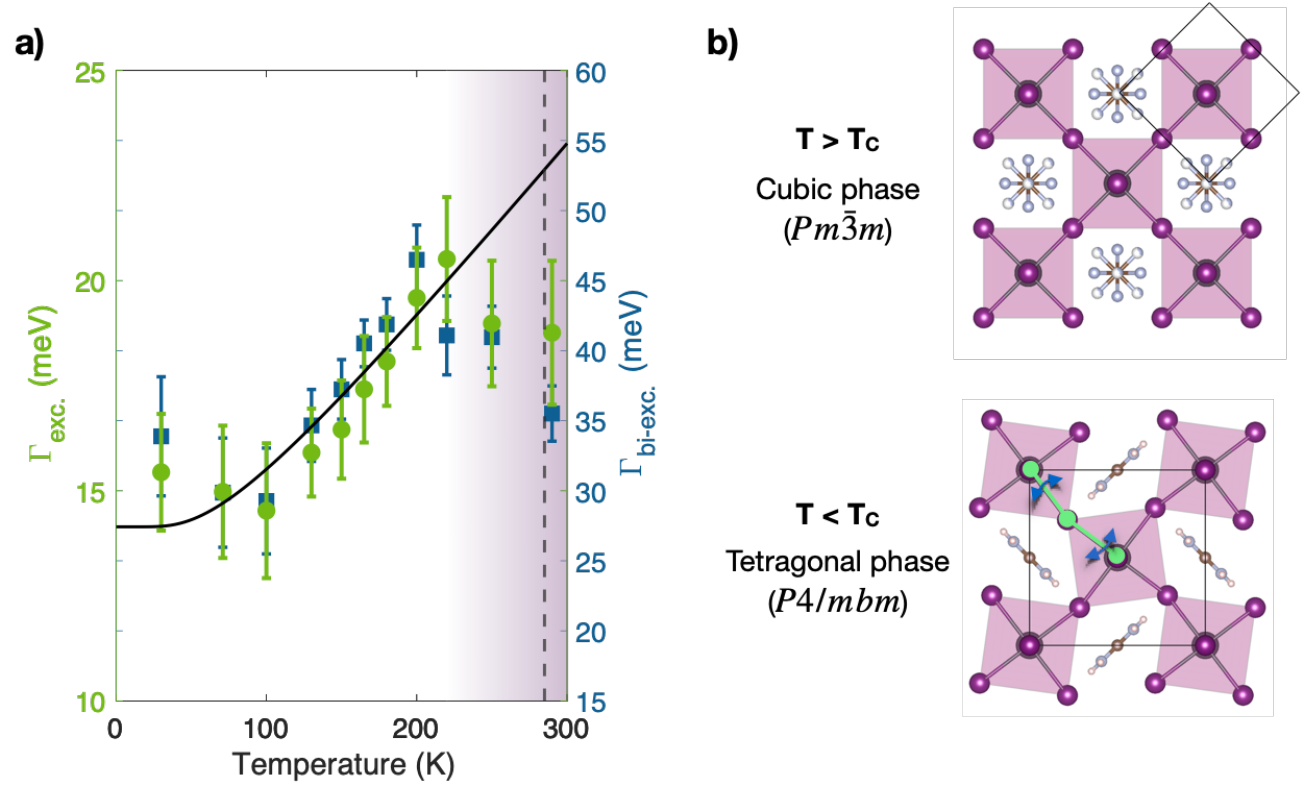}
\caption{a) Temperature dependence of the exciton (green circles, left axis) and bi-exciton (blue squares, right axis) homogeneous linewidths extracted from simultaneous fitting of diagonal and anti-diagonal profiles at short time delay t$_2$. The black solid line represents the linewidth broadening originating from interaction with an optical phonon mode at $E_{OP}$ = 17 meV. The purple shaded area represents the temperature range where FAPI nanocubes have cubic lattice structure and where the deviation from the expected temperature-dependent trend is observed. b) Top panel: FAPI crystal structure in the high-temperature cubic phase, with the organic cation subject to orientational disorder. Bottom panel: low-temperature tetragonal phase of FAPI perovskite. The black squares highlight the unit cell. The green lines and blue arrows indicate the I-Pb-I bending mode that softens in the proximity of the second-order tetragonal-to-cubic phase transition.}
\label{fig4: temperature}
\end{figure*}

The homogeneous linewidth $\Gamma_i$ of exciton and bi-exciton is estimated by lineshape analysis at short time delay t$_2$. The degree of inhomogeneity of the excitonic resonance under investigation may result in an anti-diagonal width that depends not only on the homogenous broadening but also on inhomogeneous broadening effects \cite{siemens2010resonance}. Therefore, an accurate estimate of the homogenous contribution calls for a simultaneous fit of diagonal and anti-diagonal slices of the 2D profile \cite{siemens2010resonance,russo2022dephasing}. The fitting functions for diagonal and anti-diagonal slices of the 2D maps can be obtained from the solutions of the optical Bloch equations for a two-level systems with Gaussian inhomogeneous broadening \cite{siemens2010resonance}, as discussed in Supporting Information Section S4. By considering both repahsing and non-repahsing signals, the purely absorptive lineshape for the anti-diagonal direction at $t_2 = 0$ is given by the real part of
\begin{widetext}
\begin{equation}
\begin{split}
    S^{R}_{\omega_0}(\omega) + S^{NR}_{\omega_0}(\omega) = &\frac{1}{\sigma(\Gamma-i \omega)} \exp{\left[\frac{(\Gamma-i \omega)^2}{2 \sigma^2}\right]} \erfc{\left[\frac{\Gamma - i \omega}{\sqrt{2}\sigma}\right]} \\
    &+\frac{e^{\Gamma^2/2\sigma^2}}{\sqrt{2}\sigma i}\frac{1}{\omega} e^{-\omega^2/2\sigma^2} \left[ e^{-i \omega\Gamma / \sigma^2} \erfc{\left( \frac{\Gamma - i \omega}{\sqrt{2}\sigma}\right)} - e^{i \omega\Gamma / \sigma^2} \erfc{\left( \frac{\Gamma + i \omega}{\sqrt{2}\sigma}\right)}\right]
\end{split}
\end{equation}
\end{widetext}
where $\omega$ is the frequency axis along the anti-diagonal direction of the 2D spectrum, $\omega_0$ is the absorption peak central frequency, $\Gamma$ is the dephasing rate, $\sigma$ is the inhomogeneous width, and $\erfc$ is the complementary error function. Similarly, the diagonal lineshape is the real part of
\begin{equation}
\begin{split}
    S^{R}_{\omega_0}(\omega') + S^{NR}_{\omega_0}(\omega') &= \\
     \sqrt{\frac{2}{\pi \sigma^2}} e^{-\omega'^2/2 \sigma^2} \ast \frac{1}{\Gamma^2 + \omega'^2} +  & \frac{1}{ \sigma} e^{-\omega'^2/2\sigma^2} \ast \frac{1}{(\Gamma+i\omega')^2}
    \end{split}
\end{equation}
with $\omega'$ frequency axis along the diagonal direction and $\ast$ indicating the convolution. The simultaneous fit of diagonal and anti-diagonal profiles allows to estimate the parameters $\sigma$ and $\Gamma$. Here, we employed this fitting procedure as detailed in Supporting Information Sec. S4, which allows us to disentangle, from the measured 2D spectra, the homogeneous linewidth contribution for both exciton and bi-exciton resonances.  In the following, we will adopt the notation $\Gamma_\mathrm{A}=\Gamma_{exc}$ and $\Gamma_\mathrm{B}=\Gamma_{bi-exc}$ to directly link to the physical origin of the two structures. The extracted values are plotted in Figure \ref{fig4: temperature}a as a function of sample temperature $T$. At cryogenic temperature, we estimate $\Gamma_{exc}=15 \pm 2$ meV (obtained from the average value of the three measurements at $T \leq 100$ K), which corresponds to a pure decoherence time $\tau_{{dech}}=44 \pm 5$ fs, where $\tau_{{dech}}$=$\hbar/\Gamma_{exc}$. This value is of the same order of the $\tau_{\mathrm{A}}\simeq40$ fs depopulation time, thus suggesting that the leading mechanism determining the homogenous linewidth is the decay into bi-excitons. The coherent propagation of the excitons is thus limited, even at cryogenic temperatures, by their ultra-short lifetime, preventing the development of delocalized states and superradiant coherent states similar to those observed in inorganic halide perovskite superlattices \cite{raino2018superfluorescence,mattiotti2020thermal,Milloch_NanoLett}. For the bi-excitonic state, we obtain $\Gamma_{bi-exc.} = 31 \pm 4$ meV between $T= 30$ K and $T = 100$ K, corresponding to $\tau_{{dech}}=21 \pm 3$ fs. In this case, the decoherence timescale is much shorter than the bi-exciton depopulation time, thus indicating the formation of a strongly incoherent gas of interacting bi-excitons, surviving for hundreds of femtoseconds.

The temperature dependence of both $\Gamma_{exc.}$ and $\Gamma_{bi-exc.}$ is reported in Fig. \ref{fig4: temperature}a (green circles and blue squares, respectively). We observe a similar temperature-dependent linewidth broadening, which suggests that at higher temperatures the decoherence process is accelerated by the coupling to the thermal bath, consisting of optical phonons.
The temperature-dependent increase of $\Gamma_{exc}(T)$ is compared to the thermal broadening (black solid line in Fig. \ref{fig4: temperature}a) expected for coupling to longitudinal optical phonons, which is described by the function $\Gamma_{exc}(T)=\Gamma_{exc,0} + \Gamma_{OP} / [\exp(E_{OP}/k_BT)-1]$, where $\Gamma_{exc,0}$ is the intrinsic linewidth at low-temperature, $\Gamma_{OP}$ is the exciton-optical phonon coupling coefficient and $E_{OP}$ is the optical phonon energy \cite{lee1986luminescence}. According to literature reports, FAPI displays optical phonon modes at energy $E_{OP}$ between 15 meV and 19 meV \cite{fu2018unraveling,ferreira2020direct,fang2016photoexcitation,cho2021luminescence,yuan2017high}, which have been ascribed to vibrations of the organic cation with respect to the inorganic cage \cite{quarti2013raman,ferreira2020direct}. These libration modes of the organic cation have also been reported to affect the photoluminescence intensity and the lifetime of photoexcited states in MAPI perovskite \cite{park2017critical}. Interaction with such vibrational modes fully accounts for the homogeneous broadening measured in the present work for $T <  250$ K. In this temperature range, fitting of $\Gamma_{exc}(T)$ returns $\Gamma_{exc,0} = 14 \pm 1$ meV and $\Gamma_{OP} = 8 \pm 2$ meV when $E_{OP}$ is fixed to values between 15 meV to 19 meV.  

The temperature-dependent trend suggests that at low temperatures the exciton coherence is mainly limited by the bi-exciton formation, while at higher temperatures the scattering with thermally activated optical phonons also contributes to faster decoherence.
Interestingly, a clear deviation from the trend expected for scattering with thermally activated phonons is observed for temperatures higher than $T^* \simeq 250$ K, where the measured homogeneous linewidth is smaller than what is expected due to thermal broadening in this temperature range. $T^*$ is very close to the critical temperature for the symmetry-breaking structural phase transition undergone by FAPI. The transition from the high-temperature cubic phase (space group $Pm\bar{3}m$) to the low-$T$ tetragonal phase (space group $P4/mbm$) occurs at $T_{\mathrm{c}}^{bulk} \simeq$ 285 K in the bulk material \cite{fabini2016reentrant, weber2018phase}, and at smaller temperature in perovskite nanocubes \cite{liu2019size}. For 10 nm nanocubes we can expect a $\sim 15\%$ decrease in $T_{c}$, that corresponds to $\sim$ 240 K. Therefore, the anomalous behaviour in $\Gamma_{exc}(T)$ emerging at 250 K suggests a suppression of the exciton-phonon interaction in the high-temperature structural cubic phase. More specifically, we note that the second-order structural phase transition from tetragonal to cubic at $T_{c}$ is accompanied by a symmetry increase, as shown in Fig. \ref{fig4: temperature}b.  The reduction of the unit cell in the high-$T$ cubic phase implies the reduction of the number of optical phonon modes from 36 to 18 \cite{maalej1997phase}, thus limiting the phononic channels at $E_{OP}\simeq$17 meV available for the excitonic decoherence.

 \begin{figure*}[]
\includegraphics[width=15.5cm]{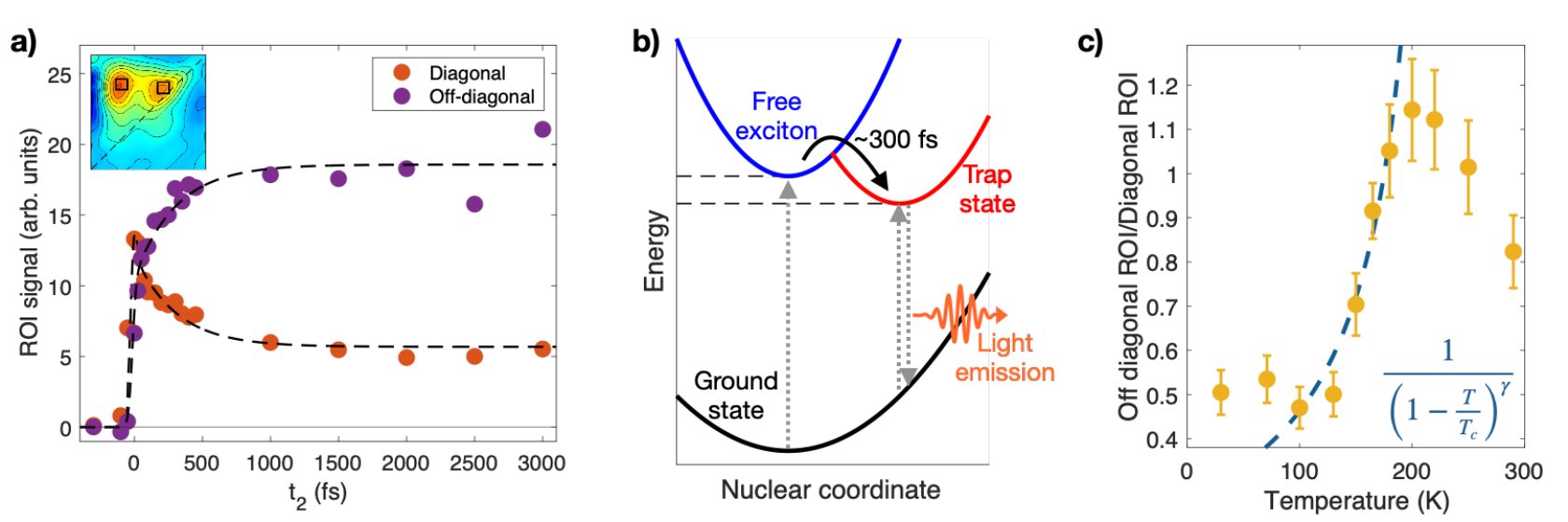}
\caption{ a) $t_2$ dynamics of the 2DES signal measured at 200 K and integrated over the two regions of interest depicted in the top left inset, which select the diagonal (orange) and off-diagonal (purple) structures.  b) Sketch of the energy level structure and population dynamics of trap states originating from defects or self-trapped excitons. c) Temperature dependence of the off-diagonal spectral feature at $t_2 = 0$ fs as compared to the diagonal excitonic resonance. The yellow markers report the ratio between the signal intensities obtained from the integration of the 2D spectra over the areas indicated in the inset of Fig. 5a. The blue dashed line indicates the divergence of a second-order phase transition order parameter scaling as $(1-T/T_c)^{-\gamma}$; here $\gamma$ = 1 and $T_c $ = 240 K.}
\label{fig5: STE}
\end{figure*}

2DES further snaps the subsequent formation of the long-lived states that are responsible for the slower emission of light under the form of fluorescence (see photoluminescence spectrum in Supporting Information Section S1-D). For this aim, we analyze the off-diagonal structure that appears in the 2D spectra shown in Fig. \ref{fig2: 2D} after a few picoseconds at $\hbar\omega_1 \simeq1.72$ eV and $\hbar\omega_3 \simeq1.61$ eV coordinates. This off-diagonal feature can not be associated with multi-exciton states such as bi-excitons because of the large energy separation from the diagonal structure (100 meV) and the absence of super-linear fluence dependence (see Supporting Information Section S5). Interpretation of the two spectral features (diagonal and off-diagonal) as originating from a charge transfer between two co-existing crystal structure phases, as previously done in Ref. \citenum{phuong2016free} for MAPI thin films, is incompatible with XRD data showing tetragonal lattice at low-$T$ and purely cubic crystal structure at room temperature (see Supporting Information Section S1-A). Significant information on the nature of this low-energy response can be obtained from its dynamics compared to the excitonic one. Fig. \ref{fig5: STE}a displays the dynamics of the two different spectral features obtained by integrating the 2D spectra over two different selected regions of interest, one centered on top of the diagonal and the other centered on the off-diagonal structures (see inset of Figure \ref{fig5: STE}a). The off-diagonal structure shows a $\sim$ 300 fs build-up time, which coincides with the decay of the diagonal signal, as evidenced by the exponential fit reported in Fig. \ref{fig5: STE}a. This finding demonstrates that the off-diagonal structure is a long-lived state that forms after the relaxation of the initial excitonic population. This long-lived state is in turn responsible for the slower light emission, which takes place several tens of picoseconds after the initial photoexcitation \cite{phuong2016free,raino2018superfluorescence}. We note that the dynamics reported in Figure \ref{fig3: Exciton}f are compatible with what is reported in Fig. \ref{fig5: STE}a because in the former case, the amplitude is obtained from integration over the whole anti-diagonal spectral range, whereas in the latter the integration was performed on a fixed squared area where the diagonal peak broadening caused by the rise of the bi-exciton component is not taken into account. This transfer of occupied states from free excitons to lower energy states is typical of trap states that get populated following the mechanism sketched in Figure \ref{fig5: STE}b \cite{wright2017band,smith2018white,grandhi2023role}. Such trap states can originate when the coupling between charge carriers and a local structural distortion of the lattice decreases the energy of the system.

One possible origin of trap states is related to the presence of static intrinsic defects, such as local lattice perturbations, structural disorder and vacancies \cite{grandhi2023role,luo2018efficient}. 
Another possibility is that the off-diagonal structure is associated to the formation of metastable STE, i.e. localized excitons dressed by a phonon mode, most likely involving the I-Pb-I bending. Indeed, we note that the build-up time (300 fs) of the off-diagonal structure is very close to half of the period of the optical modes at $\simeq7$ meV, which involve the halide-lead-halide bending (see Fig. \ref{fig4: temperature}b) \cite{quarti2013raman} that has been shown to be closely coupled with photoexcited excitons \cite{yazdani2023coupling}. In general, STE formation is facilitated in solid systems displaying soft lattice, strong exciton-phonon coupling, large lattice constants and reduced electronic dimensionality \cite{tan2022self}. These properties are often found in halide perovskites -  which typically display strong electron-phonon interactions and soft lattice \cite{yamada2022electron} - and in FAPI superlattices in particular, where the formamidinium cation represents a large cation molecule as compared to MAPI or all-metal halide perovskites \cite{tan2020temperature}, and where the nanocube synthesis reduces the dimensionality of the system to quasi 0D due to exciton quantum confinement. STE usually results in broad emission peaks \cite{tan2022self}, consistently with what is observed from room temperature photo-luminescence on the FAPI superlattices investigated in this work (see Supporting Information Section S1-D).
The formation of STE is also suggested by the temperature dependence of the off-diagonal structure, as reported in Fig. \ref{fig5: STE}c for fixed time delay $t_2$ = 0 fs. Starting from cryogenic temperature we clearly observe a significant increase of the intensity of the off-diagonal peak as the temperature is increased, with tendency to diverge when the critical temperature of the tetragonal to cubic transition is approached. The experimental data are fitted with the function (1-$T/T_{c}$)$^{\gamma}$ in order to determine the critical exponent. The data are compatible with $\gamma=1$, which is the mean-field critical exponent of compressibility across a second-order phase transition. This result suggests that the softening of the lattice and, more specifically, of the $\simeq$ 7 meV phonons responsible for the I-Pb-I bending, which softens across the tetragonal-to-cubic transition, strongly favours the formation of STE with a binding energy as large as $\simeq$ 100 meV. 
We also note that the 2D spectra contain a weak diagonal signal at $\hbar\omega_1=\hbar\omega_3 \simeq1.61$ eV (see Fig. S10 Supporting Information) which displays a temperature-dependent behaviour similar to the off-diagonal component, thus indicating the possibility of direct excitation of this phonon-mediated transition, which consists in the simultaneous absorption of a photon and the local lattice distortion related to the STE. The lower intensity of the diagonal component as compared to the off-diagonal peak could be related to the scaling of the signal intensity with the dipole strength. Since the diagonal signal scales with the fourth power of the dipole strength, while the off-diagonal cross peak scales with the second power, small dipole strength of the trap state compared to the main excitonic line can result in the suppression of diagonal features in the 2DES spectra, despite the off-diagonal peak being clearly visible \cite{tollerud2016revealing,camargo2019dark}. The spectrum of the laser excitation (see Supporting Information Fig. S8b) can also play a role in scaling down the intensity of the measured signal at small photon energies ($\hbar\omega_1 \sim$ 1.61 eV) \cite{tollerud2016revealing,camargo2017resolving}, thus further suppressing diagonal features.

\section{Conclusions}

In conclusion, our 2DES study has unveiled the complex hierarchy of the photo-physics in FAPI superlattices. The initial excitons very rapidly ($\simeq$ 40 fs) decay forming an incoherent gas of bi-excitons. Within hundreds of femtoseconds the optical excitations form trap states that originate from defects or lattice distortions causing self-trapping of excitons. The large binding energy and the spatial localization protect these new and long-lived states that are responsible for the subsequent fluorescent emission \cite{Neukirch2016,Zhu2016}. We note that a direct comparison with disordered FAPI nanocrystals cannot be performed due to the strong signal suppression in disordered films (see Supporting Information Section S7). The way the ordered arrangement of the nanocubes enhances the signal calls for further investigation that goes beyond the scope of the present work. Nevertheless, the present results suggest that the short excitonic decoherence time prevents the onset of collective superradiant emission, in contrast with observations in inorganic superlattices \cite{raino2018superfluorescence, zhou2020cooperative,krieg2020monodisperse,pashaei2023superradiance,Milloch_NanoLett}. At the same time, the strong coupling with the optical phonons, combined with the lattice softening across a continuous structural phase transition, offers a new tool to manipulate the excitonic/bi-excitonic coherent dynamics, as well as the down-conversion into trap-states, which ultimately controls the collective and coherent emission properties. 

\section{Materials and methods}
\subsection{Sample preparation}
FAPbI$_3$ (or FAPI) nanocubes (NCs) were synthesized following the procedure reported in Ref. \citenum{akkerman2018molecular} with little modifications.
A stock solution of oleylammonium iodide (OLAm-I) was prepared by loading 750 mg of I$_2$ (3 mmol), 10.5 ml of octadecene-1 (ODE, 90$\%$) and 4.5 ml of oleylamine (OLAm, 70 $\%$) into a 40 ml vial. The solution was dried under vacuum for 15 minutes at room temperature (RT), for 30 minutes at 90 °C and then heated up to 110 °C under N$_2$ flow. At this temperature, the solution turned transparent light brown, which indicates the formation of oleylammonium iodide. Before using it, the OLAm-I stock solution was vigorously stirred and heated at around 80 $\degree$C since it precipitates at RT.
In a 40 ml vial, 76 mg of lead acetate trihydrate (PbAc$_2\cdot$3H$_2$O, 0.2 mmol) and 62 mg of formamidinium acetate (FAAc, 0.6mmol) were combined with 5 ml of ODE and 3 ml of oleic acid (OA). The mixture was dried under vacuum at RT for 15 minutes and then at 125 °C for 30 minutes. Next, under N$_2$ flow, the FA-based mixture was brought down to 120 °C and 1.5 ml of pre-heated OLAm-I precursor solution were injected. The reaction mixture was cooled down at room temperature after 10 seconds with an ice-water bath.
The crude solution was centrifuged at 4000 rpm for 6 minutes and the supernatant was discarded. Then the solution was centrifuged again at 4000 rpm for 3 minutes and the remaining liquid was removed with a 100 \textmu l mechanical micropipette, and the step was repeated again, this time using a cotton swab to collect the liquid. The remaining solid was dissolved in 1 ml of toluene and centrifuged at 6000 rpm and filtered through 0.45 \textmu m hydrophobic PTFE syringe filter to eliminate any residual aggregates.
 
Nanocubes superlattices were prepared by slow solvent evaporation on 1 cm $\times$ 1 cm monocrystalline Si substrates cleaned by rinsing with methanol and acetone and dried with paper tissues and by blowing N$_2$. The substrates were placed inside a Petri dish and 30 \textmu l of stock solution of NCs were drop-casted on each of them, so to have a total amount of solution equal to $\approx$ 90 \textmu l. After the solvent was completely evaporated ($\approx$ 12 hours), films were considered ready for optical experiment and characterization (Supporting Information Section S1). 

\subsection{2DES optical setup}
A sketch of the 2DES setup is reported in Supporting Information Fig. S8a. Both pump and probe pulses are generated by a home-built non-collinear optical parametric amplifier (NOPA), seeded by a Yb:KGW laser system (Pharos by Light Conversion) emitting 300 fs pulses at 1030 nm (1.2 eV). The NOPA output signal is tuned at 1.7 eV central energy (see Fig. S7b) and compressed to 30 fs time duration (FWHM) by multiple bounces on a pair of chirped mirrors. Pump and probe beams are obtained from dividing the NOPA output by means of a beam splitter. The pair of phase-coherent excitation pulses are generated by a common-path birefringent interferometer (GEMINI 2D by NIREOS) based on the Translating-Wedge-Based Identical Pulses eNcoding System (TWINS) scheme \cite{brida2012phase}. A two-prism compressor compensates for the additional dispersion introduced in the pump pulses by the GEMINI 2D interferometer and other optical elements on the pump beam (beam splitter, half-waveplate and cryostat window). Similarly, a two-prism compressor is employed for fine compensation of dispersing elements on the probe beam, namely a half-waveplate and the cryostat window. Pump and probe pulses, orthogonally polarized in order to minimize pump scattering, are then focused onto the sample by two concave mirrors, resulting in a partially collinear 2DES scheme (collinear pump pulses and non-collinear probe propagation). The focused spot size is 170 \textmu m $\times$ 230 \textmu m for the pump beam and 70 \textmu m  $\times$ 70 \textmu m for the probe beam. The pump-probe time delay $t_2$ is controlled through a linearly motorized stage and is scanned between -500 fs and 4 ps.  The delay between the two pump pulses, $t_1$ is controlled by varying the insertion of the GEMINI 2D birefringent wedges on the beam path, and it is continuously scanned between -55 fs and 110 fs for each measured $t_2$.

The 2DES signal propagating collinearly with the probe beam is collected in reflection geometry. The transient reflectivity signal at each ($t_1$,$t_2$) is obtained by modulating the pump excitation with a mechanical chopper and performing lock-in acquisition of the reflected probe \cite{preda2016broadband}. Spectral resolution along the detection frequency axis is achieved by employing the GEMINI interferometer (NIREOS) that generates the probe interferogram and returns the transient reflectivity spectrum upon Fourier transform computation. 

The pump fluence $F$ can be continuously varied between 0 and 230 \textmu J/cm$^2$ (as measured at $t_1 = 0$) by rotating a half-waveplate positioned on the pump beam path before the GEMINI 2D, whose polarizers select the 45$\degree$ polarized component of the light. For the measurements presented in this work, the laser repetition rate was set to 40 kHz. A different repetition rate, selected by means of a pulse picker that allows keeping unvaried the energy per pulse, was employed for fluence-dependent studies where the repetition rate was increased up to 200 kHz as the excitation intensity was decreased, in order to maintain the average power at the sample constant throughout the fluence scan. 

Temperature-dependent investigations were performed by mounting the samples inside a closed-cycle helium cryostat that allows performing ultrafast optical spectroscopy experiments at temperatures between 20 K and 300 K.

\section*{acknowledgements}
C.G., P.F., A.M. and S.M. acknowledge financial support from MIUR through the PRIN 2017 (Prot. 20172H2SC4 005) and PRIN 2020 (Prot. 2020JLZ52N 003) programs and from the European Union - Next Generation EU through the MUR-PRIN2022 (Prot. 20228YCYY7) program. C.G., S.P. and G.F. acknowledge support from Universit\`a Cattolica del Sacro Cuore through D.1, D.2.2 and D.3.1 grants. S.M. acknowledges partial financial support through the grant "Finanziamenti ponte per bandi esterni" from Universit\`a Cattolica del Sacro Cuore. G.C. acknowledges financial support by the European Union’s NextGenerationEU Programme with the I-PHOQS Infrastructure [IR0000016, ID D2B8D520, CUP B53C22001750006] "Integrated infrastructure initiative in Photonic and Quantum Sciences".

\bibliography{Refs}

\begin{thebibliography}{70}%
\makeatletter
\providecommand \@ifxundefined [1]{%
 \@ifx{#1\undefined}
}%
\providecommand \@ifnum [1]{%
 \ifnum #1\expandafter \@firstoftwo
 \else \expandafter \@secondoftwo
 \fi
}%
\providecommand \@ifx [1]{%
 \ifx #1\expandafter \@firstoftwo
 \else \expandafter \@secondoftwo
 \fi
}%
\providecommand \natexlab [1]{#1}%
\providecommand \enquote  [1]{``#1''}%
\providecommand \bibnamefont  [1]{#1}%
\providecommand \bibfnamefont [1]{#1}%
\providecommand \citenamefont [1]{#1}%
\providecommand \href@noop [0]{\@secondoftwo}%
\providecommand \href [0]{\begingroup \@sanitize@url \@href}%
\providecommand \@href[1]{\@@startlink{#1}\@@href}%
\providecommand \@@href[1]{\endgroup#1\@@endlink}%
\providecommand \@sanitize@url [0]{\catcode `\\12\catcode `\$12\catcode
  `\&12\catcode `\#12\catcode `\^12\catcode `\_12\catcode `\%12\relax}%
\providecommand \@@startlink[1]{}%
\providecommand \@@endlink[0]{}%
\providecommand \url  [0]{\begingroup\@sanitize@url \@url }%
\providecommand \@url [1]{\endgroup\@href {#1}{\urlprefix }}%
\providecommand \urlprefix  [0]{URL }%
\providecommand \Eprint [0]{\href }%
\providecommand \doibase [0]{https://doi.org/}%
\providecommand \selectlanguage [0]{\@gobble}%
\providecommand \bibinfo  [0]{\@secondoftwo}%
\providecommand \bibfield  [0]{\@secondoftwo}%
\providecommand \translation [1]{[#1]}%
\providecommand \BibitemOpen [0]{}%
\providecommand \bibitemStop [0]{}%
\providecommand \bibitemNoStop [0]{.\EOS\space}%
\providecommand \EOS [0]{\spacefactor3000\relax}%
\providecommand \BibitemShut  [1]{\csname bibitem#1\endcsname}%
\let\auto@bib@innerbib\@empty
\bibitem [{\citenamefont {Herz}(2016)}]{herz2016charge}%
  \BibitemOpen
  \bibfield  {author} {\bibinfo {author} {\bibfnamefont {L.~M.}\ \bibnamefont
  {Herz}},\ }\bibfield  {title} {\bibinfo {title} {Charge-carrier dynamics in
  organic-inorganic metal halide perovskites},\ }\href@noop {} {\bibfield
  {journal} {\bibinfo  {journal} {Annual review of physical chemistry}\
  }\textbf {\bibinfo {volume} {67}},\ \bibinfo {pages} {65} (\bibinfo {year}
  {2016})}\BibitemShut {NoStop}%
\bibitem [{\citenamefont {Kanemitsu}(2017)}]{kanemitsu2017luminescence}%
  \BibitemOpen
  \bibfield  {author} {\bibinfo {author} {\bibfnamefont {Y.}~\bibnamefont
  {Kanemitsu}},\ }\bibfield  {title} {\bibinfo {title} {Luminescence
  spectroscopy of lead-halide perovskites: materials properties and application
  as photovoltaic devices},\ }\href@noop {} {\bibfield  {journal} {\bibinfo
  {journal} {Journal of Materials Chemistry C}\ }\textbf {\bibinfo {volume}
  {5}},\ \bibinfo {pages} {3427} (\bibinfo {year} {2017})}\BibitemShut
  {NoStop}%
\bibitem [{\citenamefont {Quan}\ \emph {et~al.}(2019)\citenamefont {Quan},
  \citenamefont {Rand}, \citenamefont {Friend}, \citenamefont {Mhaisalkar},
  \citenamefont {Lee},\ and\ \citenamefont {Sargent}}]{quan2019perovskites}%
  \BibitemOpen
  \bibfield  {author} {\bibinfo {author} {\bibfnamefont {L.~N.}\ \bibnamefont
  {Quan}}, \bibinfo {author} {\bibfnamefont {B.~P.}\ \bibnamefont {Rand}},
  \bibinfo {author} {\bibfnamefont {R.~H.}\ \bibnamefont {Friend}}, \bibinfo
  {author} {\bibfnamefont {S.~G.}\ \bibnamefont {Mhaisalkar}}, \bibinfo
  {author} {\bibfnamefont {T.-W.}\ \bibnamefont {Lee}},\ and\ \bibinfo {author}
  {\bibfnamefont {E.~H.}\ \bibnamefont {Sargent}},\ }\bibfield  {title}
  {\bibinfo {title} {Perovskites for next-generation optical sources},\
  }\href@noop {} {\bibfield  {journal} {\bibinfo  {journal} {Chemical reviews}\
  }\textbf {\bibinfo {volume} {119}},\ \bibinfo {pages} {7444} (\bibinfo {year}
  {2019})}\BibitemShut {NoStop}%
\bibitem [{\citenamefont {Stranks}\ and\ \citenamefont
  {Snaith}(2015)}]{stranks2015metal}%
  \BibitemOpen
  \bibfield  {author} {\bibinfo {author} {\bibfnamefont {S.~D.}\ \bibnamefont
  {Stranks}}\ and\ \bibinfo {author} {\bibfnamefont {H.~J.}\ \bibnamefont
  {Snaith}},\ }\bibfield  {title} {\bibinfo {title} {Metal-halide perovskites
  for photovoltaic and light-emitting devices},\ }\href@noop {} {\bibfield
  {journal} {\bibinfo  {journal} {Nature nanotechnology}\ }\textbf {\bibinfo
  {volume} {10}},\ \bibinfo {pages} {391} (\bibinfo {year} {2015})}\BibitemShut
  {NoStop}%
\bibitem [{\citenamefont {Shamsi}\ \emph {et~al.}(2019)\citenamefont {Shamsi},
  \citenamefont {Urban}, \citenamefont {Imran}, \citenamefont {De~Trizio},\
  and\ \citenamefont {Manna}}]{shamsi2019metal}%
  \BibitemOpen
  \bibfield  {author} {\bibinfo {author} {\bibfnamefont {J.}~\bibnamefont
  {Shamsi}}, \bibinfo {author} {\bibfnamefont {A.~S.}\ \bibnamefont {Urban}},
  \bibinfo {author} {\bibfnamefont {M.}~\bibnamefont {Imran}}, \bibinfo
  {author} {\bibfnamefont {L.}~\bibnamefont {De~Trizio}},\ and\ \bibinfo
  {author} {\bibfnamefont {L.}~\bibnamefont {Manna}},\ }\bibfield  {title}
  {\bibinfo {title} {Metal halide perovskite nanocrystals: synthesis,
  post-synthesis modifications, and their optical properties},\ }\href@noop {}
  {\bibfield  {journal} {\bibinfo  {journal} {Chemical reviews}\ }\textbf
  {\bibinfo {volume} {119}},\ \bibinfo {pages} {3296} (\bibinfo {year}
  {2019})}\BibitemShut {NoStop}%
\bibitem [{\citenamefont {Jurow}\ \emph {et~al.}(2017)\citenamefont {Jurow},
  \citenamefont {Lampe}, \citenamefont {Penzo}, \citenamefont {Kang},
  \citenamefont {Koc}, \citenamefont {Zechel}, \citenamefont {Nett},
  \citenamefont {Brady}, \citenamefont {Wang}, \citenamefont {Alivisatos} \emph
  {et~al.}}]{jurow2017tunable}%
  \BibitemOpen
  \bibfield  {author} {\bibinfo {author} {\bibfnamefont {M.~J.}\ \bibnamefont
  {Jurow}}, \bibinfo {author} {\bibfnamefont {T.}~\bibnamefont {Lampe}},
  \bibinfo {author} {\bibfnamefont {E.}~\bibnamefont {Penzo}}, \bibinfo
  {author} {\bibfnamefont {J.}~\bibnamefont {Kang}}, \bibinfo {author}
  {\bibfnamefont {M.~A.}\ \bibnamefont {Koc}}, \bibinfo {author} {\bibfnamefont
  {T.}~\bibnamefont {Zechel}}, \bibinfo {author} {\bibfnamefont
  {Z.}~\bibnamefont {Nett}}, \bibinfo {author} {\bibfnamefont {M.}~\bibnamefont
  {Brady}}, \bibinfo {author} {\bibfnamefont {L.-W.}\ \bibnamefont {Wang}},
  \bibinfo {author} {\bibfnamefont {A.~P.}\ \bibnamefont {Alivisatos}}, \emph
  {et~al.},\ }\bibfield  {title} {\bibinfo {title} {Tunable anisotropic photon
  emission from self-organized \ch{CsPbBr_3} perovskite nanocrystals},\
  }\href@noop {} {\bibfield  {journal} {\bibinfo  {journal} {Nano letters}\
  }\textbf {\bibinfo {volume} {17}},\ \bibinfo {pages} {4534} (\bibinfo {year}
  {2017})}\BibitemShut {NoStop}%
\bibitem [{\citenamefont {Tong}\ \emph {et~al.}(2018)\citenamefont {Tong},
  \citenamefont {Yao}, \citenamefont {Manzi}, \citenamefont {Bladt},
  \citenamefont {Wang}, \citenamefont {D{\"o}blinger}, \citenamefont {Bals},
  \citenamefont {M{\"u}ller-Buschbaum}, \citenamefont {Urban}, \citenamefont
  {Polavarapu} \emph {et~al.}}]{tong2018spontaneous}%
  \BibitemOpen
  \bibfield  {author} {\bibinfo {author} {\bibfnamefont {Y.}~\bibnamefont
  {Tong}}, \bibinfo {author} {\bibfnamefont {E.-P.}\ \bibnamefont {Yao}},
  \bibinfo {author} {\bibfnamefont {A.}~\bibnamefont {Manzi}}, \bibinfo
  {author} {\bibfnamefont {E.}~\bibnamefont {Bladt}}, \bibinfo {author}
  {\bibfnamefont {K.}~\bibnamefont {Wang}}, \bibinfo {author} {\bibfnamefont
  {M.}~\bibnamefont {D{\"o}blinger}}, \bibinfo {author} {\bibfnamefont
  {S.}~\bibnamefont {Bals}}, \bibinfo {author} {\bibfnamefont {P.}~\bibnamefont
  {M{\"u}ller-Buschbaum}}, \bibinfo {author} {\bibfnamefont {A.~S.}\
  \bibnamefont {Urban}}, \bibinfo {author} {\bibfnamefont {L.}~\bibnamefont
  {Polavarapu}}, \emph {et~al.},\ }\bibfield  {title} {\bibinfo {title}
  {Spontaneous self-assembly of perovskite nanocrystals into electronically
  coupled supercrystals: toward filling the green gap},\ }\href@noop {}
  {\bibfield  {journal} {\bibinfo  {journal} {Advanced Materials}\ }\textbf
  {\bibinfo {volume} {30}},\ \bibinfo {pages} {1801117} (\bibinfo {year}
  {2018})}\BibitemShut {NoStop}%
\bibitem [{\citenamefont {Baranov}\ \emph {et~al.}(2019)\citenamefont
  {Baranov}, \citenamefont {Toso}, \citenamefont {Imran},\ and\ \citenamefont
  {Manna}}]{baranov2019investigation}%
  \BibitemOpen
  \bibfield  {author} {\bibinfo {author} {\bibfnamefont {D.}~\bibnamefont
  {Baranov}}, \bibinfo {author} {\bibfnamefont {S.}~\bibnamefont {Toso}},
  \bibinfo {author} {\bibfnamefont {M.}~\bibnamefont {Imran}},\ and\ \bibinfo
  {author} {\bibfnamefont {L.}~\bibnamefont {Manna}},\ }\bibfield  {title}
  {\bibinfo {title} {Investigation into the photoluminescence red shift in
  cesium lead bromide nanocrystal superlattices},\ }\href@noop {} {\bibfield
  {journal} {\bibinfo  {journal} {The Journal of Physical Chemistry Letters}\
  }\textbf {\bibinfo {volume} {10}},\ \bibinfo {pages} {655} (\bibinfo {year}
  {2019})}\BibitemShut {NoStop}%
\bibitem [{\citenamefont {Cherniukh}\ \emph {et~al.}(2021)\citenamefont
  {Cherniukh}, \citenamefont {Rain{\`o}}, \citenamefont {St{\"o}ferle},
  \citenamefont {Burian}, \citenamefont {Travesset}, \citenamefont {Naumenko},
  \citenamefont {Amenitsch}, \citenamefont {Erni}, \citenamefont {Mahrt},
  \citenamefont {Bodnarchuk} \emph {et~al.}}]{Cherniukh2021}%
  \BibitemOpen
  \bibfield  {author} {\bibinfo {author} {\bibfnamefont {I.}~\bibnamefont
  {Cherniukh}}, \bibinfo {author} {\bibfnamefont {G.}~\bibnamefont
  {Rain{\`o}}}, \bibinfo {author} {\bibfnamefont {T.}~\bibnamefont
  {St{\"o}ferle}}, \bibinfo {author} {\bibfnamefont {M.}~\bibnamefont
  {Burian}}, \bibinfo {author} {\bibfnamefont {A.}~\bibnamefont {Travesset}},
  \bibinfo {author} {\bibfnamefont {D.}~\bibnamefont {Naumenko}}, \bibinfo
  {author} {\bibfnamefont {H.}~\bibnamefont {Amenitsch}}, \bibinfo {author}
  {\bibfnamefont {R.}~\bibnamefont {Erni}}, \bibinfo {author} {\bibfnamefont
  {R.~F.}\ \bibnamefont {Mahrt}}, \bibinfo {author} {\bibfnamefont {M.~I.}\
  \bibnamefont {Bodnarchuk}}, \emph {et~al.},\ }\bibfield  {title} {\bibinfo
  {title} {Perovskite-type superlattices from lead halide perovskite
  nanocubes},\ }\href@noop {} {\bibfield  {journal} {\bibinfo  {journal}
  {Nature}\ }\textbf {\bibinfo {volume} {593}},\ \bibinfo {pages} {535}
  (\bibinfo {year} {2021})}\BibitemShut {NoStop}%
\bibitem [{\citenamefont {Rain{\`o}}\ \emph {et~al.}(2018)\citenamefont
  {Rain{\`o}}, \citenamefont {Becker}, \citenamefont {Bodnarchuk},
  \citenamefont {Mahrt}, \citenamefont {Kovalenko},\ and\ \citenamefont
  {St{\"o}ferle}}]{raino2018superfluorescence}%
  \BibitemOpen
  \bibfield  {author} {\bibinfo {author} {\bibfnamefont {G.}~\bibnamefont
  {Rain{\`o}}}, \bibinfo {author} {\bibfnamefont {M.~A.}\ \bibnamefont
  {Becker}}, \bibinfo {author} {\bibfnamefont {M.~I.}\ \bibnamefont
  {Bodnarchuk}}, \bibinfo {author} {\bibfnamefont {R.~F.}\ \bibnamefont
  {Mahrt}}, \bibinfo {author} {\bibfnamefont {M.~V.}\ \bibnamefont
  {Kovalenko}},\ and\ \bibinfo {author} {\bibfnamefont {T.}~\bibnamefont
  {St{\"o}ferle}},\ }\bibfield  {title} {\bibinfo {title} {Superfluorescence
  from lead halide perovskite quantum dot superlattices},\ }\href@noop {}
  {\bibfield  {journal} {\bibinfo  {journal} {Nature}\ }\textbf {\bibinfo
  {volume} {563}},\ \bibinfo {pages} {671} (\bibinfo {year}
  {2018})}\BibitemShut {NoStop}%
\bibitem [{\citenamefont {Zhou}\ \emph {et~al.}(2020)\citenamefont {Zhou},
  \citenamefont {Zhong}, \citenamefont {Dong}, \citenamefont {Zheng},
  \citenamefont {Tan}, \citenamefont {Jie}, \citenamefont {Pan}, \citenamefont
  {Zhang},\ and\ \citenamefont {Xie}}]{zhou2020cooperative}%
  \BibitemOpen
  \bibfield  {author} {\bibinfo {author} {\bibfnamefont {C.}~\bibnamefont
  {Zhou}}, \bibinfo {author} {\bibfnamefont {Y.}~\bibnamefont {Zhong}},
  \bibinfo {author} {\bibfnamefont {H.}~\bibnamefont {Dong}}, \bibinfo {author}
  {\bibfnamefont {W.}~\bibnamefont {Zheng}}, \bibinfo {author} {\bibfnamefont
  {J.}~\bibnamefont {Tan}}, \bibinfo {author} {\bibfnamefont {Q.}~\bibnamefont
  {Jie}}, \bibinfo {author} {\bibfnamefont {A.}~\bibnamefont {Pan}}, \bibinfo
  {author} {\bibfnamefont {L.}~\bibnamefont {Zhang}},\ and\ \bibinfo {author}
  {\bibfnamefont {W.}~\bibnamefont {Xie}},\ }\bibfield  {title} {\bibinfo
  {title} {Cooperative excitonic quantum ensemble in perovskite-assembly
  superlattice microcavities},\ }\href@noop {} {\bibfield  {journal} {\bibinfo
  {journal} {Nature Communications}\ }\textbf {\bibinfo {volume} {11}},\
  \bibinfo {pages} {329} (\bibinfo {year} {2020})}\BibitemShut {NoStop}%
\bibitem [{\citenamefont {Krieg}\ \emph {et~al.}(2020)\citenamefont {Krieg},
  \citenamefont {Sercel}, \citenamefont {Burian}, \citenamefont {Andrusiv},
  \citenamefont {Bodnarchuk}, \citenamefont {St{\"o}ferle}, \citenamefont
  {Mahrt}, \citenamefont {Naumenko}, \citenamefont {Amenitsch}, \citenamefont
  {Rain{\`o}} \emph {et~al.}}]{krieg2020monodisperse}%
  \BibitemOpen
  \bibfield  {author} {\bibinfo {author} {\bibfnamefont {F.}~\bibnamefont
  {Krieg}}, \bibinfo {author} {\bibfnamefont {P.~C.}\ \bibnamefont {Sercel}},
  \bibinfo {author} {\bibfnamefont {M.}~\bibnamefont {Burian}}, \bibinfo
  {author} {\bibfnamefont {H.}~\bibnamefont {Andrusiv}}, \bibinfo {author}
  {\bibfnamefont {M.~I.}\ \bibnamefont {Bodnarchuk}}, \bibinfo {author}
  {\bibfnamefont {T.}~\bibnamefont {St{\"o}ferle}}, \bibinfo {author}
  {\bibfnamefont {R.~F.}\ \bibnamefont {Mahrt}}, \bibinfo {author}
  {\bibfnamefont {D.}~\bibnamefont {Naumenko}}, \bibinfo {author}
  {\bibfnamefont {H.}~\bibnamefont {Amenitsch}}, \bibinfo {author}
  {\bibfnamefont {G.}~\bibnamefont {Rain{\`o}}}, \emph {et~al.},\ }\bibfield
  {title} {\bibinfo {title} {Monodisperse long-chain sulfobetaine-capped
  \ch{CsPbBr_3} nanocrystals and their superfluorescent assemblies},\
  }\href@noop {} {\bibfield  {journal} {\bibinfo  {journal} {ACS central
  science}\ }\textbf {\bibinfo {volume} {7}},\ \bibinfo {pages} {135} (\bibinfo
  {year} {2020})}\BibitemShut {NoStop}%
\bibitem [{\citenamefont {Pashaei~Adl}\ \emph {et~al.}(2023)\citenamefont
  {Pashaei~Adl}, \citenamefont {Gorji}, \citenamefont {Mu{\~n}oz-Matutano},
  \citenamefont {Gualdr{\'o}n-Reyes}, \citenamefont {Su{\'a}rez}, \citenamefont
  {Chirvony}, \citenamefont {Mora-Ser{\'o}},\ and\ \citenamefont
  {Mart{\'\i}nez-Pastor}}]{pashaei2023superradiance}%
  \BibitemOpen
  \bibfield  {author} {\bibinfo {author} {\bibfnamefont {H.}~\bibnamefont
  {Pashaei~Adl}}, \bibinfo {author} {\bibfnamefont {S.}~\bibnamefont {Gorji}},
  \bibinfo {author} {\bibfnamefont {G.}~\bibnamefont {Mu{\~n}oz-Matutano}},
  \bibinfo {author} {\bibfnamefont {A.~F.}\ \bibnamefont {Gualdr{\'o}n-Reyes}},
  \bibinfo {author} {\bibfnamefont {I.}~\bibnamefont {Su{\'a}rez}}, \bibinfo
  {author} {\bibfnamefont {V.~S.}\ \bibnamefont {Chirvony}}, \bibinfo {author}
  {\bibfnamefont {I.}~\bibnamefont {Mora-Ser{\'o}}},\ and\ \bibinfo {author}
  {\bibfnamefont {J.~P.}\ \bibnamefont {Mart{\'\i}nez-Pastor}},\ }\bibfield
  {title} {\bibinfo {title} {Superradiance emission and its thermal decoherence
  in lead halide perovskites superlattices},\ }\href@noop {} {\bibfield
  {journal} {\bibinfo  {journal} {Advanced Optical Materials}\ ,\ \bibinfo
  {pages} {2202497}} (\bibinfo {year} {2023})}\BibitemShut {NoStop}%
\bibitem [{\citenamefont {Milloch}\ \emph {et~al.}(2023)\citenamefont
  {Milloch}, \citenamefont {Filippi}, \citenamefont {Franceschini},
  \citenamefont {Galvani}, \citenamefont {Mor}, \citenamefont {Pagliara},
  \citenamefont {Ferrini}, \citenamefont {Banfi}, \citenamefont {Capone},
  \citenamefont {Baranov}, \citenamefont {Manna},\ and\ \citenamefont
  {Giannetti}}]{Milloch_NanoLett}%
  \BibitemOpen
  \bibfield  {author} {\bibinfo {author} {\bibfnamefont {A.}~\bibnamefont
  {Milloch}}, \bibinfo {author} {\bibfnamefont {U.}~\bibnamefont {Filippi}},
  \bibinfo {author} {\bibfnamefont {P.}~\bibnamefont {Franceschini}}, \bibinfo
  {author} {\bibfnamefont {M.}~\bibnamefont {Galvani}}, \bibinfo {author}
  {\bibfnamefont {S.}~\bibnamefont {Mor}}, \bibinfo {author} {\bibfnamefont
  {S.}~\bibnamefont {Pagliara}}, \bibinfo {author} {\bibfnamefont
  {G.}~\bibnamefont {Ferrini}}, \bibinfo {author} {\bibfnamefont
  {F.}~\bibnamefont {Banfi}}, \bibinfo {author} {\bibfnamefont
  {M.}~\bibnamefont {Capone}}, \bibinfo {author} {\bibfnamefont
  {D.}~\bibnamefont {Baranov}}, \bibinfo {author} {\bibfnamefont
  {L.}~\bibnamefont {Manna}},\ and\ \bibinfo {author} {\bibfnamefont
  {C.}~\bibnamefont {Giannetti}},\ }\bibfield  {title} {\bibinfo {title}
  {Halide perovskite artificial solids as a new platform to simulate collective
  phenomena in doped mott insulators},\ }\href
  {https://doi.org/10.1021/acs.nanolett.3c03715} {\bibfield  {journal}
  {\bibinfo  {journal} {Nano Letters}\ }\textbf {\bibinfo {volume} {23}},\
  \bibinfo {pages} {10617} (\bibinfo {year} {2023})},\ \Eprint
  {https://arxiv.org/abs/https://doi.org/10.1021/acs.nanolett.3c03715}
  {https://doi.org/10.1021/acs.nanolett.3c03715} \BibitemShut {NoStop}%
\bibitem [{\citenamefont {Cong}\ \emph {et~al.}(2016)\citenamefont {Cong},
  \citenamefont {Zhang}, \citenamefont {Wang}, \citenamefont {Noe},
  \citenamefont {Belyanin},\ and\ \citenamefont {Kono}}]{cong2016dicke}%
  \BibitemOpen
  \bibfield  {author} {\bibinfo {author} {\bibfnamefont {K.}~\bibnamefont
  {Cong}}, \bibinfo {author} {\bibfnamefont {Q.}~\bibnamefont {Zhang}},
  \bibinfo {author} {\bibfnamefont {Y.}~\bibnamefont {Wang}}, \bibinfo {author}
  {\bibfnamefont {G.~T.}\ \bibnamefont {Noe}}, \bibinfo {author} {\bibfnamefont
  {A.}~\bibnamefont {Belyanin}},\ and\ \bibinfo {author} {\bibfnamefont
  {J.}~\bibnamefont {Kono}},\ }\bibfield  {title} {\bibinfo {title} {Dicke
  superradiance in solids},\ }\href@noop {} {\bibfield  {journal} {\bibinfo
  {journal} {JOSA B}\ }\textbf {\bibinfo {volume} {33}},\ \bibinfo {pages}
  {C80} (\bibinfo {year} {2016})}\BibitemShut {NoStop}%
\bibitem [{\citenamefont {Mattiotti}\ \emph {et~al.}(2020)\citenamefont
  {Mattiotti}, \citenamefont {Kuno}, \citenamefont {Borgonovi}, \citenamefont
  {Jank{\'o}},\ and\ \citenamefont {Celardo}}]{mattiotti2020thermal}%
  \BibitemOpen
  \bibfield  {author} {\bibinfo {author} {\bibfnamefont {F.}~\bibnamefont
  {Mattiotti}}, \bibinfo {author} {\bibfnamefont {M.}~\bibnamefont {Kuno}},
  \bibinfo {author} {\bibfnamefont {F.}~\bibnamefont {Borgonovi}}, \bibinfo
  {author} {\bibfnamefont {B.}~\bibnamefont {Jank{\'o}}},\ and\ \bibinfo
  {author} {\bibfnamefont {G.~L.}\ \bibnamefont {Celardo}},\ }\bibfield
  {title} {\bibinfo {title} {Thermal decoherence of superradiance in lead
  halide perovskite nanocrystal superlattices},\ }\href@noop {} {\bibfield
  {journal} {\bibinfo  {journal} {Nano Letters}\ }\textbf {\bibinfo {volume}
  {20}},\ \bibinfo {pages} {7382} (\bibinfo {year} {2020})}\BibitemShut
  {NoStop}%
\bibitem [{\citenamefont {Ghonge}\ \emph {et~al.}(2023)\citenamefont {Ghonge},
  \citenamefont {Engel}, \citenamefont {Mattiotti}, \citenamefont {Celardo},
  \citenamefont {Kuno},\ and\ \citenamefont {Jank{\'o}}}]{ghonge2023enhanced}%
  \BibitemOpen
  \bibfield  {author} {\bibinfo {author} {\bibfnamefont {S.}~\bibnamefont
  {Ghonge}}, \bibinfo {author} {\bibfnamefont {D.}~\bibnamefont {Engel}},
  \bibinfo {author} {\bibfnamefont {F.}~\bibnamefont {Mattiotti}}, \bibinfo
  {author} {\bibfnamefont {G.~L.}\ \bibnamefont {Celardo}}, \bibinfo {author}
  {\bibfnamefont {M.}~\bibnamefont {Kuno}},\ and\ \bibinfo {author}
  {\bibfnamefont {B.}~\bibnamefont {Jank{\'o}}},\ }\bibfield  {title} {\bibinfo
  {title} {Enhanced robustness and dimensional crossover of superradiance in
  cuboidal nanocrystal superlattices},\ }\href@noop {} {\bibfield  {journal}
  {\bibinfo  {journal} {Physical Review Research}\ }\textbf {\bibinfo {volume}
  {5}},\ \bibinfo {pages} {023068} (\bibinfo {year} {2023})}\BibitemShut
  {NoStop}%
\bibitem [{\citenamefont {Blach}\ \emph {et~al.}(2022)\citenamefont {Blach},
  \citenamefont {Lumsargis}, \citenamefont {Clark}, \citenamefont {Chuang},
  \citenamefont {Wang}, \citenamefont {Dou}, \citenamefont {Schaller},
  \citenamefont {Cao}, \citenamefont {Li},\ and\ \citenamefont
  {Huang}}]{blach2022superradiance}%
  \BibitemOpen
  \bibfield  {author} {\bibinfo {author} {\bibfnamefont {D.~D.}\ \bibnamefont
  {Blach}}, \bibinfo {author} {\bibfnamefont {V.~A.}\ \bibnamefont
  {Lumsargis}}, \bibinfo {author} {\bibfnamefont {D.~E.}\ \bibnamefont
  {Clark}}, \bibinfo {author} {\bibfnamefont {C.}~\bibnamefont {Chuang}},
  \bibinfo {author} {\bibfnamefont {K.}~\bibnamefont {Wang}}, \bibinfo {author}
  {\bibfnamefont {L.}~\bibnamefont {Dou}}, \bibinfo {author} {\bibfnamefont
  {R.~D.}\ \bibnamefont {Schaller}}, \bibinfo {author} {\bibfnamefont
  {J.}~\bibnamefont {Cao}}, \bibinfo {author} {\bibfnamefont {C.~W.}\
  \bibnamefont {Li}},\ and\ \bibinfo {author} {\bibfnamefont {L.}~\bibnamefont
  {Huang}},\ }\bibfield  {title} {\bibinfo {title} {Superradiance and exciton
  delocalization in perovskite quantum dot superlattices},\ }\href@noop {}
  {\bibfield  {journal} {\bibinfo  {journal} {Nano letters}\ }\textbf {\bibinfo
  {volume} {22}},\ \bibinfo {pages} {7811} (\bibinfo {year}
  {2022})}\BibitemShut {NoStop}%
\bibitem [{\citenamefont {Smith}\ and\ \citenamefont
  {Karunadasa}(2018)}]{smith2018white}%
  \BibitemOpen
  \bibfield  {author} {\bibinfo {author} {\bibfnamefont {M.~D.}\ \bibnamefont
  {Smith}}\ and\ \bibinfo {author} {\bibfnamefont {H.~I.}\ \bibnamefont
  {Karunadasa}},\ }\bibfield  {title} {\bibinfo {title} {White-light emission
  from layered halide perovskites},\ }\href@noop {} {\bibfield  {journal}
  {\bibinfo  {journal} {Accounts of chemical research}\ }\textbf {\bibinfo
  {volume} {51}},\ \bibinfo {pages} {619} (\bibinfo {year} {2018})}\BibitemShut
  {NoStop}%
\bibitem [{\citenamefont {Li}\ \emph {et~al.}(2019)\citenamefont {Li},
  \citenamefont {Luo}, \citenamefont {Liu},\ and\ \citenamefont
  {Tang}}]{li2019self}%
  \BibitemOpen
  \bibfield  {author} {\bibinfo {author} {\bibfnamefont {S.}~\bibnamefont
  {Li}}, \bibinfo {author} {\bibfnamefont {J.}~\bibnamefont {Luo}}, \bibinfo
  {author} {\bibfnamefont {J.}~\bibnamefont {Liu}},\ and\ \bibinfo {author}
  {\bibfnamefont {J.}~\bibnamefont {Tang}},\ }\bibfield  {title} {\bibinfo
  {title} {Self-trapped excitons in all-inorganic halide perovskites:
  fundamentals, status, and potential applications},\ }\href@noop {} {\bibfield
   {journal} {\bibinfo  {journal} {The journal of physical chemistry letters}\
  }\textbf {\bibinfo {volume} {10}},\ \bibinfo {pages} {1999} (\bibinfo {year}
  {2019})}\BibitemShut {NoStop}%
\bibitem [{\citenamefont {Manser}\ \emph {et~al.}(2016)\citenamefont {Manser},
  \citenamefont {Christians},\ and\ \citenamefont
  {Kamat}}]{manser2016intriguing}%
  \BibitemOpen
  \bibfield  {author} {\bibinfo {author} {\bibfnamefont {J.~S.}\ \bibnamefont
  {Manser}}, \bibinfo {author} {\bibfnamefont {J.~A.}\ \bibnamefont
  {Christians}},\ and\ \bibinfo {author} {\bibfnamefont {P.~V.}\ \bibnamefont
  {Kamat}},\ }\bibfield  {title} {\bibinfo {title} {Intriguing optoelectronic
  properties of metal halide perovskites},\ }\href@noop {} {\bibfield
  {journal} {\bibinfo  {journal} {Chemical reviews}\ }\textbf {\bibinfo
  {volume} {116}},\ \bibinfo {pages} {12956} (\bibinfo {year}
  {2016})}\BibitemShut {NoStop}%
\bibitem [{\citenamefont {Chen}\ \emph {et~al.}(2015)\citenamefont {Chen},
  \citenamefont {De~Marco}, \citenamefont {Yang}, \citenamefont {Song},
  \citenamefont {Chen}, \citenamefont {Zhao}, \citenamefont {Hong},
  \citenamefont {Zhou},\ and\ \citenamefont {Yang}}]{chen2015under}%
  \BibitemOpen
  \bibfield  {author} {\bibinfo {author} {\bibfnamefont {Q.}~\bibnamefont
  {Chen}}, \bibinfo {author} {\bibfnamefont {N.}~\bibnamefont {De~Marco}},
  \bibinfo {author} {\bibfnamefont {Y.~M.}\ \bibnamefont {Yang}}, \bibinfo
  {author} {\bibfnamefont {T.-B.}\ \bibnamefont {Song}}, \bibinfo {author}
  {\bibfnamefont {C.-C.}\ \bibnamefont {Chen}}, \bibinfo {author}
  {\bibfnamefont {H.}~\bibnamefont {Zhao}}, \bibinfo {author} {\bibfnamefont
  {Z.}~\bibnamefont {Hong}}, \bibinfo {author} {\bibfnamefont {H.}~\bibnamefont
  {Zhou}},\ and\ \bibinfo {author} {\bibfnamefont {Y.}~\bibnamefont {Yang}},\
  }\bibfield  {title} {\bibinfo {title} {Under the spotlight: The
  organic--inorganic hybrid halide perovskite for optoelectronic
  applications},\ }\href@noop {} {\bibfield  {journal} {\bibinfo  {journal}
  {Nano Today}\ }\textbf {\bibinfo {volume} {10}},\ \bibinfo {pages} {355}
  (\bibinfo {year} {2015})}\BibitemShut {NoStop}%
\bibitem [{\citenamefont {Fang}\ \emph {et~al.}(2017)\citenamefont {Fang},
  \citenamefont {Protesescu}, \citenamefont {Balazs}, \citenamefont
  {Adjokatse}, \citenamefont {Kovalenko},\ and\ \citenamefont
  {Loi}}]{Fang2017}%
  \BibitemOpen
  \bibfield  {author} {\bibinfo {author} {\bibfnamefont {H.-H.}\ \bibnamefont
  {Fang}}, \bibinfo {author} {\bibfnamefont {L.}~\bibnamefont {Protesescu}},
  \bibinfo {author} {\bibfnamefont {D.~M.}\ \bibnamefont {Balazs}}, \bibinfo
  {author} {\bibfnamefont {S.}~\bibnamefont {Adjokatse}}, \bibinfo {author}
  {\bibfnamefont {M.~V.}\ \bibnamefont {Kovalenko}},\ and\ \bibinfo {author}
  {\bibfnamefont {M.~A.}\ \bibnamefont {Loi}},\ }\bibfield  {title} {\bibinfo
  {title} {Exciton recombination in formamidinium lead triiodide: Nanocrystals
  versus thin films},\ }\href
  {https://doi.org/https://doi.org/10.1002/smll.201700673} {\bibfield
  {journal} {\bibinfo  {journal} {Small}\ }\textbf {\bibinfo {volume} {13}},\
  \bibinfo {pages} {1700673} (\bibinfo {year} {2017})}\BibitemShut {NoStop}%
\bibitem [{\citenamefont {Levchuk}\ \emph {et~al.}(2017)\citenamefont
  {Levchuk}, \citenamefont {Osvet}, \citenamefont {Tang}, \citenamefont
  {Brandl}, \citenamefont {Perea}, \citenamefont {Hoegl}, \citenamefont {Matt},
  \citenamefont {Hock}, \citenamefont {Batentschuk},\ and\ \citenamefont
  {Brabec}}]{levchuk2017brightly}%
  \BibitemOpen
  \bibfield  {author} {\bibinfo {author} {\bibfnamefont {I.}~\bibnamefont
  {Levchuk}}, \bibinfo {author} {\bibfnamefont {A.}~\bibnamefont {Osvet}},
  \bibinfo {author} {\bibfnamefont {X.}~\bibnamefont {Tang}}, \bibinfo {author}
  {\bibfnamefont {M.}~\bibnamefont {Brandl}}, \bibinfo {author} {\bibfnamefont
  {J.~D.}\ \bibnamefont {Perea}}, \bibinfo {author} {\bibfnamefont
  {F.}~\bibnamefont {Hoegl}}, \bibinfo {author} {\bibfnamefont {G.~J.}\
  \bibnamefont {Matt}}, \bibinfo {author} {\bibfnamefont {R.}~\bibnamefont
  {Hock}}, \bibinfo {author} {\bibfnamefont {M.}~\bibnamefont {Batentschuk}},\
  and\ \bibinfo {author} {\bibfnamefont {C.~J.}\ \bibnamefont {Brabec}},\
  }\bibfield  {title} {\bibinfo {title} {Brightly luminescent and color-tunable
  formamidinium lead halide perovskite \ch{FAPbX_3} (\ch{X}= \ch{Cl}, \ch{Br},
  \ch{I}) colloidal nanocrystals},\ }\href@noop {} {\bibfield  {journal}
  {\bibinfo  {journal} {Nano letters}\ }\textbf {\bibinfo {volume} {17}},\
  \bibinfo {pages} {2765} (\bibinfo {year} {2017})}\BibitemShut {NoStop}%
\bibitem [{\citenamefont {Tan}\ \emph {et~al.}(2020)\citenamefont {Tan},
  \citenamefont {Chen}, \citenamefont {Zhang}, \citenamefont {Ni},
  \citenamefont {Wang}, \citenamefont {Zhang}, \citenamefont {Zhou},
  \citenamefont {Bao},\ and\ \citenamefont {Wang}}]{tan2020temperature}%
  \BibitemOpen
  \bibfield  {author} {\bibinfo {author} {\bibfnamefont {M.}~\bibnamefont
  {Tan}}, \bibinfo {author} {\bibfnamefont {B.}~\bibnamefont {Chen}}, \bibinfo
  {author} {\bibfnamefont {Y.}~\bibnamefont {Zhang}}, \bibinfo {author}
  {\bibfnamefont {M.}~\bibnamefont {Ni}}, \bibinfo {author} {\bibfnamefont
  {W.}~\bibnamefont {Wang}}, \bibinfo {author} {\bibfnamefont {H.}~\bibnamefont
  {Zhang}}, \bibinfo {author} {\bibfnamefont {Q.}~\bibnamefont {Zhou}},
  \bibinfo {author} {\bibfnamefont {Y.}~\bibnamefont {Bao}},\ and\ \bibinfo
  {author} {\bibfnamefont {Y.}~\bibnamefont {Wang}},\ }\bibfield  {title}
  {\bibinfo {title} {Temperature-dependent dynamic carrier process of
  \ch{FAPbI_3} nanocrystals’ film},\ }\href@noop {} {\bibfield  {journal}
  {\bibinfo  {journal} {The Journal of Physical Chemistry C}\ }\textbf
  {\bibinfo {volume} {124}},\ \bibinfo {pages} {5093} (\bibinfo {year}
  {2020})}\BibitemShut {NoStop}%
\bibitem [{\citenamefont {Fang}\ \emph {et~al.}(2016)\citenamefont {Fang},
  \citenamefont {Wang}, \citenamefont {Adjokatse}, \citenamefont {Zhao},
  \citenamefont {Even},\ and\ \citenamefont
  {Antonietta~Loi}}]{fang2016photoexcitation}%
  \BibitemOpen
  \bibfield  {author} {\bibinfo {author} {\bibfnamefont {H.-H.}\ \bibnamefont
  {Fang}}, \bibinfo {author} {\bibfnamefont {F.}~\bibnamefont {Wang}}, \bibinfo
  {author} {\bibfnamefont {S.}~\bibnamefont {Adjokatse}}, \bibinfo {author}
  {\bibfnamefont {N.}~\bibnamefont {Zhao}}, \bibinfo {author} {\bibfnamefont
  {J.}~\bibnamefont {Even}},\ and\ \bibinfo {author} {\bibfnamefont
  {M.}~\bibnamefont {Antonietta~Loi}},\ }\bibfield  {title} {\bibinfo {title}
  {Photoexcitation dynamics in solution-processed formamidinium lead iodide
  perovskite thin films for solar cell applications},\ }\href@noop {}
  {\bibfield  {journal} {\bibinfo  {journal} {Light: Science \& Applications}\
  }\textbf {\bibinfo {volume} {5}},\ \bibinfo {pages} {e16056} (\bibinfo {year}
  {2016})}\BibitemShut {NoStop}%
\bibitem [{\citenamefont {Yamada}\ \emph {et~al.}(2014)\citenamefont {Yamada},
  \citenamefont {Nakamura}, \citenamefont {Endo}, \citenamefont {Wakamiya},\
  and\ \citenamefont {Kanemitsu}}]{yamada2014photocarrier}%
  \BibitemOpen
  \bibfield  {author} {\bibinfo {author} {\bibfnamefont {Y.}~\bibnamefont
  {Yamada}}, \bibinfo {author} {\bibfnamefont {T.}~\bibnamefont {Nakamura}},
  \bibinfo {author} {\bibfnamefont {M.}~\bibnamefont {Endo}}, \bibinfo {author}
  {\bibfnamefont {A.}~\bibnamefont {Wakamiya}},\ and\ \bibinfo {author}
  {\bibfnamefont {Y.}~\bibnamefont {Kanemitsu}},\ }\bibfield  {title} {\bibinfo
  {title} {Photocarrier recombination dynamics in perovskite \ch{CH_3NH_3PbI_3}
  for solar cell applications},\ }\href@noop {} {\bibfield  {journal} {\bibinfo
   {journal} {Journal of the American Chemical Society}\ }\textbf {\bibinfo
  {volume} {136}},\ \bibinfo {pages} {11610} (\bibinfo {year}
  {2014})}\BibitemShut {NoStop}%
\bibitem [{\citenamefont {Phuong}\ \emph {et~al.}(2016)\citenamefont {Phuong},
  \citenamefont {Yamada}, \citenamefont {Nagai}, \citenamefont {Maruyama},
  \citenamefont {Wakamiya},\ and\ \citenamefont {Kanemitsu}}]{phuong2016free}%
  \BibitemOpen
  \bibfield  {author} {\bibinfo {author} {\bibfnamefont {L.~Q.}\ \bibnamefont
  {Phuong}}, \bibinfo {author} {\bibfnamefont {Y.}~\bibnamefont {Yamada}},
  \bibinfo {author} {\bibfnamefont {M.}~\bibnamefont {Nagai}}, \bibinfo
  {author} {\bibfnamefont {N.}~\bibnamefont {Maruyama}}, \bibinfo {author}
  {\bibfnamefont {A.}~\bibnamefont {Wakamiya}},\ and\ \bibinfo {author}
  {\bibfnamefont {Y.}~\bibnamefont {Kanemitsu}},\ }\bibfield  {title} {\bibinfo
  {title} {Free carriers versus excitons in \ch{CH_3NH_3PbI_3} perovskite thin
  films at low temperatures: charge transfer from the orthorhombic phase to the
  tetragonal phase},\ }\href@noop {} {\bibfield  {journal} {\bibinfo  {journal}
  {The journal of physical chemistry letters}\ }\textbf {\bibinfo {volume}
  {7}},\ \bibinfo {pages} {2316} (\bibinfo {year} {2016})}\BibitemShut
  {NoStop}%
\bibitem [{\citenamefont {Tahara}\ \emph {et~al.}(2016)\citenamefont {Tahara},
  \citenamefont {Endo}, \citenamefont {Wakamiya},\ and\ \citenamefont
  {Kanemitsu}}]{tahara2016experimental}%
  \BibitemOpen
  \bibfield  {author} {\bibinfo {author} {\bibfnamefont {H.}~\bibnamefont
  {Tahara}}, \bibinfo {author} {\bibfnamefont {M.}~\bibnamefont {Endo}},
  \bibinfo {author} {\bibfnamefont {A.}~\bibnamefont {Wakamiya}},\ and\
  \bibinfo {author} {\bibfnamefont {Y.}~\bibnamefont {Kanemitsu}},\ }\bibfield
  {title} {\bibinfo {title} {Experimental evidence of localized shallow states
  in orthorhombic phase of \ch{CH_3NH_3PbI_3} perovskite thin films revealed by
  photocurrent beat spectroscopy},\ }\href@noop {} {\bibfield  {journal}
  {\bibinfo  {journal} {The Journal of Physical Chemistry C}\ }\textbf
  {\bibinfo {volume} {120}},\ \bibinfo {pages} {5347} (\bibinfo {year}
  {2016})}\BibitemShut {NoStop}%
\bibitem [{\citenamefont {Li}\ \emph {et~al.}(2021)\citenamefont {Li},
  \citenamefont {Allegro}, \citenamefont {Kaiser}, \citenamefont {Malla},
  \citenamefont {Richards}, \citenamefont {Lemmer}, \citenamefont {Paetzold},\
  and\ \citenamefont {Howard}}]{li2021exciton}%
  \BibitemOpen
  \bibfield  {author} {\bibinfo {author} {\bibfnamefont {Y.}~\bibnamefont
  {Li}}, \bibinfo {author} {\bibfnamefont {I.}~\bibnamefont {Allegro}},
  \bibinfo {author} {\bibfnamefont {M.}~\bibnamefont {Kaiser}}, \bibinfo
  {author} {\bibfnamefont {A.~J.}\ \bibnamefont {Malla}}, \bibinfo {author}
  {\bibfnamefont {B.~S.}\ \bibnamefont {Richards}}, \bibinfo {author}
  {\bibfnamefont {U.}~\bibnamefont {Lemmer}}, \bibinfo {author} {\bibfnamefont
  {U.~W.}\ \bibnamefont {Paetzold}},\ and\ \bibinfo {author} {\bibfnamefont
  {I.~A.}\ \bibnamefont {Howard}},\ }\bibfield  {title} {\bibinfo {title}
  {Exciton versus free carrier emission: Implications for photoluminescence
  efficiency and amplified spontaneous emission thresholds in quasi-2{D} and
  3{D} perovskites},\ }\href@noop {} {\bibfield  {journal} {\bibinfo  {journal}
  {Materials Today}\ }\textbf {\bibinfo {volume} {49}},\ \bibinfo {pages} {35}
  (\bibinfo {year} {2021})}\BibitemShut {NoStop}%
\bibitem [{\citenamefont {Sutherland}\ and\ \citenamefont
  {Sargent}(2016)}]{sutherland2016perovskite}%
  \BibitemOpen
  \bibfield  {author} {\bibinfo {author} {\bibfnamefont {B.~R.}\ \bibnamefont
  {Sutherland}}\ and\ \bibinfo {author} {\bibfnamefont {E.~H.}\ \bibnamefont
  {Sargent}},\ }\bibfield  {title} {\bibinfo {title} {Perovskite photonic
  sources},\ }\href@noop {} {\bibfield  {journal} {\bibinfo  {journal} {Nature
  Photonics}\ }\textbf {\bibinfo {volume} {10}},\ \bibinfo {pages} {295}
  (\bibinfo {year} {2016})}\BibitemShut {NoStop}%
\bibitem [{\citenamefont {Zhang}\ \emph {et~al.}(2024)\citenamefont {Zhang},
  \citenamefont {Jin}, \citenamefont {Lu}, \citenamefont {Zhang}, \citenamefont
  {Yang}, \citenamefont {Zhao}, \citenamefont {Sun}, \citenamefont {Thompson},
  \citenamefont {Yuan}, \citenamefont {Ma} \emph {et~al.}}]{zhang2024moire}%
  \BibitemOpen
  \bibfield  {author} {\bibinfo {author} {\bibfnamefont {S.}~\bibnamefont
  {Zhang}}, \bibinfo {author} {\bibfnamefont {L.}~\bibnamefont {Jin}}, \bibinfo
  {author} {\bibfnamefont {Y.}~\bibnamefont {Lu}}, \bibinfo {author}
  {\bibfnamefont {L.}~\bibnamefont {Zhang}}, \bibinfo {author} {\bibfnamefont
  {J.}~\bibnamefont {Yang}}, \bibinfo {author} {\bibfnamefont {Q.}~\bibnamefont
  {Zhao}}, \bibinfo {author} {\bibfnamefont {D.}~\bibnamefont {Sun}}, \bibinfo
  {author} {\bibfnamefont {J.~J.}\ \bibnamefont {Thompson}}, \bibinfo {author}
  {\bibfnamefont {B.}~\bibnamefont {Yuan}}, \bibinfo {author} {\bibfnamefont
  {K.}~\bibnamefont {Ma}}, \emph {et~al.},\ }\bibfield  {title} {\bibinfo
  {title} {Moir{\'e} superlattices in twisted two-dimensional halide
  perovskites},\ }\href@noop {} {\bibfield  {journal} {\bibinfo  {journal}
  {Nature Materials}\ ,\ \bibinfo {pages} {1}} (\bibinfo {year}
  {2024})}\BibitemShut {NoStop}%
\bibitem [{\citenamefont {Jonas}(2003)}]{jonas2003two}%
  \BibitemOpen
  \bibfield  {author} {\bibinfo {author} {\bibfnamefont {D.~M.}\ \bibnamefont
  {Jonas}},\ }\bibfield  {title} {\bibinfo {title} {Two-dimensional femtosecond
  spectroscopy},\ }\href@noop {} {\bibfield  {journal} {\bibinfo  {journal}
  {Annual review of physical chemistry}\ }\textbf {\bibinfo {volume} {54}},\
  \bibinfo {pages} {425} (\bibinfo {year} {2003})}\BibitemShut {NoStop}%
\bibitem [{\citenamefont {Cho}(2008)}]{cho2008coherent}%
  \BibitemOpen
  \bibfield  {author} {\bibinfo {author} {\bibfnamefont {M.}~\bibnamefont
  {Cho}},\ }\bibfield  {title} {\bibinfo {title} {Coherent two-dimensional
  optical spectroscopy},\ }\href@noop {} {\bibfield  {journal} {\bibinfo
  {journal} {Chemical reviews}\ }\textbf {\bibinfo {volume} {108}},\ \bibinfo
  {pages} {1331} (\bibinfo {year} {2008})}\BibitemShut {NoStop}%
\bibitem [{\citenamefont {Mukamel}(2000)}]{mukamel2000multidimensional}%
  \BibitemOpen
  \bibfield  {author} {\bibinfo {author} {\bibfnamefont {S.}~\bibnamefont
  {Mukamel}},\ }\bibfield  {title} {\bibinfo {title} {Multidimensional
  femtosecond correlation spectroscopies of electronic and vibrational
  excitations},\ }\href@noop {} {\bibfield  {journal} {\bibinfo  {journal}
  {Annual review of physical chemistry}\ }\textbf {\bibinfo {volume} {51}},\
  \bibinfo {pages} {691} (\bibinfo {year} {2000})}\BibitemShut {NoStop}%
\bibitem [{\citenamefont {Fuller}\ and\ \citenamefont
  {Ogilvie}(2015)}]{fuller2015experimental}%
  \BibitemOpen
  \bibfield  {author} {\bibinfo {author} {\bibfnamefont {F.~D.}\ \bibnamefont
  {Fuller}}\ and\ \bibinfo {author} {\bibfnamefont {J.~P.}\ \bibnamefont
  {Ogilvie}},\ }\bibfield  {title} {\bibinfo {title} {Experimental
  implementations of two-dimensional fourier transform electronic
  spectroscopy},\ }\href@noop {} {\bibfield  {journal} {\bibinfo  {journal}
  {Annual review of physical chemistry}\ }\textbf {\bibinfo {volume} {66}},\
  \bibinfo {pages} {667} (\bibinfo {year} {2015})}\BibitemShut {NoStop}%
\bibitem [{\citenamefont {Tollerud}\ and\ \citenamefont
  {Davis}(2017)}]{tollerud2017coherent}%
  \BibitemOpen
  \bibfield  {author} {\bibinfo {author} {\bibfnamefont {J.~O.}\ \bibnamefont
  {Tollerud}}\ and\ \bibinfo {author} {\bibfnamefont {J.~A.}\ \bibnamefont
  {Davis}},\ }\bibfield  {title} {\bibinfo {title} {Coherent multi-dimensional
  spectroscopy: Experimental considerations, direct comparisons and new
  capabilities},\ }\href@noop {} {\bibfield  {journal} {\bibinfo  {journal}
  {Progress in Quantum Electronics}\ }\textbf {\bibinfo {volume} {55}},\
  \bibinfo {pages} {1} (\bibinfo {year} {2017})}\BibitemShut {NoStop}%
\bibitem [{\citenamefont {Oriana}\ \emph {et~al.}(2016)\citenamefont {Oriana},
  \citenamefont {R{\'e}hault}, \citenamefont {Preda}, \citenamefont {Polli},\
  and\ \citenamefont {Cerullo}}]{oriana2016scanning}%
  \BibitemOpen
  \bibfield  {author} {\bibinfo {author} {\bibfnamefont {A.}~\bibnamefont
  {Oriana}}, \bibinfo {author} {\bibfnamefont {J.}~\bibnamefont {R{\'e}hault}},
  \bibinfo {author} {\bibfnamefont {F.}~\bibnamefont {Preda}}, \bibinfo
  {author} {\bibfnamefont {D.}~\bibnamefont {Polli}},\ and\ \bibinfo {author}
  {\bibfnamefont {G.}~\bibnamefont {Cerullo}},\ }\bibfield  {title} {\bibinfo
  {title} {Scanning fourier transform spectrometer in the visible range based
  on birefringent wedges},\ }\href@noop {} {\bibfield  {journal} {\bibinfo
  {journal} {JOSA A}\ }\textbf {\bibinfo {volume} {33}},\ \bibinfo {pages}
  {1415} (\bibinfo {year} {2016})}\BibitemShut {NoStop}%
\bibitem [{\citenamefont {Brida}\ \emph {et~al.}(2012)\citenamefont {Brida},
  \citenamefont {Manzoni},\ and\ \citenamefont {Cerullo}}]{brida2012phase}%
  \BibitemOpen
  \bibfield  {author} {\bibinfo {author} {\bibfnamefont {D.}~\bibnamefont
  {Brida}}, \bibinfo {author} {\bibfnamefont {C.}~\bibnamefont {Manzoni}},\
  and\ \bibinfo {author} {\bibfnamefont {G.}~\bibnamefont {Cerullo}},\
  }\bibfield  {title} {\bibinfo {title} {Phase-locked pulses for
  two-dimensional spectroscopy by a birefringent delay line},\ }\href@noop {}
  {\bibfield  {journal} {\bibinfo  {journal} {Optics letters}\ }\textbf
  {\bibinfo {volume} {37}},\ \bibinfo {pages} {3027} (\bibinfo {year}
  {2012})}\BibitemShut {NoStop}%
\bibitem [{\citenamefont {Preda}\ \emph {et~al.}(2016)\citenamefont {Preda},
  \citenamefont {Kumar}, \citenamefont {Crisafi}, \citenamefont {Del~Valle},
  \citenamefont {Cerullo},\ and\ \citenamefont {Polli}}]{preda2016broadband}%
  \BibitemOpen
  \bibfield  {author} {\bibinfo {author} {\bibfnamefont {F.}~\bibnamefont
  {Preda}}, \bibinfo {author} {\bibfnamefont {V.}~\bibnamefont {Kumar}},
  \bibinfo {author} {\bibfnamefont {F.}~\bibnamefont {Crisafi}}, \bibinfo
  {author} {\bibfnamefont {D.~G.~F.}\ \bibnamefont {Del~Valle}}, \bibinfo
  {author} {\bibfnamefont {G.}~\bibnamefont {Cerullo}},\ and\ \bibinfo {author}
  {\bibfnamefont {D.}~\bibnamefont {Polli}},\ }\bibfield  {title} {\bibinfo
  {title} {Broadband pump-probe spectroscopy at 20-{MH}z modulation
  frequency},\ }\href@noop {} {\bibfield  {journal} {\bibinfo  {journal}
  {Optics Letters}\ }\textbf {\bibinfo {volume} {41}},\ \bibinfo {pages} {2970}
  (\bibinfo {year} {2016})}\BibitemShut {NoStop}%
\bibitem [{\citenamefont {Aneesh}\ \emph {et~al.}(2017)\citenamefont {Aneesh},
  \citenamefont {Swarnkar}, \citenamefont {Kumar~Ravi}, \citenamefont {Sharma},
  \citenamefont {Nag},\ and\ \citenamefont {Adarsh}}]{aneesh2017ultrafast}%
  \BibitemOpen
  \bibfield  {author} {\bibinfo {author} {\bibfnamefont {J.}~\bibnamefont
  {Aneesh}}, \bibinfo {author} {\bibfnamefont {A.}~\bibnamefont {Swarnkar}},
  \bibinfo {author} {\bibfnamefont {V.}~\bibnamefont {Kumar~Ravi}}, \bibinfo
  {author} {\bibfnamefont {R.}~\bibnamefont {Sharma}}, \bibinfo {author}
  {\bibfnamefont {A.}~\bibnamefont {Nag}},\ and\ \bibinfo {author}
  {\bibfnamefont {K.}~\bibnamefont {Adarsh}},\ }\bibfield  {title} {\bibinfo
  {title} {Ultrafast exciton dynamics in colloidal \ch{CsPbBr3} perovskite
  nanocrystals: Biexciton effect and auger recombination},\ }\href@noop {}
  {\bibfield  {journal} {\bibinfo  {journal} {The Journal of Physical Chemistry
  C}\ }\textbf {\bibinfo {volume} {121}},\ \bibinfo {pages} {4734} (\bibinfo
  {year} {2017})}\BibitemShut {NoStop}%
\bibitem [{\citenamefont {Cho}\ \emph {et~al.}(2021)\citenamefont {Cho},
  \citenamefont {Yamada}, \citenamefont {Tahara}, \citenamefont {Tadano},
  \citenamefont {Suzuura}, \citenamefont {Saruyama}, \citenamefont {Sato},
  \citenamefont {Teranishi},\ and\ \citenamefont
  {Kanemitsu}}]{cho2021luminescence}%
  \BibitemOpen
  \bibfield  {author} {\bibinfo {author} {\bibfnamefont {K.}~\bibnamefont
  {Cho}}, \bibinfo {author} {\bibfnamefont {T.}~\bibnamefont {Yamada}},
  \bibinfo {author} {\bibfnamefont {H.}~\bibnamefont {Tahara}}, \bibinfo
  {author} {\bibfnamefont {T.}~\bibnamefont {Tadano}}, \bibinfo {author}
  {\bibfnamefont {H.}~\bibnamefont {Suzuura}}, \bibinfo {author} {\bibfnamefont
  {M.}~\bibnamefont {Saruyama}}, \bibinfo {author} {\bibfnamefont
  {R.}~\bibnamefont {Sato}}, \bibinfo {author} {\bibfnamefont {T.}~\bibnamefont
  {Teranishi}},\ and\ \bibinfo {author} {\bibfnamefont {Y.}~\bibnamefont
  {Kanemitsu}},\ }\bibfield  {title} {\bibinfo {title} {Luminescence fine
  structures in single lead halide perovskite nanocrystals: size dependence of
  the exciton--phonon coupling},\ }\href@noop {} {\bibfield  {journal}
  {\bibinfo  {journal} {Nano Letters}\ }\textbf {\bibinfo {volume} {21}},\
  \bibinfo {pages} {7206} (\bibinfo {year} {2021})}\BibitemShut {NoStop}%
\bibitem [{\citenamefont {Huang}\ \emph {et~al.}(2020)\citenamefont {Huang},
  \citenamefont {Chen}, \citenamefont {Zhang}, \citenamefont {Qin},
  \citenamefont {Yu}, \citenamefont {Wang},\ and\ \citenamefont
  {Xiao}}]{huang2020inhomogeneous}%
  \BibitemOpen
  \bibfield  {author} {\bibinfo {author} {\bibfnamefont {X.}~\bibnamefont
  {Huang}}, \bibinfo {author} {\bibfnamefont {L.}~\bibnamefont {Chen}},
  \bibinfo {author} {\bibfnamefont {C.}~\bibnamefont {Zhang}}, \bibinfo
  {author} {\bibfnamefont {Z.}~\bibnamefont {Qin}}, \bibinfo {author}
  {\bibfnamefont {B.}~\bibnamefont {Yu}}, \bibinfo {author} {\bibfnamefont
  {X.}~\bibnamefont {Wang}},\ and\ \bibinfo {author} {\bibfnamefont
  {M.}~\bibnamefont {Xiao}},\ }\bibfield  {title} {\bibinfo {title}
  {Inhomogeneous biexciton binding in perovskite semiconductor nanocrystals
  measured with two-dimensional spectroscopy},\ }\href@noop {} {\bibfield
  {journal} {\bibinfo  {journal} {The Journal of Physical Chemistry Letters}\
  }\textbf {\bibinfo {volume} {11}},\ \bibinfo {pages} {10173} (\bibinfo {year}
  {2020})}\BibitemShut {NoStop}%
\bibitem [{\citenamefont {Shulenberger}\ \emph {et~al.}(2019)\citenamefont
  {Shulenberger}, \citenamefont {Ashner}, \citenamefont {Ha}, \citenamefont
  {Krieg}, \citenamefont {Kovalenko}, \citenamefont {Tisdale},\ and\
  \citenamefont {Bawendi}}]{shulenberger2019setting}%
  \BibitemOpen
  \bibfield  {author} {\bibinfo {author} {\bibfnamefont {K.~E.}\ \bibnamefont
  {Shulenberger}}, \bibinfo {author} {\bibfnamefont {M.~N.}\ \bibnamefont
  {Ashner}}, \bibinfo {author} {\bibfnamefont {S.~K.}\ \bibnamefont {Ha}},
  \bibinfo {author} {\bibfnamefont {F.}~\bibnamefont {Krieg}}, \bibinfo
  {author} {\bibfnamefont {M.~V.}\ \bibnamefont {Kovalenko}}, \bibinfo {author}
  {\bibfnamefont {W.~A.}\ \bibnamefont {Tisdale}},\ and\ \bibinfo {author}
  {\bibfnamefont {M.~G.}\ \bibnamefont {Bawendi}},\ }\bibfield  {title}
  {\bibinfo {title} {Setting an upper bound to the biexciton binding energy in
  \ch{CsPbBr_3} perovskite nanocrystals},\ }\href@noop {} {\bibfield  {journal}
  {\bibinfo  {journal} {The Journal of Physical Chemistry Letters}\ }\textbf
  {\bibinfo {volume} {10}},\ \bibinfo {pages} {5680} (\bibinfo {year}
  {2019})}\BibitemShut {NoStop}%
\bibitem [{\citenamefont {Dost{\'a}l}\ \emph {et~al.}(2018)\citenamefont
  {Dost{\'a}l}, \citenamefont {Fennel}, \citenamefont {Koch}, \citenamefont
  {Herbst}, \citenamefont {W{\"u}rthner},\ and\ \citenamefont
  {Brixner}}]{dostal2018direct}%
  \BibitemOpen
  \bibfield  {author} {\bibinfo {author} {\bibfnamefont {J.}~\bibnamefont
  {Dost{\'a}l}}, \bibinfo {author} {\bibfnamefont {F.}~\bibnamefont {Fennel}},
  \bibinfo {author} {\bibfnamefont {F.}~\bibnamefont {Koch}}, \bibinfo {author}
  {\bibfnamefont {S.}~\bibnamefont {Herbst}}, \bibinfo {author} {\bibfnamefont
  {F.}~\bibnamefont {W{\"u}rthner}},\ and\ \bibinfo {author} {\bibfnamefont
  {T.}~\bibnamefont {Brixner}},\ }\bibfield  {title} {\bibinfo {title} {Direct
  observation of exciton--exciton interactions},\ }\href@noop {} {\bibfield
  {journal} {\bibinfo  {journal} {Nature Communications}\ }\textbf {\bibinfo
  {volume} {9}},\ \bibinfo {pages} {2466} (\bibinfo {year} {2018})}\BibitemShut
  {NoStop}%
\bibitem [{\citenamefont {Kriete}\ \emph {et~al.}(2019)\citenamefont {Kriete},
  \citenamefont {L{\"u}ttig}, \citenamefont {Kunsel}, \citenamefont {Mal{\`y}},
  \citenamefont {Jansen}, \citenamefont {Knoester}, \citenamefont {Brixner},\
  and\ \citenamefont {Pshenichnikov}}]{kriete2019interplay}%
  \BibitemOpen
  \bibfield  {author} {\bibinfo {author} {\bibfnamefont {B.}~\bibnamefont
  {Kriete}}, \bibinfo {author} {\bibfnamefont {J.}~\bibnamefont {L{\"u}ttig}},
  \bibinfo {author} {\bibfnamefont {T.}~\bibnamefont {Kunsel}}, \bibinfo
  {author} {\bibfnamefont {P.}~\bibnamefont {Mal{\`y}}}, \bibinfo {author}
  {\bibfnamefont {T.~L.}\ \bibnamefont {Jansen}}, \bibinfo {author}
  {\bibfnamefont {J.}~\bibnamefont {Knoester}}, \bibinfo {author}
  {\bibfnamefont {T.}~\bibnamefont {Brixner}},\ and\ \bibinfo {author}
  {\bibfnamefont {M.~S.}\ \bibnamefont {Pshenichnikov}},\ }\bibfield  {title}
  {\bibinfo {title} {Interplay between structural hierarchy and exciton
  diffusion in artificial light harvesting},\ }\href@noop {} {\bibfield
  {journal} {\bibinfo  {journal} {Nature Communications}\ }\textbf {\bibinfo
  {volume} {10}},\ \bibinfo {pages} {4615} (\bibinfo {year}
  {2019})}\BibitemShut {NoStop}%
\bibitem [{\citenamefont {Siemens}\ \emph {et~al.}(2010)\citenamefont
  {Siemens}, \citenamefont {Moody}, \citenamefont {Li}, \citenamefont
  {Bristow},\ and\ \citenamefont {Cundiff}}]{siemens2010resonance}%
  \BibitemOpen
  \bibfield  {author} {\bibinfo {author} {\bibfnamefont {M.~E.}\ \bibnamefont
  {Siemens}}, \bibinfo {author} {\bibfnamefont {G.}~\bibnamefont {Moody}},
  \bibinfo {author} {\bibfnamefont {H.}~\bibnamefont {Li}}, \bibinfo {author}
  {\bibfnamefont {A.~D.}\ \bibnamefont {Bristow}},\ and\ \bibinfo {author}
  {\bibfnamefont {S.~T.}\ \bibnamefont {Cundiff}},\ }\bibfield  {title}
  {\bibinfo {title} {Resonance lineshapes in two-dimensional fourier transform
  spectroscopy},\ }\href@noop {} {\bibfield  {journal} {\bibinfo  {journal}
  {Optics express}\ }\textbf {\bibinfo {volume} {18}},\ \bibinfo {pages}
  {17699} (\bibinfo {year} {2010})}\BibitemShut {NoStop}%
\bibitem [{\citenamefont {Russo}\ \emph {et~al.}(2022)\citenamefont {Russo},
  \citenamefont {McGhee}, \citenamefont {Virgili}, \citenamefont {Lidzey},
  \citenamefont {Cerullo},\ and\ \citenamefont {Maiuri}}]{russo2022dephasing}%
  \BibitemOpen
  \bibfield  {author} {\bibinfo {author} {\bibfnamefont {M.}~\bibnamefont
  {Russo}}, \bibinfo {author} {\bibfnamefont {K.~E.}\ \bibnamefont {McGhee}},
  \bibinfo {author} {\bibfnamefont {T.}~\bibnamefont {Virgili}}, \bibinfo
  {author} {\bibfnamefont {D.~G.}\ \bibnamefont {Lidzey}}, \bibinfo {author}
  {\bibfnamefont {G.}~\bibnamefont {Cerullo}},\ and\ \bibinfo {author}
  {\bibfnamefont {M.}~\bibnamefont {Maiuri}},\ }\bibfield  {title} {\bibinfo
  {title} {Dephasing processes in the molecular dye lumogen-f orange
  characterized by two-dimensional electronic spectroscopy},\ }\href@noop {}
  {\bibfield  {journal} {\bibinfo  {journal} {Molecules}\ }\textbf {\bibinfo
  {volume} {27}},\ \bibinfo {pages} {7095} (\bibinfo {year}
  {2022})}\BibitemShut {NoStop}%
\bibitem [{\citenamefont {Lee}\ \emph {et~al.}(1986)\citenamefont {Lee},
  \citenamefont {Koteles},\ and\ \citenamefont
  {Vassell}}]{lee1986luminescence}%
  \BibitemOpen
  \bibfield  {author} {\bibinfo {author} {\bibfnamefont {J.}~\bibnamefont
  {Lee}}, \bibinfo {author} {\bibfnamefont {E.~S.}\ \bibnamefont {Koteles}},\
  and\ \bibinfo {author} {\bibfnamefont {M.}~\bibnamefont {Vassell}},\
  }\bibfield  {title} {\bibinfo {title} {Luminescence linewidths of excitons in
  gaas quantum wells below 150 {K}},\ }\href@noop {} {\bibfield  {journal}
  {\bibinfo  {journal} {Physical Review B}\ }\textbf {\bibinfo {volume} {33}},\
  \bibinfo {pages} {5512} (\bibinfo {year} {1986})}\BibitemShut {NoStop}%
\bibitem [{\citenamefont {Fu}\ \emph {et~al.}(2018)\citenamefont {Fu},
  \citenamefont {Tamarat}, \citenamefont {Trebbia}, \citenamefont {Bodnarchuk},
  \citenamefont {Kovalenko}, \citenamefont {Even},\ and\ \citenamefont
  {Lounis}}]{fu2018unraveling}%
  \BibitemOpen
  \bibfield  {author} {\bibinfo {author} {\bibfnamefont {M.}~\bibnamefont
  {Fu}}, \bibinfo {author} {\bibfnamefont {P.}~\bibnamefont {Tamarat}},
  \bibinfo {author} {\bibfnamefont {J.-B.}\ \bibnamefont {Trebbia}}, \bibinfo
  {author} {\bibfnamefont {M.~I.}\ \bibnamefont {Bodnarchuk}}, \bibinfo
  {author} {\bibfnamefont {M.~V.}\ \bibnamefont {Kovalenko}}, \bibinfo {author}
  {\bibfnamefont {J.}~\bibnamefont {Even}},\ and\ \bibinfo {author}
  {\bibfnamefont {B.}~\bibnamefont {Lounis}},\ }\bibfield  {title} {\bibinfo
  {title} {Unraveling exciton--phonon coupling in individual \ch{FAPbI_3}
  nanocrystals emitting near-infrared single photons},\ }\href@noop {}
  {\bibfield  {journal} {\bibinfo  {journal} {Nature communications}\ }\textbf
  {\bibinfo {volume} {9}},\ \bibinfo {pages} {3318} (\bibinfo {year}
  {2018})}\BibitemShut {NoStop}%
\bibitem [{\citenamefont {Ferreira}\ \emph {et~al.}(2020)\citenamefont
  {Ferreira}, \citenamefont {Paofai}, \citenamefont {L{\'e}toublon},
  \citenamefont {Ollivier}, \citenamefont {Raymond}, \citenamefont {Hehlen},
  \citenamefont {Ruffl{\'e}}, \citenamefont {Cordier}, \citenamefont {Katan},
  \citenamefont {Even} \emph {et~al.}}]{ferreira2020direct}%
  \BibitemOpen
  \bibfield  {author} {\bibinfo {author} {\bibfnamefont {A.}~\bibnamefont
  {Ferreira}}, \bibinfo {author} {\bibfnamefont {S.}~\bibnamefont {Paofai}},
  \bibinfo {author} {\bibfnamefont {A.}~\bibnamefont {L{\'e}toublon}}, \bibinfo
  {author} {\bibfnamefont {J.}~\bibnamefont {Ollivier}}, \bibinfo {author}
  {\bibfnamefont {S.}~\bibnamefont {Raymond}}, \bibinfo {author} {\bibfnamefont
  {B.}~\bibnamefont {Hehlen}}, \bibinfo {author} {\bibfnamefont
  {B.}~\bibnamefont {Ruffl{\'e}}}, \bibinfo {author} {\bibfnamefont
  {S.}~\bibnamefont {Cordier}}, \bibinfo {author} {\bibfnamefont
  {C.}~\bibnamefont {Katan}}, \bibinfo {author} {\bibfnamefont
  {J.}~\bibnamefont {Even}}, \emph {et~al.},\ }\bibfield  {title} {\bibinfo
  {title} {Direct evidence of weakly dispersed and strongly anharmonic optical
  phonons in hybrid perovskites},\ }\href@noop {} {\bibfield  {journal}
  {\bibinfo  {journal} {Communications Physics}\ }\textbf {\bibinfo {volume}
  {3}},\ \bibinfo {pages} {48} (\bibinfo {year} {2020})}\BibitemShut {NoStop}%
\bibitem [{\citenamefont {Yuan}\ \emph {et~al.}(2017)\citenamefont {Yuan},
  \citenamefont {Wu}, \citenamefont {Dong}, \citenamefont {Xi}, \citenamefont
  {Xi}, \citenamefont {Divitini}, \citenamefont {Jiao}, \citenamefont {Hou},
  \citenamefont {Wang},\ and\ \citenamefont {Gong}}]{yuan2017high}%
  \BibitemOpen
  \bibfield  {author} {\bibinfo {author} {\bibfnamefont {F.}~\bibnamefont
  {Yuan}}, \bibinfo {author} {\bibfnamefont {Z.}~\bibnamefont {Wu}}, \bibinfo
  {author} {\bibfnamefont {H.}~\bibnamefont {Dong}}, \bibinfo {author}
  {\bibfnamefont {J.}~\bibnamefont {Xi}}, \bibinfo {author} {\bibfnamefont
  {K.}~\bibnamefont {Xi}}, \bibinfo {author} {\bibfnamefont {G.}~\bibnamefont
  {Divitini}}, \bibinfo {author} {\bibfnamefont {B.}~\bibnamefont {Jiao}},
  \bibinfo {author} {\bibfnamefont {X.}~\bibnamefont {Hou}}, \bibinfo {author}
  {\bibfnamefont {S.}~\bibnamefont {Wang}},\ and\ \bibinfo {author}
  {\bibfnamefont {Q.}~\bibnamefont {Gong}},\ }\bibfield  {title} {\bibinfo
  {title} {High stability and ultralow threshold amplified spontaneous emission
  from formamidinium lead halide perovskite films},\ }\href@noop {} {\bibfield
  {journal} {\bibinfo  {journal} {The Journal of Physical Chemistry C}\
  }\textbf {\bibinfo {volume} {121}},\ \bibinfo {pages} {15318} (\bibinfo
  {year} {2017})}\BibitemShut {NoStop}%
\bibitem [{\citenamefont {Quarti}\ \emph {et~al.}(2013)\citenamefont {Quarti},
  \citenamefont {Grancini}, \citenamefont {Mosconi}, \citenamefont {Bruno},
  \citenamefont {Ball}, \citenamefont {Lee}, \citenamefont {Snaith},
  \citenamefont {Petrozza},\ and\ \citenamefont
  {De~Angelis}}]{quarti2013raman}%
  \BibitemOpen
  \bibfield  {author} {\bibinfo {author} {\bibfnamefont {C.}~\bibnamefont
  {Quarti}}, \bibinfo {author} {\bibfnamefont {G.}~\bibnamefont {Grancini}},
  \bibinfo {author} {\bibfnamefont {E.}~\bibnamefont {Mosconi}}, \bibinfo
  {author} {\bibfnamefont {P.}~\bibnamefont {Bruno}}, \bibinfo {author}
  {\bibfnamefont {J.~M.}\ \bibnamefont {Ball}}, \bibinfo {author}
  {\bibfnamefont {M.~M.}\ \bibnamefont {Lee}}, \bibinfo {author} {\bibfnamefont
  {H.~J.}\ \bibnamefont {Snaith}}, \bibinfo {author} {\bibfnamefont
  {A.}~\bibnamefont {Petrozza}},\ and\ \bibinfo {author} {\bibfnamefont
  {F.}~\bibnamefont {De~Angelis}},\ }\bibfield  {title} {\bibinfo {title} {The
  raman spectrum of the \ch{CH_3NH_3PbI_3} hybrid perovskite: interplay of
  theory and experiment},\ }\href@noop {} {\bibfield  {journal} {\bibinfo
  {journal} {The journal of physical chemistry letters}\ }\textbf {\bibinfo
  {volume} {5}},\ \bibinfo {pages} {279} (\bibinfo {year} {2013})}\BibitemShut
  {NoStop}%
\bibitem [{\citenamefont {Park}\ \emph {et~al.}(2017)\citenamefont {Park},
  \citenamefont {Kornienko}, \citenamefont {Reyes-Lillo}, \citenamefont {Lai},
  \citenamefont {Neaton}, \citenamefont {Yang},\ and\ \citenamefont
  {Mathies}}]{park2017critical}%
  \BibitemOpen
  \bibfield  {author} {\bibinfo {author} {\bibfnamefont {M.}~\bibnamefont
  {Park}}, \bibinfo {author} {\bibfnamefont {N.}~\bibnamefont {Kornienko}},
  \bibinfo {author} {\bibfnamefont {S.~E.}\ \bibnamefont {Reyes-Lillo}},
  \bibinfo {author} {\bibfnamefont {M.}~\bibnamefont {Lai}}, \bibinfo {author}
  {\bibfnamefont {J.~B.}\ \bibnamefont {Neaton}}, \bibinfo {author}
  {\bibfnamefont {P.}~\bibnamefont {Yang}},\ and\ \bibinfo {author}
  {\bibfnamefont {R.~A.}\ \bibnamefont {Mathies}},\ }\bibfield  {title}
  {\bibinfo {title} {Critical role of methylammonium librational motion in
  methylammonium lead iodide (\ch{CH_3NH_3PbI_3}) perovskite photochemistry},\
  }\href@noop {} {\bibfield  {journal} {\bibinfo  {journal} {Nano letters}\
  }\textbf {\bibinfo {volume} {17}},\ \bibinfo {pages} {4151} (\bibinfo {year}
  {2017})}\BibitemShut {NoStop}%
\bibitem [{\citenamefont {Fabini}\ \emph {et~al.}(2016)\citenamefont {Fabini},
  \citenamefont {Stoumpos}, \citenamefont {Laurita}, \citenamefont
  {Kaltzoglou}, \citenamefont {Kontos}, \citenamefont {Falaras}, \citenamefont
  {Kanatzidis},\ and\ \citenamefont {Seshadri}}]{fabini2016reentrant}%
  \BibitemOpen
  \bibfield  {author} {\bibinfo {author} {\bibfnamefont {D.~H.}\ \bibnamefont
  {Fabini}}, \bibinfo {author} {\bibfnamefont {C.~C.}\ \bibnamefont
  {Stoumpos}}, \bibinfo {author} {\bibfnamefont {G.}~\bibnamefont {Laurita}},
  \bibinfo {author} {\bibfnamefont {A.}~\bibnamefont {Kaltzoglou}}, \bibinfo
  {author} {\bibfnamefont {A.~G.}\ \bibnamefont {Kontos}}, \bibinfo {author}
  {\bibfnamefont {P.}~\bibnamefont {Falaras}}, \bibinfo {author} {\bibfnamefont
  {M.~G.}\ \bibnamefont {Kanatzidis}},\ and\ \bibinfo {author} {\bibfnamefont
  {R.}~\bibnamefont {Seshadri}},\ }\bibfield  {title} {\bibinfo {title}
  {Reentrant structural and optical properties and large positive thermal
  expansion in perovskite formamidinium lead iodide},\ }\href@noop {}
  {\bibfield  {journal} {\bibinfo  {journal} {Angewandte Chemie}\ }\textbf
  {\bibinfo {volume} {128}},\ \bibinfo {pages} {15618} (\bibinfo {year}
  {2016})}\BibitemShut {NoStop}%
\bibitem [{\citenamefont {Weber}\ \emph {et~al.}(2018)\citenamefont {Weber},
  \citenamefont {Ghosh}, \citenamefont {Gaines}, \citenamefont {Henry},
  \citenamefont {Walker}, \citenamefont {Islam},\ and\ \citenamefont
  {Weller}}]{weber2018phase}%
  \BibitemOpen
  \bibfield  {author} {\bibinfo {author} {\bibfnamefont {O.~J.}\ \bibnamefont
  {Weber}}, \bibinfo {author} {\bibfnamefont {D.}~\bibnamefont {Ghosh}},
  \bibinfo {author} {\bibfnamefont {S.}~\bibnamefont {Gaines}}, \bibinfo
  {author} {\bibfnamefont {P.~F.}\ \bibnamefont {Henry}}, \bibinfo {author}
  {\bibfnamefont {A.~B.}\ \bibnamefont {Walker}}, \bibinfo {author}
  {\bibfnamefont {M.~S.}\ \bibnamefont {Islam}},\ and\ \bibinfo {author}
  {\bibfnamefont {M.~T.}\ \bibnamefont {Weller}},\ }\bibfield  {title}
  {\bibinfo {title} {Phase behavior and polymorphism of formamidinium lead
  iodide},\ }\href@noop {} {\bibfield  {journal} {\bibinfo  {journal}
  {Chemistry of Materials}\ }\textbf {\bibinfo {volume} {30}},\ \bibinfo
  {pages} {3768} (\bibinfo {year} {2018})}\BibitemShut {NoStop}%
\bibitem [{\citenamefont {Liu}\ \emph {et~al.}(2019)\citenamefont {Liu},
  \citenamefont {Zhao}, \citenamefont {Xiao}, \citenamefont {Zhang},
  \citenamefont {Pevere}, \citenamefont {Shi}, \citenamefont {Huang},
  \citenamefont {Zhong},\ and\ \citenamefont {Sychugov}}]{liu2019size}%
  \BibitemOpen
  \bibfield  {author} {\bibinfo {author} {\bibfnamefont {L.}~\bibnamefont
  {Liu}}, \bibinfo {author} {\bibfnamefont {R.}~\bibnamefont {Zhao}}, \bibinfo
  {author} {\bibfnamefont {C.}~\bibnamefont {Xiao}}, \bibinfo {author}
  {\bibfnamefont {F.}~\bibnamefont {Zhang}}, \bibinfo {author} {\bibfnamefont
  {F.}~\bibnamefont {Pevere}}, \bibinfo {author} {\bibfnamefont
  {K.}~\bibnamefont {Shi}}, \bibinfo {author} {\bibfnamefont {H.}~\bibnamefont
  {Huang}}, \bibinfo {author} {\bibfnamefont {H.}~\bibnamefont {Zhong}},\ and\
  \bibinfo {author} {\bibfnamefont {I.}~\bibnamefont {Sychugov}},\ }\bibfield
  {title} {\bibinfo {title} {Size-dependent phase transition in perovskite
  nanocrystals},\ }\href@noop {} {\bibfield  {journal} {\bibinfo  {journal}
  {The Journal of Physical Chemistry Letters}\ }\textbf {\bibinfo {volume}
  {10}},\ \bibinfo {pages} {5451} (\bibinfo {year} {2019})}\BibitemShut
  {NoStop}%
\bibitem [{\citenamefont {Maalej}\ \emph {et~al.}(1997)\citenamefont {Maalej},
  \citenamefont {Abid}, \citenamefont {Kallel}, \citenamefont {Daoud},
  \citenamefont {Lauti{\'e}},\ and\ \citenamefont {Romain}}]{maalej1997phase}%
  \BibitemOpen
  \bibfield  {author} {\bibinfo {author} {\bibfnamefont {A.}~\bibnamefont
  {Maalej}}, \bibinfo {author} {\bibfnamefont {Y.}~\bibnamefont {Abid}},
  \bibinfo {author} {\bibfnamefont {A.}~\bibnamefont {Kallel}}, \bibinfo
  {author} {\bibfnamefont {A.}~\bibnamefont {Daoud}}, \bibinfo {author}
  {\bibfnamefont {A.}~\bibnamefont {Lauti{\'e}}},\ and\ \bibinfo {author}
  {\bibfnamefont {F.}~\bibnamefont {Romain}},\ }\bibfield  {title} {\bibinfo
  {title} {Phase transitions and crystal dynamics in the cubic perovskite
  \ch{CH_3NH_3PbCl_3}},\ }\href@noop {} {\bibfield  {journal} {\bibinfo
  {journal} {Solid state communications}\ }\textbf {\bibinfo {volume} {103}},\
  \bibinfo {pages} {279} (\bibinfo {year} {1997})}\BibitemShut {NoStop}%
\bibitem [{\citenamefont {Wright}\ \emph {et~al.}(2017)\citenamefont {Wright},
  \citenamefont {Milot}, \citenamefont {Eperon}, \citenamefont {Snaith},
  \citenamefont {Johnston},\ and\ \citenamefont {Herz}}]{wright2017band}%
  \BibitemOpen
  \bibfield  {author} {\bibinfo {author} {\bibfnamefont {A.~D.}\ \bibnamefont
  {Wright}}, \bibinfo {author} {\bibfnamefont {R.~L.}\ \bibnamefont {Milot}},
  \bibinfo {author} {\bibfnamefont {G.~E.}\ \bibnamefont {Eperon}}, \bibinfo
  {author} {\bibfnamefont {H.~J.}\ \bibnamefont {Snaith}}, \bibinfo {author}
  {\bibfnamefont {M.~B.}\ \bibnamefont {Johnston}},\ and\ \bibinfo {author}
  {\bibfnamefont {L.~M.}\ \bibnamefont {Herz}},\ }\bibfield  {title} {\bibinfo
  {title} {Band-tail recombination in hybrid lead iodide perovskite},\
  }\href@noop {} {\bibfield  {journal} {\bibinfo  {journal} {Advanced
  Functional Materials}\ }\textbf {\bibinfo {volume} {27}},\ \bibinfo {pages}
  {1700860} (\bibinfo {year} {2017})}\BibitemShut {NoStop}%
\bibitem [{\citenamefont {Grandhi}\ \emph {et~al.}(2023)\citenamefont
  {Grandhi}, \citenamefont {Dhama}, \citenamefont {Viswanath}, \citenamefont
  {Lisitsyna}, \citenamefont {Al-Anesi}, \citenamefont {Dana}, \citenamefont
  {Sugathan}, \citenamefont {Caglayan},\ and\ \citenamefont
  {Vivo}}]{grandhi2023role}%
  \BibitemOpen
  \bibfield  {author} {\bibinfo {author} {\bibfnamefont {G.~K.}\ \bibnamefont
  {Grandhi}}, \bibinfo {author} {\bibfnamefont {R.}~\bibnamefont {Dhama}},
  \bibinfo {author} {\bibfnamefont {N.~S.~M.}\ \bibnamefont {Viswanath}},
  \bibinfo {author} {\bibfnamefont {E.~S.}\ \bibnamefont {Lisitsyna}}, \bibinfo
  {author} {\bibfnamefont {B.}~\bibnamefont {Al-Anesi}}, \bibinfo {author}
  {\bibfnamefont {J.}~\bibnamefont {Dana}}, \bibinfo {author} {\bibfnamefont
  {V.}~\bibnamefont {Sugathan}}, \bibinfo {author} {\bibfnamefont
  {H.}~\bibnamefont {Caglayan}},\ and\ \bibinfo {author} {\bibfnamefont
  {P.}~\bibnamefont {Vivo}},\ }\bibfield  {title} {\bibinfo {title} {Role of
  self-trapped excitons in the broadband emission of lead-free
  perovskite-inspired \ch{Cu2AgBiI6}},\ }\href@noop {} {\bibfield  {journal}
  {\bibinfo  {journal} {The journal of physical chemistry letters}\ }\textbf
  {\bibinfo {volume} {14}},\ \bibinfo {pages} {4192} (\bibinfo {year}
  {2023})}\BibitemShut {NoStop}%
\bibitem [{\citenamefont {Luo}\ \emph {et~al.}(2018)\citenamefont {Luo},
  \citenamefont {Wang}, \citenamefont {Li}, \citenamefont {Liu}, \citenamefont
  {Guo}, \citenamefont {Niu}, \citenamefont {Yao}, \citenamefont {Fu},
  \citenamefont {Gao}, \citenamefont {Dong} \emph {et~al.}}]{luo2018efficient}%
  \BibitemOpen
  \bibfield  {author} {\bibinfo {author} {\bibfnamefont {J.}~\bibnamefont
  {Luo}}, \bibinfo {author} {\bibfnamefont {X.}~\bibnamefont {Wang}}, \bibinfo
  {author} {\bibfnamefont {S.}~\bibnamefont {Li}}, \bibinfo {author}
  {\bibfnamefont {J.}~\bibnamefont {Liu}}, \bibinfo {author} {\bibfnamefont
  {Y.}~\bibnamefont {Guo}}, \bibinfo {author} {\bibfnamefont {G.}~\bibnamefont
  {Niu}}, \bibinfo {author} {\bibfnamefont {L.}~\bibnamefont {Yao}}, \bibinfo
  {author} {\bibfnamefont {Y.}~\bibnamefont {Fu}}, \bibinfo {author}
  {\bibfnamefont {L.}~\bibnamefont {Gao}}, \bibinfo {author} {\bibfnamefont
  {Q.}~\bibnamefont {Dong}}, \emph {et~al.},\ }\bibfield  {title} {\bibinfo
  {title} {Efficient and stable emission of warm-white light from lead-free
  halide double perovskites},\ }\href@noop {} {\bibfield  {journal} {\bibinfo
  {journal} {Nature}\ }\textbf {\bibinfo {volume} {563}},\ \bibinfo {pages}
  {541} (\bibinfo {year} {2018})}\BibitemShut {NoStop}%
\bibitem [{\citenamefont {Yazdani}\ \emph {et~al.}(2024)\citenamefont
  {Yazdani}, \citenamefont {Bodnarchuk}, \citenamefont {Bertolotti},
  \citenamefont {Masciocchi}, \citenamefont {Fureraj}, \citenamefont
  {Guzelturk}, \citenamefont {Cotts}, \citenamefont {Zajac}, \citenamefont
  {Rain{\`o}}, \citenamefont {Jansen} \emph {et~al.}}]{yazdani2023coupling}%
  \BibitemOpen
  \bibfield  {author} {\bibinfo {author} {\bibfnamefont {N.}~\bibnamefont
  {Yazdani}}, \bibinfo {author} {\bibfnamefont {M.~I.}\ \bibnamefont
  {Bodnarchuk}}, \bibinfo {author} {\bibfnamefont {F.}~\bibnamefont
  {Bertolotti}}, \bibinfo {author} {\bibfnamefont {N.}~\bibnamefont
  {Masciocchi}}, \bibinfo {author} {\bibfnamefont {I.}~\bibnamefont {Fureraj}},
  \bibinfo {author} {\bibfnamefont {B.}~\bibnamefont {Guzelturk}}, \bibinfo
  {author} {\bibfnamefont {B.~L.}\ \bibnamefont {Cotts}}, \bibinfo {author}
  {\bibfnamefont {M.}~\bibnamefont {Zajac}}, \bibinfo {author} {\bibfnamefont
  {G.}~\bibnamefont {Rain{\`o}}}, \bibinfo {author} {\bibfnamefont
  {M.}~\bibnamefont {Jansen}}, \emph {et~al.},\ }\bibfield  {title} {\bibinfo
  {title} {Coupling to octahedral tilts in halide perovskite nanocrystals
  induces phonon-mediated attractive interactions between excitons},\
  }\href@noop {} {\bibfield  {journal} {\bibinfo  {journal} {Nature Physics}\
  }\textbf {\bibinfo {volume} {20}},\ \bibinfo {pages} {47–53} (\bibinfo
  {year} {2024})}\BibitemShut {NoStop}%
\bibitem [{\citenamefont {Tan}\ \emph {et~al.}(2022)\citenamefont {Tan},
  \citenamefont {Li}, \citenamefont {Zhu}, \citenamefont {Han}, \citenamefont
  {Gong},\ and\ \citenamefont {Zhang}}]{tan2022self}%
  \BibitemOpen
  \bibfield  {author} {\bibinfo {author} {\bibfnamefont {J.}~\bibnamefont
  {Tan}}, \bibinfo {author} {\bibfnamefont {D.}~\bibnamefont {Li}}, \bibinfo
  {author} {\bibfnamefont {J.}~\bibnamefont {Zhu}}, \bibinfo {author}
  {\bibfnamefont {N.}~\bibnamefont {Han}}, \bibinfo {author} {\bibfnamefont
  {Y.}~\bibnamefont {Gong}},\ and\ \bibinfo {author} {\bibfnamefont
  {Y.}~\bibnamefont {Zhang}},\ }\bibfield  {title} {\bibinfo {title}
  {Self-trapped excitons in soft semiconductors},\ }\href@noop {} {\bibfield
  {journal} {\bibinfo  {journal} {Nanoscale}\ }\textbf {\bibinfo {volume}
  {14}},\ \bibinfo {pages} {16394} (\bibinfo {year} {2022})}\BibitemShut
  {NoStop}%
\bibitem [{\citenamefont {Yamada}\ and\ \citenamefont
  {Kanemitsu}(2022)}]{yamada2022electron}%
  \BibitemOpen
  \bibfield  {author} {\bibinfo {author} {\bibfnamefont {Y.}~\bibnamefont
  {Yamada}}\ and\ \bibinfo {author} {\bibfnamefont {Y.}~\bibnamefont
  {Kanemitsu}},\ }\bibfield  {title} {\bibinfo {title} {Electron-phonon
  interactions in halide perovskites},\ }\href@noop {} {\bibfield  {journal}
  {\bibinfo  {journal} {NPG Asia Materials}\ }\textbf {\bibinfo {volume}
  {14}},\ \bibinfo {pages} {48} (\bibinfo {year} {2022})}\BibitemShut {NoStop}%
\bibitem [{\citenamefont {Tollerud}\ \emph {et~al.}(2016)\citenamefont
  {Tollerud}, \citenamefont {Cundiff},\ and\ \citenamefont
  {Davis}}]{tollerud2016revealing}%
  \BibitemOpen
  \bibfield  {author} {\bibinfo {author} {\bibfnamefont {J.~O.}\ \bibnamefont
  {Tollerud}}, \bibinfo {author} {\bibfnamefont {S.~T.}\ \bibnamefont
  {Cundiff}},\ and\ \bibinfo {author} {\bibfnamefont {J.~A.}\ \bibnamefont
  {Davis}},\ }\bibfield  {title} {\bibinfo {title} {Revealing and
  characterizing dark excitons through coherent multidimensional
  spectroscopy},\ }\href@noop {} {\bibfield  {journal} {\bibinfo  {journal}
  {Physical review letters}\ }\textbf {\bibinfo {volume} {117}},\ \bibinfo
  {pages} {097401} (\bibinfo {year} {2016})}\BibitemShut {NoStop}%
\bibitem [{\citenamefont {Camargo}\ \emph {et~al.}(2019)\citenamefont
  {Camargo}, \citenamefont {Nagahara}, \citenamefont {Feldmann}, \citenamefont
  {Richter}, \citenamefont {Friend}, \citenamefont {Cerullo},\ and\
  \citenamefont {Deschler}}]{camargo2019dark}%
  \BibitemOpen
  \bibfield  {author} {\bibinfo {author} {\bibfnamefont {F.~V.}\ \bibnamefont
  {Camargo}}, \bibinfo {author} {\bibfnamefont {T.}~\bibnamefont {Nagahara}},
  \bibinfo {author} {\bibfnamefont {S.}~\bibnamefont {Feldmann}}, \bibinfo
  {author} {\bibfnamefont {J.~M.}\ \bibnamefont {Richter}}, \bibinfo {author}
  {\bibfnamefont {R.~H.}\ \bibnamefont {Friend}}, \bibinfo {author}
  {\bibfnamefont {G.}~\bibnamefont {Cerullo}},\ and\ \bibinfo {author}
  {\bibfnamefont {F.}~\bibnamefont {Deschler}},\ }\bibfield  {title} {\bibinfo
  {title} {Dark subgap states in metal-halide perovskites revealed by coherent
  multidimensional spectroscopy},\ }\href@noop {} {\bibfield  {journal}
  {\bibinfo  {journal} {Journal of the American Chemical Society}\ }\textbf
  {\bibinfo {volume} {142}},\ \bibinfo {pages} {777} (\bibinfo {year}
  {2019})}\BibitemShut {NoStop}%
\bibitem [{\citenamefont {Camargo}\ \emph {et~al.}(2017)\citenamefont
  {Camargo}, \citenamefont {Grimmelsmann}, \citenamefont {Anderson},
  \citenamefont {Meech},\ and\ \citenamefont {Heisler}}]{camargo2017resolving}%
  \BibitemOpen
  \bibfield  {author} {\bibinfo {author} {\bibfnamefont {F.~V. d.~A.}\
  \bibnamefont {Camargo}}, \bibinfo {author} {\bibfnamefont {L.}~\bibnamefont
  {Grimmelsmann}}, \bibinfo {author} {\bibfnamefont {H.~L.}\ \bibnamefont
  {Anderson}}, \bibinfo {author} {\bibfnamefont {S.~R.}\ \bibnamefont
  {Meech}},\ and\ \bibinfo {author} {\bibfnamefont {I.~A.}\ \bibnamefont
  {Heisler}},\ }\bibfield  {title} {\bibinfo {title} {Resolving vibrational
  from electronic coherences in two-dimensional electronic spectroscopy: The
  role of the laser spectrum},\ }\href@noop {} {\bibfield  {journal} {\bibinfo
  {journal} {Physical Review Letters}\ }\textbf {\bibinfo {volume} {118}},\
  \bibinfo {pages} {033001} (\bibinfo {year} {2017})}\BibitemShut {NoStop}%
\bibitem [{\citenamefont {Neukirch}\ \emph {et~al.}(2016)\citenamefont
  {Neukirch}, \citenamefont {Nie}, \citenamefont {Blancon}, \citenamefont
  {Appavoo}, \citenamefont {Tsai}, \citenamefont {Sfeir}, \citenamefont
  {Katan}, \citenamefont {Pedesseau}, \citenamefont {Even}, \citenamefont
  {Crochet}, \citenamefont {Gupta}, \citenamefont {Mohite},\ and\ \citenamefont
  {Tretiak}}]{Neukirch2016}%
  \BibitemOpen
  \bibfield  {author} {\bibinfo {author} {\bibfnamefont {A.~J.}\ \bibnamefont
  {Neukirch}}, \bibinfo {author} {\bibfnamefont {W.}~\bibnamefont {Nie}},
  \bibinfo {author} {\bibfnamefont {J.-C.}\ \bibnamefont {Blancon}}, \bibinfo
  {author} {\bibfnamefont {K.}~\bibnamefont {Appavoo}}, \bibinfo {author}
  {\bibfnamefont {H.}~\bibnamefont {Tsai}}, \bibinfo {author} {\bibfnamefont
  {M.~Y.}\ \bibnamefont {Sfeir}}, \bibinfo {author} {\bibfnamefont
  {C.}~\bibnamefont {Katan}}, \bibinfo {author} {\bibfnamefont
  {L.}~\bibnamefont {Pedesseau}}, \bibinfo {author} {\bibfnamefont
  {J.}~\bibnamefont {Even}}, \bibinfo {author} {\bibfnamefont {J.~J.}\
  \bibnamefont {Crochet}}, \bibinfo {author} {\bibfnamefont {G.}~\bibnamefont
  {Gupta}}, \bibinfo {author} {\bibfnamefont {A.~D.}\ \bibnamefont {Mohite}},\
  and\ \bibinfo {author} {\bibfnamefont {S.}~\bibnamefont {Tretiak}},\
  }\bibfield  {title} {\bibinfo {title} {{Polaron Stabilization by Cooperative
  Lattice Distortion and Cation Rotations in Hybrid Perovskite Materials}},\
  }\href {https://doi.org/10.1021/acs.nanolett.6b01218} {\bibfield  {journal}
  {\bibinfo  {journal} {Nano Letters}\ }\textbf {\bibinfo {volume} {16}},\
  \bibinfo {pages} {3809} (\bibinfo {year} {2016})}\BibitemShut {NoStop}%
\bibitem [{\citenamefont {Zhu}\ \emph {et~al.}(2016)\citenamefont {Zhu},
  \citenamefont {Miyata}, \citenamefont {Fu}, \citenamefont {Wang},
  \citenamefont {Joshi}, \citenamefont {Niesner}, \citenamefont {Williams},
  \citenamefont {Jin},\ and\ \citenamefont {Zhu}}]{Zhu2016}%
  \BibitemOpen
  \bibfield  {author} {\bibinfo {author} {\bibfnamefont {H.}~\bibnamefont
  {Zhu}}, \bibinfo {author} {\bibfnamefont {K.}~\bibnamefont {Miyata}},
  \bibinfo {author} {\bibfnamefont {Y.}~\bibnamefont {Fu}}, \bibinfo {author}
  {\bibfnamefont {J.}~\bibnamefont {Wang}}, \bibinfo {author} {\bibfnamefont
  {P.~P.}\ \bibnamefont {Joshi}}, \bibinfo {author} {\bibfnamefont
  {D.}~\bibnamefont {Niesner}}, \bibinfo {author} {\bibfnamefont {K.~W.}\
  \bibnamefont {Williams}}, \bibinfo {author} {\bibfnamefont {S.}~\bibnamefont
  {Jin}},\ and\ \bibinfo {author} {\bibfnamefont {X.-Y.}\ \bibnamefont {Zhu}},\
  }\bibfield  {title} {\bibinfo {title} {Screening in crystalline liquids
  protects energetic carriers in hybrid perovskites},\ }\href
  {https://doi.org/10.1126/science.aaf9570} {\bibfield  {journal} {\bibinfo
  {journal} {Science}\ }\textbf {\bibinfo {volume} {353}},\ \bibinfo {pages}
  {1409} (\bibinfo {year} {2016})}\BibitemShut {NoStop}%
\bibitem [{\citenamefont {Akkerman}\ \emph {et~al.}(2018)\citenamefont
  {Akkerman}, \citenamefont {Martinez-Sarti}, \citenamefont {Goldoni},
  \citenamefont {Imran}, \citenamefont {Baranov}, \citenamefont {Bolink},
  \citenamefont {Palazon},\ and\ \citenamefont
  {Manna}}]{akkerman2018molecular}%
  \BibitemOpen
  \bibfield  {author} {\bibinfo {author} {\bibfnamefont {Q.~A.}\ \bibnamefont
  {Akkerman}}, \bibinfo {author} {\bibfnamefont {L.}~\bibnamefont
  {Martinez-Sarti}}, \bibinfo {author} {\bibfnamefont {L.}~\bibnamefont
  {Goldoni}}, \bibinfo {author} {\bibfnamefont {M.}~\bibnamefont {Imran}},
  \bibinfo {author} {\bibfnamefont {D.}~\bibnamefont {Baranov}}, \bibinfo
  {author} {\bibfnamefont {H.~J.}\ \bibnamefont {Bolink}}, \bibinfo {author}
  {\bibfnamefont {F.}~\bibnamefont {Palazon}},\ and\ \bibinfo {author}
  {\bibfnamefont {L.}~\bibnamefont {Manna}},\ }\bibfield  {title} {\bibinfo
  {title} {Molecular iodine for a general synthesis of binary and ternary
  inorganic and hybrid organic--inorganic iodide nanocrystals},\ }\href@noop {}
  {\bibfield  {journal} {\bibinfo  {journal} {Chemistry of Materials}\ }\textbf
  {\bibinfo {volume} {30}},\ \bibinfo {pages} {6915} (\bibinfo {year}
  {2018})}\BibitemShut {NoStop}%
\end{thebibliography}%


\begin{thebibliography}{15}%
\makeatletter
\providecommand \@ifxundefined [1]{%
 \@ifx{#1\undefined}
}%
\providecommand \@ifnum [1]{%
 \ifnum #1\expandafter \@firstoftwo
 \else \expandafter \@secondoftwo
 \fi
}%
\providecommand \@ifx [1]{%
 \ifx #1\expandafter \@firstoftwo
 \else \expandafter \@secondoftwo
 \fi
}%
\providecommand \natexlab [1]{#1}%
\providecommand \enquote  [1]{``#1''}%
\providecommand \bibnamefont  [1]{#1}%
\providecommand \bibfnamefont [1]{#1}%
\providecommand \citenamefont [1]{#1}%
\providecommand \href@noop [0]{\@secondoftwo}%
\providecommand \href [0]{\begingroup \@sanitize@url \@href}%
\providecommand \@href[1]{\@@startlink{#1}\@@href}%
\providecommand \@@href[1]{\endgroup#1\@@endlink}%
\providecommand \@sanitize@url [0]{\catcode `\\12\catcode `\$12\catcode
  `\&12\catcode `\#12\catcode `\^12\catcode `\_12\catcode `\%12\relax}%
\providecommand \@@startlink[1]{}%
\providecommand \@@endlink[0]{}%
\providecommand \url  [0]{\begingroup\@sanitize@url \@url }%
\providecommand \@url [1]{\endgroup\@href {#1}{\urlprefix }}%
\providecommand \urlprefix  [0]{URL }%
\providecommand \Eprint [0]{\href }%
\providecommand \doibase [0]{https://doi.org/}%
\providecommand \selectlanguage [0]{\@gobble}%
\providecommand \bibinfo  [0]{\@secondoftwo}%
\providecommand \bibfield  [0]{\@secondoftwo}%
\providecommand \translation [1]{[#1]}%
\providecommand \BibitemOpen [0]{}%
\providecommand \bibitemStop [0]{}%
\providecommand \bibitemNoStop [0]{.\EOS\space}%
\providecommand \EOS [0]{\spacefactor3000\relax}%
\providecommand \BibitemShut  [1]{\csname bibitem#1\endcsname}%
\let\auto@bib@innerbib\@empty
\bibitem [{\citenamefont {Toso}\ \emph {et~al.}(2021)\citenamefont {Toso},
  \citenamefont {Baranov}, \citenamefont {Altamura}, \citenamefont
  {Scattarella}, \citenamefont {Dahl}, \citenamefont {Wang}, \citenamefont
  {Marras}, \citenamefont {Alivisatos}, \citenamefont {Singer}, \citenamefont
  {Giannini} \emph {et~al.}}]{toso2021multilayer}%
  \BibitemOpen
  \bibfield  {author} {\bibinfo {author} {\bibfnamefont {S.}~\bibnamefont
  {Toso}}, \bibinfo {author} {\bibfnamefont {D.}~\bibnamefont {Baranov}},
  \bibinfo {author} {\bibfnamefont {D.}~\bibnamefont {Altamura}}, \bibinfo
  {author} {\bibfnamefont {F.}~\bibnamefont {Scattarella}}, \bibinfo {author}
  {\bibfnamefont {J.}~\bibnamefont {Dahl}}, \bibinfo {author} {\bibfnamefont
  {X.}~\bibnamefont {Wang}}, \bibinfo {author} {\bibfnamefont {S.}~\bibnamefont
  {Marras}}, \bibinfo {author} {\bibfnamefont {A.~P.}\ \bibnamefont
  {Alivisatos}}, \bibinfo {author} {\bibfnamefont {A.}~\bibnamefont {Singer}},
  \bibinfo {author} {\bibfnamefont {C.}~\bibnamefont {Giannini}}, \emph
  {et~al.},\ }\bibfield  {title} {\bibinfo {title} {Multilayer diffraction
  reveals that colloidal superlattices approach the structural perfection of
  single crystals},\ }\href@noop {} {\bibfield  {journal} {\bibinfo  {journal}
  {ACS nano}\ }\textbf {\bibinfo {volume} {15}},\ \bibinfo {pages} {6243}
  (\bibinfo {year} {2021})}\BibitemShut {NoStop}%
\bibitem [{\citenamefont {Toso}\ \emph {et~al.}(2022)\citenamefont {Toso},
  \citenamefont {Baranov}, \citenamefont {Filippi}, \citenamefont {Giannini},\
  and\ \citenamefont {Manna}}]{toso2022collective}%
  \BibitemOpen
  \bibfield  {author} {\bibinfo {author} {\bibfnamefont {S.}~\bibnamefont
  {Toso}}, \bibinfo {author} {\bibfnamefont {D.}~\bibnamefont {Baranov}},
  \bibinfo {author} {\bibfnamefont {U.}~\bibnamefont {Filippi}}, \bibinfo
  {author} {\bibfnamefont {C.}~\bibnamefont {Giannini}},\ and\ \bibinfo
  {author} {\bibfnamefont {L.}~\bibnamefont {Manna}},\ }\bibfield  {title}
  {\bibinfo {title} {Collective diffraction effects in perovskite nanocrystal
  superlattices},\ }\href@noop {} {\bibfield  {journal} {\bibinfo  {journal}
  {Accounts of Chemical Research}\ }\textbf {\bibinfo {volume} {56}},\ \bibinfo
  {pages} {66} (\bibinfo {year} {2022})}\BibitemShut {NoStop}%
\bibitem [{\citenamefont {Fabini}\ \emph {et~al.}(2016)\citenamefont {Fabini},
  \citenamefont {Stoumpos}, \citenamefont {Laurita}, \citenamefont
  {Kaltzoglou}, \citenamefont {Kontos}, \citenamefont {Falaras}, \citenamefont
  {Kanatzidis},\ and\ \citenamefont {Seshadri}}]{fabini2016reentrant}%
  \BibitemOpen
  \bibfield  {author} {\bibinfo {author} {\bibfnamefont {D.~H.}\ \bibnamefont
  {Fabini}}, \bibinfo {author} {\bibfnamefont {C.~C.}\ \bibnamefont
  {Stoumpos}}, \bibinfo {author} {\bibfnamefont {G.}~\bibnamefont {Laurita}},
  \bibinfo {author} {\bibfnamefont {A.}~\bibnamefont {Kaltzoglou}}, \bibinfo
  {author} {\bibfnamefont {A.~G.}\ \bibnamefont {Kontos}}, \bibinfo {author}
  {\bibfnamefont {P.}~\bibnamefont {Falaras}}, \bibinfo {author} {\bibfnamefont
  {M.~G.}\ \bibnamefont {Kanatzidis}},\ and\ \bibinfo {author} {\bibfnamefont
  {R.}~\bibnamefont {Seshadri}},\ }\bibfield  {title} {\bibinfo {title}
  {Reentrant structural and optical properties and large positive thermal
  expansion in perovskite formamidinium lead iodide},\ }\href@noop {}
  {\bibfield  {journal} {\bibinfo  {journal} {Angewandte Chemie}\ }\textbf
  {\bibinfo {volume} {128}},\ \bibinfo {pages} {15618} (\bibinfo {year}
  {2016})}\BibitemShut {NoStop}%
\bibitem [{\citenamefont {Weber}\ \emph {et~al.}(2018)\citenamefont {Weber},
  \citenamefont {Ghosh}, \citenamefont {Gaines}, \citenamefont {Henry},
  \citenamefont {Walker}, \citenamefont {Islam},\ and\ \citenamefont
  {Weller}}]{weber2018phase}%
  \BibitemOpen
  \bibfield  {author} {\bibinfo {author} {\bibfnamefont {O.~J.}\ \bibnamefont
  {Weber}}, \bibinfo {author} {\bibfnamefont {D.}~\bibnamefont {Ghosh}},
  \bibinfo {author} {\bibfnamefont {S.}~\bibnamefont {Gaines}}, \bibinfo
  {author} {\bibfnamefont {P.~F.}\ \bibnamefont {Henry}}, \bibinfo {author}
  {\bibfnamefont {A.~B.}\ \bibnamefont {Walker}}, \bibinfo {author}
  {\bibfnamefont {M.~S.}\ \bibnamefont {Islam}},\ and\ \bibinfo {author}
  {\bibfnamefont {M.~T.}\ \bibnamefont {Weller}},\ }\bibfield  {title}
  {\bibinfo {title} {Phase behavior and polymorphism of formamidinium lead
  iodide},\ }\href@noop {} {\bibfield  {journal} {\bibinfo  {journal}
  {Chemistry of Materials}\ }\textbf {\bibinfo {volume} {30}},\ \bibinfo
  {pages} {3768} (\bibinfo {year} {2018})}\BibitemShut {NoStop}%
\bibitem [{\citenamefont {Berg}\ \emph {et~al.}(2019)\citenamefont {Berg},
  \citenamefont {Kutra}, \citenamefont {Kroeger}, \citenamefont {Straehle},
  \citenamefont {Kausler}, \citenamefont {Haubold}, \citenamefont {Schiegg},
  \citenamefont {Ales}, \citenamefont {Beier}, \citenamefont {Rudy},
  \citenamefont {Eren}, \citenamefont {Cervantes}, \citenamefont {Xu},
  \citenamefont {Beuttenmueller}, \citenamefont {Wolny}, \citenamefont {Zhang},
  \citenamefont {Koethe}, \citenamefont {Hamprecht},\ and\ \citenamefont
  {Kreshuk}}]{berg2019}%
  \BibitemOpen
  \bibfield  {author} {\bibinfo {author} {\bibfnamefont {S.}~\bibnamefont
  {Berg}}, \bibinfo {author} {\bibfnamefont {D.}~\bibnamefont {Kutra}},
  \bibinfo {author} {\bibfnamefont {T.}~\bibnamefont {Kroeger}}, \bibinfo
  {author} {\bibfnamefont {C.~N.}\ \bibnamefont {Straehle}}, \bibinfo {author}
  {\bibfnamefont {B.~X.}\ \bibnamefont {Kausler}}, \bibinfo {author}
  {\bibfnamefont {C.}~\bibnamefont {Haubold}}, \bibinfo {author} {\bibfnamefont
  {M.}~\bibnamefont {Schiegg}}, \bibinfo {author} {\bibfnamefont
  {J.}~\bibnamefont {Ales}}, \bibinfo {author} {\bibfnamefont {T.}~\bibnamefont
  {Beier}}, \bibinfo {author} {\bibfnamefont {M.}~\bibnamefont {Rudy}},
  \bibinfo {author} {\bibfnamefont {K.}~\bibnamefont {Eren}}, \bibinfo {author}
  {\bibfnamefont {J.~I.}\ \bibnamefont {Cervantes}}, \bibinfo {author}
  {\bibfnamefont {B.}~\bibnamefont {Xu}}, \bibinfo {author} {\bibfnamefont
  {F.}~\bibnamefont {Beuttenmueller}}, \bibinfo {author} {\bibfnamefont
  {A.}~\bibnamefont {Wolny}}, \bibinfo {author} {\bibfnamefont
  {C.}~\bibnamefont {Zhang}}, \bibinfo {author} {\bibfnamefont
  {U.}~\bibnamefont {Koethe}}, \bibinfo {author} {\bibfnamefont {F.~A.}\
  \bibnamefont {Hamprecht}},\ and\ \bibinfo {author} {\bibfnamefont
  {A.}~\bibnamefont {Kreshuk}},\ }\bibfield  {title} {\bibinfo {title}
  {ilastik: interactive machine learning for (bio)image analysis},\ }\bibfield
  {journal} {\bibinfo  {journal} {Nature Methods}\ }\href
  {https://doi.org/10.1038/s41592-019-0582-9} {10.1038/s41592-019-0582-9}
  (\bibinfo {year} {2019})\BibitemShut {NoStop}%
\bibitem [{\citenamefont {Elliott}(1957)}]{elliott1957intensity}%
  \BibitemOpen
  \bibfield  {author} {\bibinfo {author} {\bibfnamefont {R.}~\bibnamefont
  {Elliott}},\ }\bibfield  {title} {\bibinfo {title} {Intensity of optical
  absorption by excitons},\ }\href@noop {} {\bibfield  {journal} {\bibinfo
  {journal} {Physical Review}\ }\textbf {\bibinfo {volume} {108}},\ \bibinfo
  {pages} {1384} (\bibinfo {year} {1957})}\BibitemShut {NoStop}%
\bibitem [{\citenamefont {Yang}\ \emph {et~al.}(2016)\citenamefont {Yang},
  \citenamefont {Yang}, \citenamefont {Zhu}, \citenamefont {Johnson},
  \citenamefont {Berry}, \citenamefont {Van De~Lagemaat},\ and\ \citenamefont
  {Beard}}]{yang2016large}%
  \BibitemOpen
  \bibfield  {author} {\bibinfo {author} {\bibfnamefont {Y.}~\bibnamefont
  {Yang}}, \bibinfo {author} {\bibfnamefont {M.}~\bibnamefont {Yang}}, \bibinfo
  {author} {\bibfnamefont {K.}~\bibnamefont {Zhu}}, \bibinfo {author}
  {\bibfnamefont {J.~C.}\ \bibnamefont {Johnson}}, \bibinfo {author}
  {\bibfnamefont {J.~J.}\ \bibnamefont {Berry}}, \bibinfo {author}
  {\bibfnamefont {J.}~\bibnamefont {Van De~Lagemaat}},\ and\ \bibinfo {author}
  {\bibfnamefont {M.~C.}\ \bibnamefont {Beard}},\ }\bibfield  {title} {\bibinfo
  {title} {Large polarization-dependent exciton optical stark effect in lead
  iodide perovskites},\ }\href@noop {} {\bibfield  {journal} {\bibinfo
  {journal} {Nature communications}\ }\textbf {\bibinfo {volume} {7}},\
  \bibinfo {pages} {12613} (\bibinfo {year} {2016})}\BibitemShut {NoStop}%
\bibitem [{\citenamefont {Saba}\ \emph {et~al.}(2014)\citenamefont {Saba},
  \citenamefont {Cadelano}, \citenamefont {Marongiu}, \citenamefont {Chen},
  \citenamefont {Sarritzu}, \citenamefont {Sestu}, \citenamefont {Figus},
  \citenamefont {Aresti}, \citenamefont {Piras}, \citenamefont {Geddo~Lehmann}
  \emph {et~al.}}]{saba2014correlated}%
  \BibitemOpen
  \bibfield  {author} {\bibinfo {author} {\bibfnamefont {M.}~\bibnamefont
  {Saba}}, \bibinfo {author} {\bibfnamefont {M.}~\bibnamefont {Cadelano}},
  \bibinfo {author} {\bibfnamefont {D.}~\bibnamefont {Marongiu}}, \bibinfo
  {author} {\bibfnamefont {F.}~\bibnamefont {Chen}}, \bibinfo {author}
  {\bibfnamefont {V.}~\bibnamefont {Sarritzu}}, \bibinfo {author}
  {\bibfnamefont {N.}~\bibnamefont {Sestu}}, \bibinfo {author} {\bibfnamefont
  {C.}~\bibnamefont {Figus}}, \bibinfo {author} {\bibfnamefont
  {M.}~\bibnamefont {Aresti}}, \bibinfo {author} {\bibfnamefont
  {R.}~\bibnamefont {Piras}}, \bibinfo {author} {\bibfnamefont
  {A.}~\bibnamefont {Geddo~Lehmann}}, \emph {et~al.},\ }\bibfield  {title}
  {\bibinfo {title} {Correlated electron-hole plasma in organometal
  perovskites},\ }\href@noop {} {\bibfield  {journal} {\bibinfo  {journal}
  {Nature communications}\ }\textbf {\bibinfo {volume} {5}},\ \bibinfo {pages}
  {5049} (\bibinfo {year} {2014})}\BibitemShut {NoStop}%
\bibitem [{\citenamefont {Yang}\ \emph {et~al.}(2015)\citenamefont {Yang},
  \citenamefont {Yan}, \citenamefont {Yang}, \citenamefont {Choi},
  \citenamefont {Zhu}, \citenamefont {Luther},\ and\ \citenamefont
  {Beard}}]{yang2015low}%
  \BibitemOpen
  \bibfield  {author} {\bibinfo {author} {\bibfnamefont {Y.}~\bibnamefont
  {Yang}}, \bibinfo {author} {\bibfnamefont {Y.}~\bibnamefont {Yan}}, \bibinfo
  {author} {\bibfnamefont {M.}~\bibnamefont {Yang}}, \bibinfo {author}
  {\bibfnamefont {S.}~\bibnamefont {Choi}}, \bibinfo {author} {\bibfnamefont
  {K.}~\bibnamefont {Zhu}}, \bibinfo {author} {\bibfnamefont {J.~M.}\
  \bibnamefont {Luther}},\ and\ \bibinfo {author} {\bibfnamefont {M.~C.}\
  \bibnamefont {Beard}},\ }\bibfield  {title} {\bibinfo {title} {Low surface
  recombination velocity in solution-grown \ch{CH3NH3PbBr3} perovskite single
  crystal},\ }\href@noop {} {\bibfield  {journal} {\bibinfo  {journal} {Nature
  communications}\ }\textbf {\bibinfo {volume} {6}},\ \bibinfo {pages} {7961}
  (\bibinfo {year} {2015})}\BibitemShut {NoStop}%
\bibitem [{\citenamefont {Yamamoto}\ \emph {et~al.}(2001)\citenamefont
  {Yamamoto}, \citenamefont {Miyajima}, \citenamefont {Goto}, \citenamefont
  {Ju~Ko},\ and\ \citenamefont {Yao}}]{yamamoto2001biexciton}%
  \BibitemOpen
  \bibfield  {author} {\bibinfo {author} {\bibfnamefont {A.}~\bibnamefont
  {Yamamoto}}, \bibinfo {author} {\bibfnamefont {K.}~\bibnamefont {Miyajima}},
  \bibinfo {author} {\bibfnamefont {T.}~\bibnamefont {Goto}}, \bibinfo {author}
  {\bibfnamefont {H.}~\bibnamefont {Ju~Ko}},\ and\ \bibinfo {author}
  {\bibfnamefont {T.}~\bibnamefont {Yao}},\ }\bibfield  {title} {\bibinfo
  {title} {Biexciton luminescence in high-quality \ch{ZnO} epitaxial thin
  films},\ }\href@noop {} {\bibfield  {journal} {\bibinfo  {journal} {Journal
  of Applied Physics}\ }\textbf {\bibinfo {volume} {90}},\ \bibinfo {pages}
  {4973} (\bibinfo {year} {2001})}\BibitemShut {NoStop}%
\bibitem [{\citenamefont {You}\ \emph {et~al.}(2015)\citenamefont {You},
  \citenamefont {Zhang}, \citenamefont {Berkelbach}, \citenamefont {Hybertsen},
  \citenamefont {Reichman},\ and\ \citenamefont {Heinz}}]{you2015observation}%
  \BibitemOpen
  \bibfield  {author} {\bibinfo {author} {\bibfnamefont {Y.}~\bibnamefont
  {You}}, \bibinfo {author} {\bibfnamefont {X.-X.}\ \bibnamefont {Zhang}},
  \bibinfo {author} {\bibfnamefont {T.~C.}\ \bibnamefont {Berkelbach}},
  \bibinfo {author} {\bibfnamefont {M.~S.}\ \bibnamefont {Hybertsen}}, \bibinfo
  {author} {\bibfnamefont {D.~R.}\ \bibnamefont {Reichman}},\ and\ \bibinfo
  {author} {\bibfnamefont {T.~F.}\ \bibnamefont {Heinz}},\ }\bibfield  {title}
  {\bibinfo {title} {Observation of biexcitons in monolayer \ch{WSe_2}},\
  }\href@noop {} {\bibfield  {journal} {\bibinfo  {journal} {Nature Physics}\
  }\textbf {\bibinfo {volume} {11}},\ \bibinfo {pages} {477} (\bibinfo {year}
  {2015})}\BibitemShut {NoStop}%
\bibitem [{\citenamefont {Siemens}\ \emph {et~al.}(2010)\citenamefont
  {Siemens}, \citenamefont {Moody}, \citenamefont {Li}, \citenamefont
  {Bristow},\ and\ \citenamefont {Cundiff}}]{siemens2010resonance}%
  \BibitemOpen
  \bibfield  {author} {\bibinfo {author} {\bibfnamefont {M.~E.}\ \bibnamefont
  {Siemens}}, \bibinfo {author} {\bibfnamefont {G.}~\bibnamefont {Moody}},
  \bibinfo {author} {\bibfnamefont {H.}~\bibnamefont {Li}}, \bibinfo {author}
  {\bibfnamefont {A.~D.}\ \bibnamefont {Bristow}},\ and\ \bibinfo {author}
  {\bibfnamefont {S.~T.}\ \bibnamefont {Cundiff}},\ }\bibfield  {title}
  {\bibinfo {title} {Resonance lineshapes in two-dimensional fourier transform
  spectroscopy},\ }\href@noop {} {\bibfield  {journal} {\bibinfo  {journal}
  {Optics express}\ }\textbf {\bibinfo {volume} {18}},\ \bibinfo {pages}
  {17699} (\bibinfo {year} {2010})}\BibitemShut {NoStop}%
\bibitem [{\citenamefont {Mukamel}(1995)}]{mukamel1995principles}%
  \BibitemOpen
  \bibfield  {author} {\bibinfo {author} {\bibfnamefont {S.}~\bibnamefont
  {Mukamel}},\ }\href {https://books.google.be/books?id=k_7uAAAAMAAJ} {\emph
  {\bibinfo {title} {Principles of Nonlinear Optical Spectroscopy}}},\ Oxford
  series in optical and imaging sciences\ (\bibinfo  {publisher} {Oxford
  University Press},\ \bibinfo {year} {1995})\BibitemShut {NoStop}%
\bibitem [{\citenamefont {Russo}\ \emph {et~al.}(2022)\citenamefont {Russo},
  \citenamefont {McGhee}, \citenamefont {Virgili}, \citenamefont {Lidzey},
  \citenamefont {Cerullo},\ and\ \citenamefont {Maiuri}}]{russo2022dephasing}%
  \BibitemOpen
  \bibfield  {author} {\bibinfo {author} {\bibfnamefont {M.}~\bibnamefont
  {Russo}}, \bibinfo {author} {\bibfnamefont {K.~E.}\ \bibnamefont {McGhee}},
  \bibinfo {author} {\bibfnamefont {T.}~\bibnamefont {Virgili}}, \bibinfo
  {author} {\bibfnamefont {D.~G.}\ \bibnamefont {Lidzey}}, \bibinfo {author}
  {\bibfnamefont {G.}~\bibnamefont {Cerullo}},\ and\ \bibinfo {author}
  {\bibfnamefont {M.}~\bibnamefont {Maiuri}},\ }\bibfield  {title} {\bibinfo
  {title} {Dephasing processes in the molecular dye lumogen-f orange
  characterized by two-dimensional electronic spectroscopy},\ }\href@noop {}
  {\bibfield  {journal} {\bibinfo  {journal} {Molecules}\ }\textbf {\bibinfo
  {volume} {27}},\ \bibinfo {pages} {7095} (\bibinfo {year}
  {2022})}\BibitemShut {NoStop}%
\bibitem [{\citenamefont {Alonso}\ \emph {et~al.}(2019)\citenamefont {Alonso},
  \citenamefont {Charles}, \citenamefont {Francisco-L{\'o}pez}, \citenamefont
  {Garriga}, \citenamefont {Weller},\ and\ \citenamefont
  {Go{\~n}i}}]{alonso2019spectroscopic}%
  \BibitemOpen
  \bibfield  {author} {\bibinfo {author} {\bibfnamefont {M.~I.}\ \bibnamefont
  {Alonso}}, \bibinfo {author} {\bibfnamefont {B.}~\bibnamefont {Charles}},
  \bibinfo {author} {\bibfnamefont {A.}~\bibnamefont {Francisco-L{\'o}pez}},
  \bibinfo {author} {\bibfnamefont {M.}~\bibnamefont {Garriga}}, \bibinfo
  {author} {\bibfnamefont {M.~T.}\ \bibnamefont {Weller}},\ and\ \bibinfo
  {author} {\bibfnamefont {A.~R.}\ \bibnamefont {Go{\~n}i}},\ }\bibfield
  {title} {\bibinfo {title} {Spectroscopic ellipsometry study of \ch{FA_xMA_{1-
  x}PbI_3} hybrid perovskite single crystals},\ }\href@noop {} {\bibfield
  {journal} {\bibinfo  {journal} {Journal of Vacuum Science \& Technology B}\
  }\textbf {\bibinfo {volume} {37}} (\bibinfo {year} {2019})}\BibitemShut
  {NoStop}%
\end{thebibliography}%
\end{document}


\author{Alessandra Milloch}
\email{alessandra.milloch@unicatt.it}
\affiliation{Department of Mathematics and Physics, Università Cattolica del Sacro Cuore, Brescia I-25133, Italy}
\affiliation{ILAMP (Interdisciplinary Laboratories for Advanced
Materials Physics), Università Cattolica del Sacro Cuore, Brescia I-25133, Italy}
\affiliation{Department of Physics and Astronomy, KU Leuven, B-3001 Leuven, Belgium}

\author{Umberto Filippi}
\affiliation{Italian Institute of Technology (IIT), Genova 16163, Italy}

\author{Paolo Franceschini}
\affiliation{CNR-INO (National Institute of Optics), via Branze 45, 25123 Brescia, Italy}
\affiliation{Department of Information Engineering, University of Brescia, Brescia I-25123, Italy}

\author{Selene Mor}
\affiliation{Department of Mathematics and Physics, Università Cattolica del Sacro Cuore, Brescia I-25133, Italy}
\affiliation{ILAMP (Interdisciplinary Laboratories for Advanced
Materials Physics), Università Cattolica del Sacro Cuore, Brescia I-25133, Italy}

\author{Stefania Pagliara}
\affiliation{Department of Mathematics and Physics, Università Cattolica del Sacro Cuore, Brescia I-25133, Italy}
\affiliation{ILAMP (Interdisciplinary Laboratories for Advanced
Materials Physics), Università Cattolica del Sacro Cuore, Brescia I-25133, Italy}

\author{Gabriele Ferrini}
\affiliation{Department of Mathematics and Physics, Università Cattolica del Sacro Cuore, Brescia I-25133, Italy}
\affiliation{ILAMP (Interdisciplinary Laboratories for Advanced
Materials Physics), Università Cattolica del Sacro Cuore, Brescia I-25133, Italy}

\author{Franco V. A. Camargo}
\affiliation{IFN-CNR, Piazza Leonardo da Vinci 32, I-20133, Milano, Italy}

\author{Giulio Cerullo}
\affiliation{IFN-CNR, Piazza Leonardo da Vinci 32, I-20133, Milano, Italy}
\affiliation{Department of Physics, Politecnico di Milano, Piazza Leonardo da Vinci 32, I-20133 Milano, Italy}

\author{Dmitry Baranov}
\affiliation{Italian Institute of Technology (IIT), Genova 16163, Italy}
\affiliation{Division of Chemical Physics, Department of Chemistry, Lund University, P.O. Box 124, SE-221 00 Lund, Sweden}
 
\author{Liberato Manna}
\affiliation{Italian Institute of Technology (IIT), Genova 16163, Italy}

\author{Claudio Giannetti}
\email{claudio.giannetti@unicatt.it}
\affiliation{Department of Mathematics and Physics, Università Cattolica del Sacro Cuore, Brescia I-25133, Italy}
\affiliation{ILAMP (Interdisciplinary Laboratories for Advanced
Materials Physics), Università Cattolica del Sacro Cuore, Brescia I-25133, Italy}
\affiliation{CNR-INO (National Institute of Optics), via Branze 45, 25123 Brescia, Italy}

\title{Supporting Information for \\ The fate of optical excitons in \ch{FAPbI_3} nanocube superlattices}

\maketitle

\section{Sample characterization}

\subsection{X-ray diffraction}
\begin{center}
\begin{figure}[h!]
    \centering    
    \includegraphics[width=0.9\textwidth]{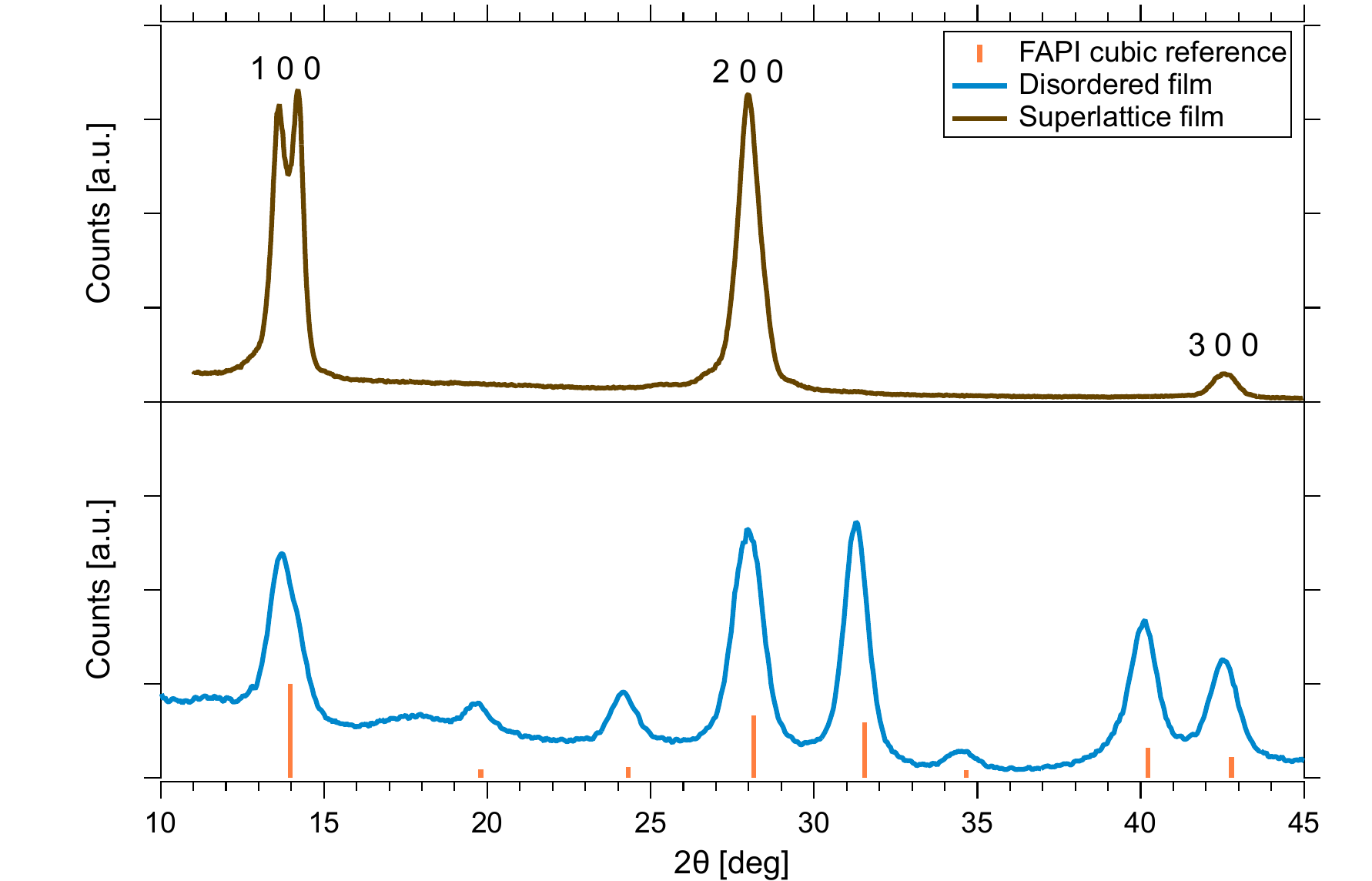} 
    \caption{XRD pattern of a nanocrystal superlattice film (brown solid line, top panel) and of a disordered film composed of randomly oriented nanocubes (blue solid line, bottom panel). The reference for the cubic structure (orange lines) is taken from Ref. \citenum{fabini2016reentrant}.}
    \label{SI - fig: XRD_SL}
\end{figure}
\end{center}

X-ray diffraction (XRD) patterns were acquired with a Panalytical Empyrean diffractometer in a parallel beam configuration, equipped with a Cu K$\alpha$ ($\lambda$ = 1.5406 \AA) ceramic X-ray tube operating at 45 kV and 40 mA, 1 mm wide incident and receiving slits, and a 40 mA PIXcel3D 2×2 two-dimensional detector. The XRD pattern of a film of FAPI NC superlattices is shown in Figure \ref{SI - fig: XRD_SL} (top panel) and compared to the XRD pattern of a film of randomly oriented nanocubes (bottom panel). This disordered sample is prepared by collecting the precipitate after the third centrifugation step (before dissolving it in toluene) and by spreading it on a silicon substrate with a plastic scoop.

The first peculiar feature that characterizes the XRD pattern of superlattices in Figure \ref{SI - fig: XRD_SL} is the enhancement of some reflections with respect to others which belong to the same cubic phase of FAPI. This is a consequence of preferential orientation which comes from the fact that nanocrystals assemble with the (100) planes parallel to the substrate. Conversely, XRD pattern of the disordered sample displays all the peaks of the perovskite cubic structure. 
The second feature that marks the presence of superlattices is the splitting of the first Bragg reflection (2$\theta$ = 13.87$^{\circ}$): as explained in detail in previous works (Refs. \citenum{toso2021multilayer} and \citenum{toso2022collective}), it originates from an additional interference which comes from the exact periodicity of NCs inside SLs and results in the fine structure observable in Figure \ref{SI - fig: XRD_SL}.

\begin{figure}[h!]
    \centering    
    \includegraphics[width=0.8\textwidth]{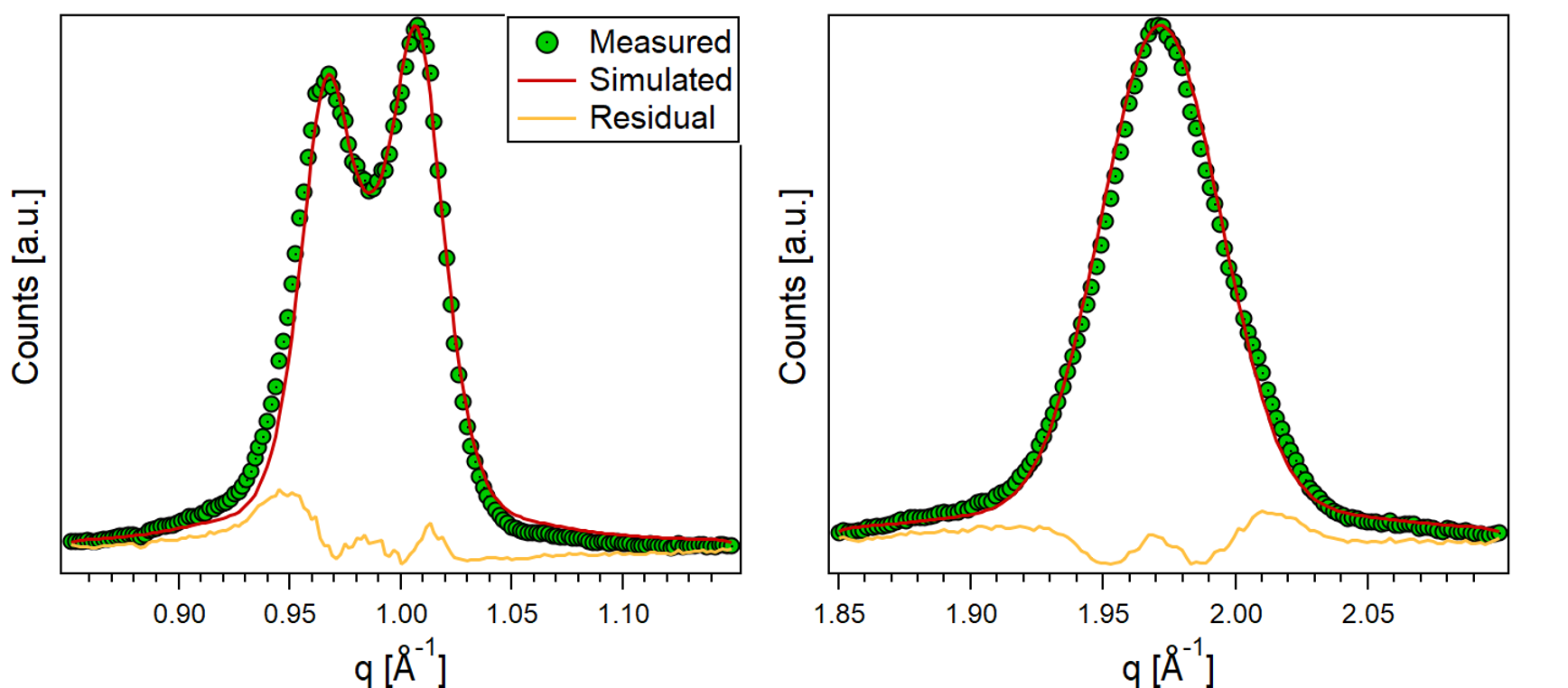} 
    \caption{Fitting of the first and second Bragg peak from the XRD pattern of the FAPI NC SL film by means of the multilayer diffraction model described in Ref. \cite{toso2021multilayer}.}
    \label{SI - fig: fit_XRD}
\end{figure}

By exploiting the multilayer diffraction fitting routine introduced in Ref. \citenum{toso2021multilayer}, it is possible to extract important information about nanocrystals and their packing order. The fitted data are plotted in Figure \ref{SI - fig: fit_XRD} and fitting results are listed in table \ref{SI - tab: fit}. We observe that our nanocrystals have a broad size distribution (NC edge length $= 7.9 \pm2.9 $ nm).  Despite this, there is a high level of packing order as indicated by the parameter $\sigma_L$, which represents the statistical fluctuation of the nanocrystal-to-nanocrystal distance ($\sigma_L = 1.688\, $\r{A}).

\FloatBarrier
\begin{center}
\begin{table}[t]
\caption{\textbf{Results obtained by fitting the XRD pattern shown in Figure \ref{SI - fig: fit_XRD}.}}
    \centering
\begin{tabular}{|c | c | c | c}
\toprule[1.25pt]
    \bf{Parameter} & \bf{Definition} & \bf{Fit value}   
    \\ \midrule[1.25pt] \rowcolor[gray]{.95}
d [\r{A}]  & Lattice constant &  6.389    \\ 
\midrule
L [\r{A}]  &  Interparticle spacing  &   28.920                              \\\rowcolor[gray]{.95}
\midrule
$\sigma_L$ [\r{A}]  & Interparticle spacing fluctuation &  1.688                              \\
\midrule
N [atomic planes]  & Average number of atomic planes per NC &  13.42                           \\ \rowcolor[gray]{.95}
\midrule
$\sigma_N$ [atomic planes]  & Size distribution of NCs & 4.53                               \\  
\midrule
q-zero correction [\r{A}]  & Correction of diffractometer misalignment&  -0.004                               \\ \rowcolor[gray]{.95}
\midrule
NC edge [nm]  & (N-1)$\times$d &  7.9 $\pm$ 2.9\\
\bottomrule[1.25pt]
\end{tabular}
\label{SI - tab: fit}
\end{table}
\end{center}

\begin{figure}[h!]
    \centering    
    \includegraphics[width=0.95\textwidth]{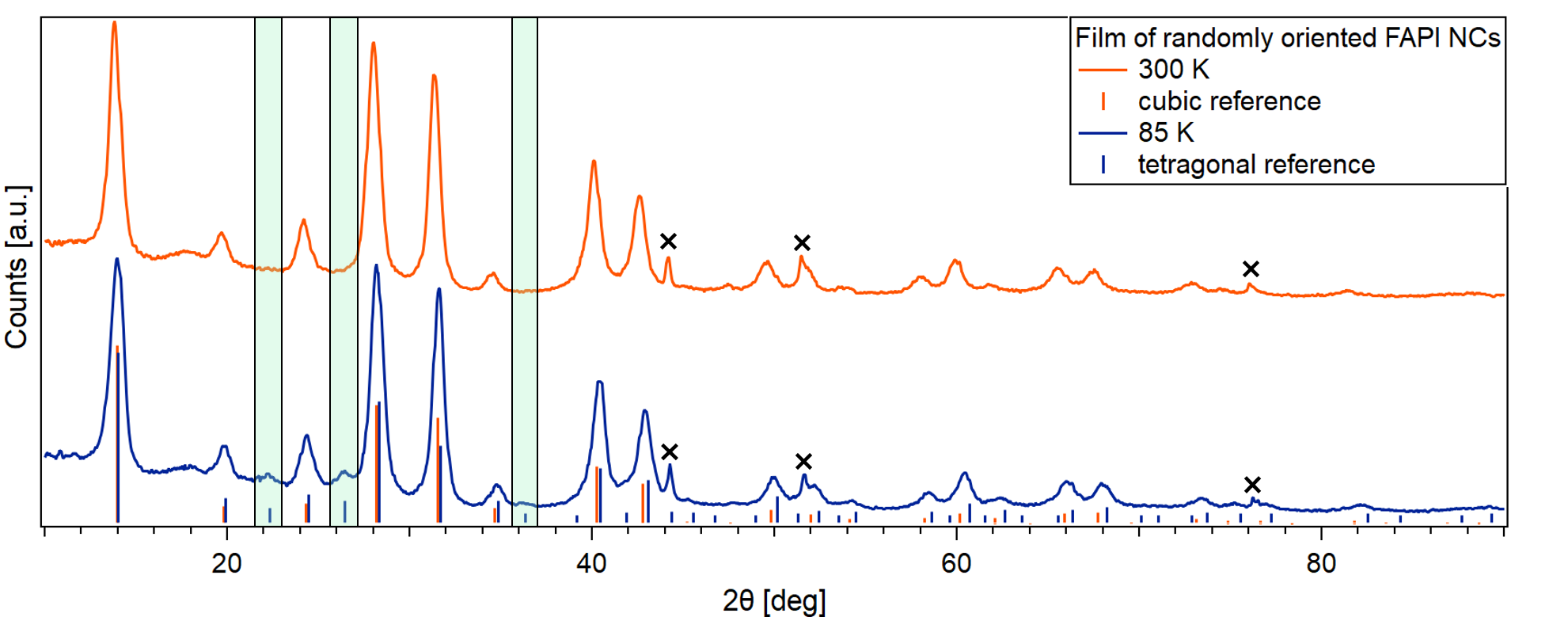} 
    \caption{XRD patterns at 300 K and at 85 K of a film of randomly oriented FAPI NCs (Ref. \citenum{fabini2016reentrant}). The transparent light blue columns emphasize the new peaks arising with the phase transition. The "\textbf{$\times$}" symbol indicates the impurity peaks belonging to the FeNi$_3$ sample holder (ICSD number 188242).}
    \label{SI - fig: lowT_XRD}
\end{figure}

\newpage
In order to confirm the continuous phase transition from the room temperature cubic $\alpha$-phase ($Pm\bar3m$) to the low temperature tetragonal $\beta$-phase ($P4/mbm$) expected for the FAPI NCs (occurring around 285 K in the bulk counterpart \cite{fabini2016reentrant,weber2018phase}), we performed low-temperature XRD measurements in vacuum atmosphere. In order to appreciate all the Bragg reflections belonging to the different crystal phases we performed the measurement on a film of randomly oriented nanocrystals. The resulting XRD patterns are shown in Figure \ref{SI - fig: lowT_XRD}: it is possible to notice that upon cooling we observe a shifting of the peaks towards higher angles, consistent with a shrinking of the lattice parameters expected with the decreasing of temperature, and the appearing of three new peaks, highlighted by transparent light blue columns, at $2\theta =$ 22.2$^\circ$, 26.3$^\circ$$^\circ$ and 36.2$^\circ$, which belong to the low temperature tetragonal $\beta$-phase \cite{fabini2016reentrant}, and confirm the transition from the high symmetry cubic phase to the lower symmetry tetragonal phase.

\FloatBarrier

\newpage
\subsection{Optical microscopy}

\begin{figure}[h!]
    \centering    
    \includegraphics[width=0.95\textwidth]{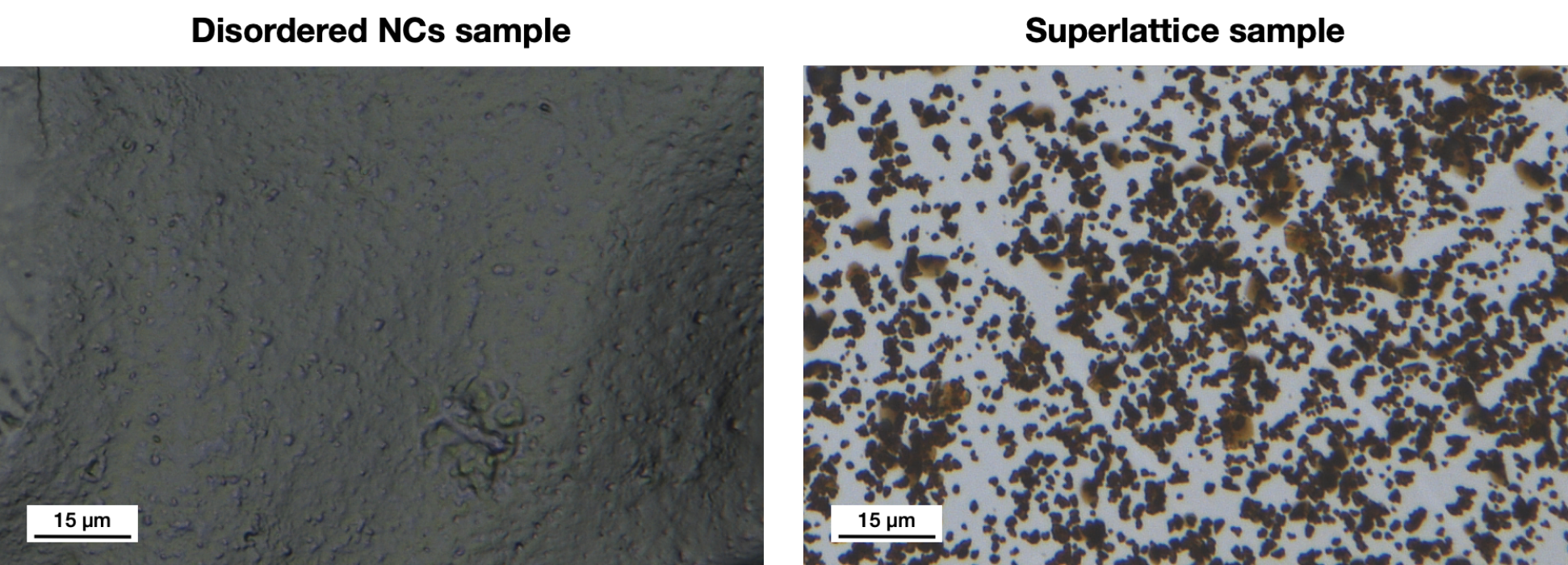} 
    \caption{Optical microscopy images of a film of randomly oriented nanocubes (left panel) and of a nanocube superlattices sample (right panel).}
    \label{SI - fig: optical microscopy}
\end{figure}

Optical microscope images of both the superlattice and the disordered samples are acquired on a ZETA-20 true color 3D optical profiler and are displayed in Figure \ref{SI - fig: optical microscopy}. For the disordered NCs film (left panel), the image shows a roughly homogeneous spatial distribution of deposited material, suggesting that mechanical scrambling produces largely disordered NC film. Conversely, optical microscopy images of the superlattice sample (right panel), display $\sim 1 $ \textmu m aggregates each corresponding to a superlattice of FAPI nanocubes.

\subsection{Transmission electron microscopy}

\begin{figure}[h!]
    \centering    
    \includegraphics[width=0.8\textwidth]{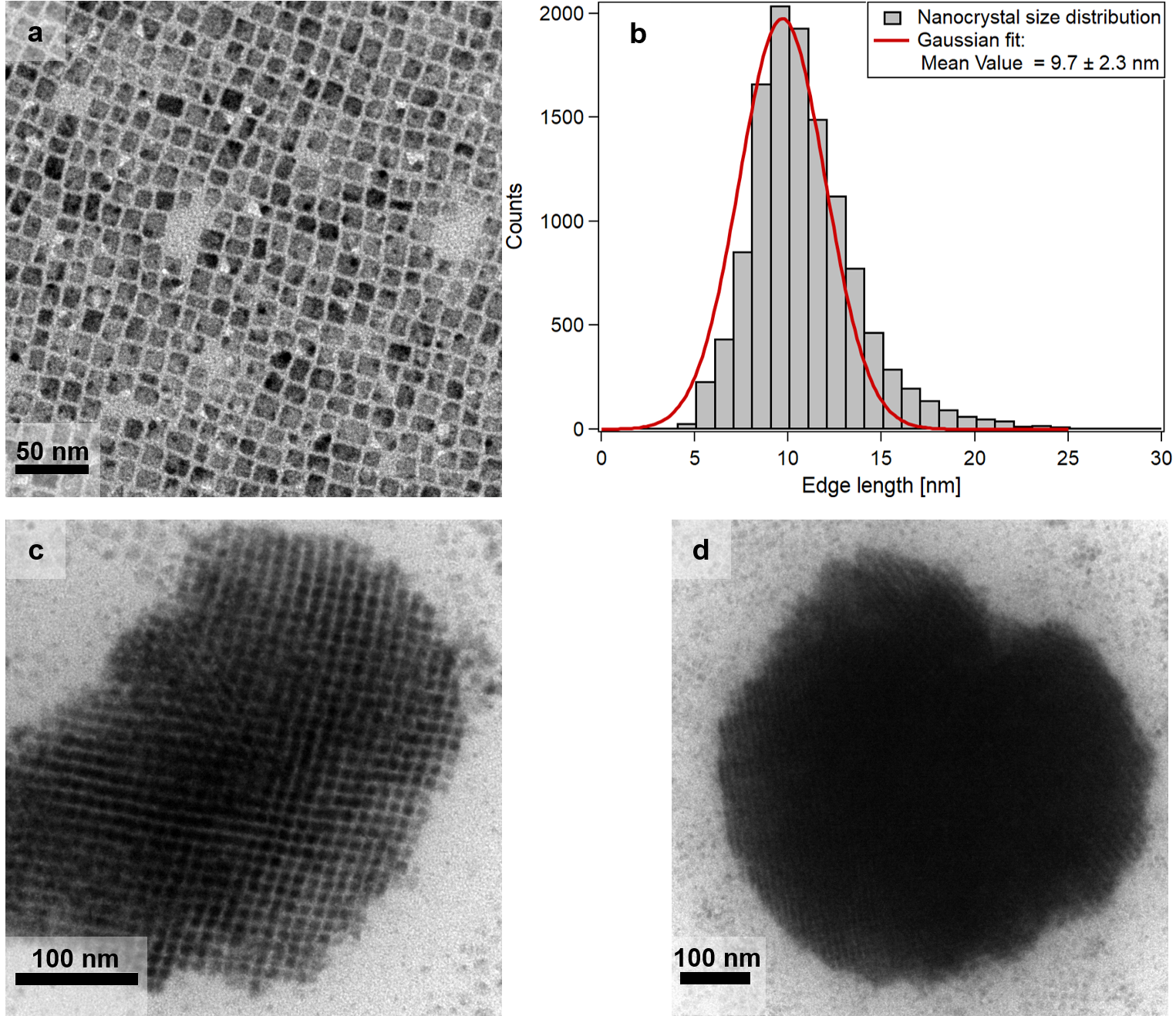} 
    \caption{a) TEM picture of FAPI NCs. b) NC size distribution calculated with ilastik software. c,d) TEM pictures of FAPI NCs superlattices.}
    \label{SI - fig: TEM_images}
\end{figure}

Transmission electron microscopy (TEM) characterization of FAPI nanocrystals and nanocrystal superlattices was performed
on a JEOL JEM 1400-Plus microscope operating at 120 kV
accelerating voltage. The samples for TEM were prepared by drop casting sample dispersion in toluene on top of a carbon-coated copper grid. To measure the size distribution of the FAPI nanocrystals 10 TEM images (similar to the one shown in Figure \ref{SI - fig: TEM_images}a) were acquired and analyzed by mean of the open source software ilastik \cite{berg2019}, which allows to select the NCs in the image and quantify their size by treating their area as a perfect square. The resulting edge length of $9.7 \pm 2.3$ nm (for a total number of measured NCs $\approx$ 10000) differs from the result obtained through the XRD analysis ($7.9 \pm 2.9$ nm). The reason could be that the NCs are not perfect cubes but rather a rectangular cuboid with a slightly shorter edge (see Fig. \ref{SI - fig: TEM_images}a), and when they self-assemble in SLs the short edge results to be in the direction perpendicular to the substrate, which is the direction of detection for X-rays.

\FloatBarrier

\clearpage
\subsection{Photoluminescence and absorbance}
\begin{figure}[h!]
\includegraphics[width=10cm]{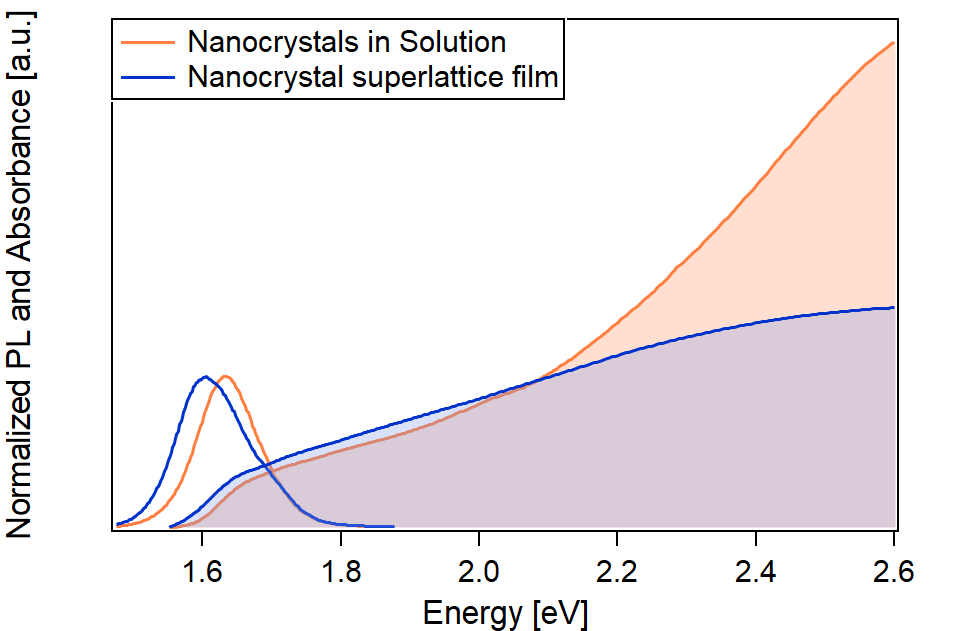}
\caption{Photoluminescence (solid lines) and absorption (filled areas) spectra of NCs in toluene dispersion and NCs superlattices.}
\label{SI fig: PL&abs}
\end{figure}

Absorption and photoluminescence (PL) spectra were acquired at room temperature with
a Cary 300 spectrophotometer and with a Cary Eclipse spectrofluorometer, respectively. The equilibrium PL spectrum, measured under 350 nm excitation, and the absorption spectrum of NCs dispersed in toluene and of NC superlattices deposited on a glass substrate are displayed in Figure \ref{SI fig: PL&abs}. We observe a red-shift of the PL of NCs assembled into superlattices, compared to \if on glass substrate with respect \fi  NCs in toluene dispersion (max ${PL_{superlattice}} - $max $PL_{solution} = 1605 $ meV $ - 1635 $ meV $ = - 30 $ meV) together with a PL broadening (FWHM$_{superlattice} -$ FWHM$_{solution} = 122$ meV $- 102$ meV $= 20$ meV). 
\FloatBarrier

\begin{figure}[h!]
\includegraphics[width=10cm]{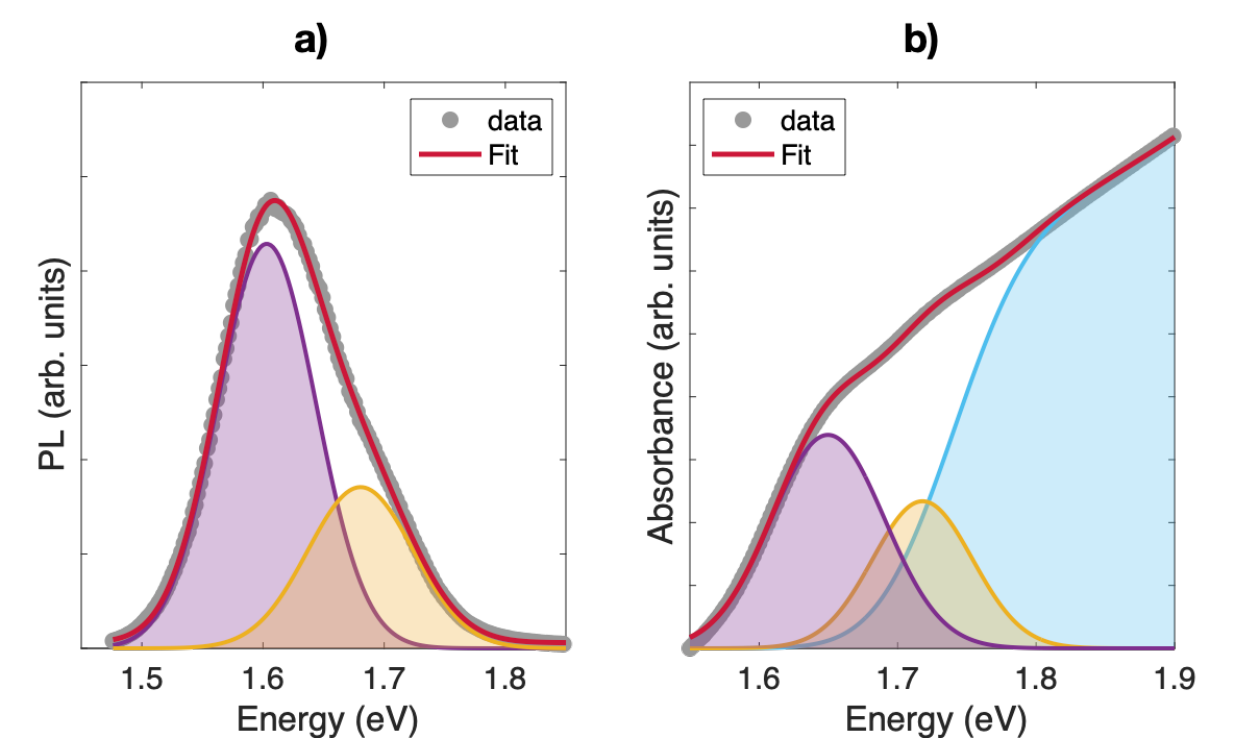}
\caption{Fit of photoluminescence (a) and absorption (b) spectra of NCs superlattice.}
\label{SI fig: PL&abs fit}
\end{figure}

In order to better characterize the equilibrium optical properties of the superlattice sample employed for 2D spectroscopy, we fit a multi-peak model to the PL and absorbance spectra (Figure \ref{SI fig: PL&abs fit}). 
The photoluminescence spectrum is well described by two Gaussian peaks centred at $1.681 \pm 0.004$ eV and $1.603 \pm 0.002$ eV, corresponding to an energy separation $\Delta = 78 \pm 6$ meV. The full width at half maximum (FWHM) of the two peaks is $96 \pm 2$ meV and $103 \pm 5$ meV, respectively.

The absorbance spectrum hints at the presence of two excitonic-like resonances in the energy range between 1.6 and 1.8 eV, which are better resolved by 2DES. While an accurate analysis of the linear spectrum is hindered by the presence of broad and overlapping features, the data are compatible with the presence of two excitonic-like peaks described by the Elliott model \cite{elliott1957intensity}. For each of the two contributions, the absorption spectrum is described by an equation of the form
\begin{equation}
    A_0 \theta(\hbar \omega-E_g) \frac{\pi e^{\pi x}}{\sinh(\pi x)} \frac{1}{1-b(\hbar \omega -E_g)} + A_0 R_{ex} 4 \pi \delta(\hbar \omega - E_g + R_{ex}) 
    \label{eq: Elliot}
\end{equation}
where $E_g$ is the bandage, $R_{ex}$ is the exciton binding energy, $A_0$ is an amplitude parameter, $b$ is a parameter taking into account the non-parabolicity of the bands, $\delta$ is the Dirac delta function and $\theta$ is the Heaviside step function \cite{yang2016large,saba2014correlated,yang2015low}.  Equation \ref{eq: Elliot} is then convoluted with a Gaussian to account for inhomogeneous broadening. According to the best fit (red solid line in Fig \ref{SI fig: PL&abs fit}b), the conduction band edge is located at $1.750 \pm 0.007$ eV, while the two peaks are centred at $1.65 \pm 0.01$ eV and $1.72 \pm 0.01$ eV and have a FWHM of $94 \pm 5$ meV and $90 \pm 9$ meV, respectively. The presence of two structures agrees with the 2D spectroscopy data discussed in the main text. Based on the 2DES spectra, these two structures can therefore be assigned to a trap state and an exciton state for the lower and higher energy peak, respectively.

\FloatBarrier

\section{2DES optical setup}

\begin{figure}[h!]
\includegraphics[width=16cm]{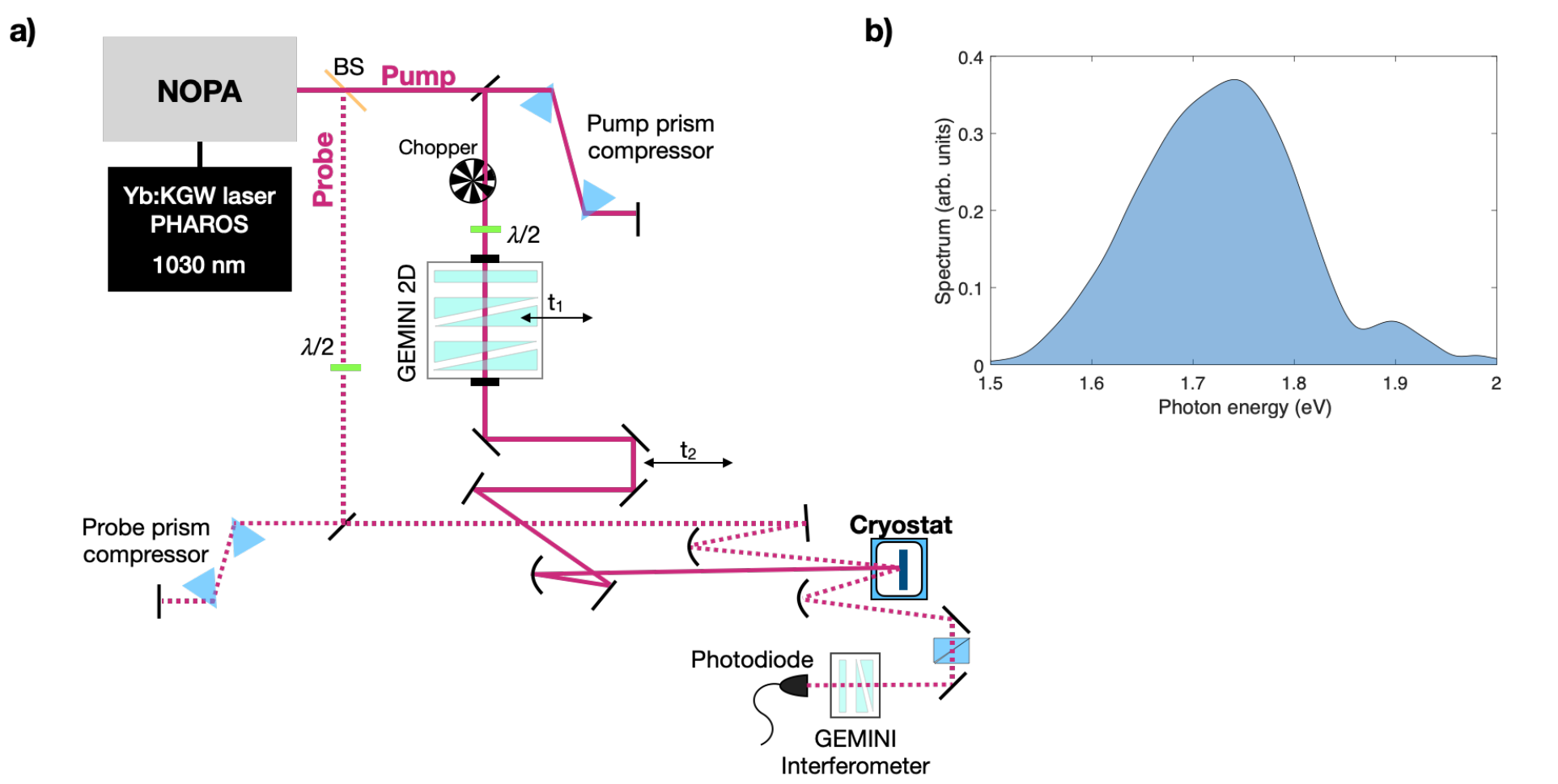}
\caption{a) Sketch of the 2DES experimental setup. b) Spectrum of the NOPA signal, employed for pump and probe beams.}
\label{SI fig: setup}
\end{figure}

\clearpage
\section{Exciton-biexciton rate equations}
\begin{figure}[]
\includegraphics[width = 12 cm]{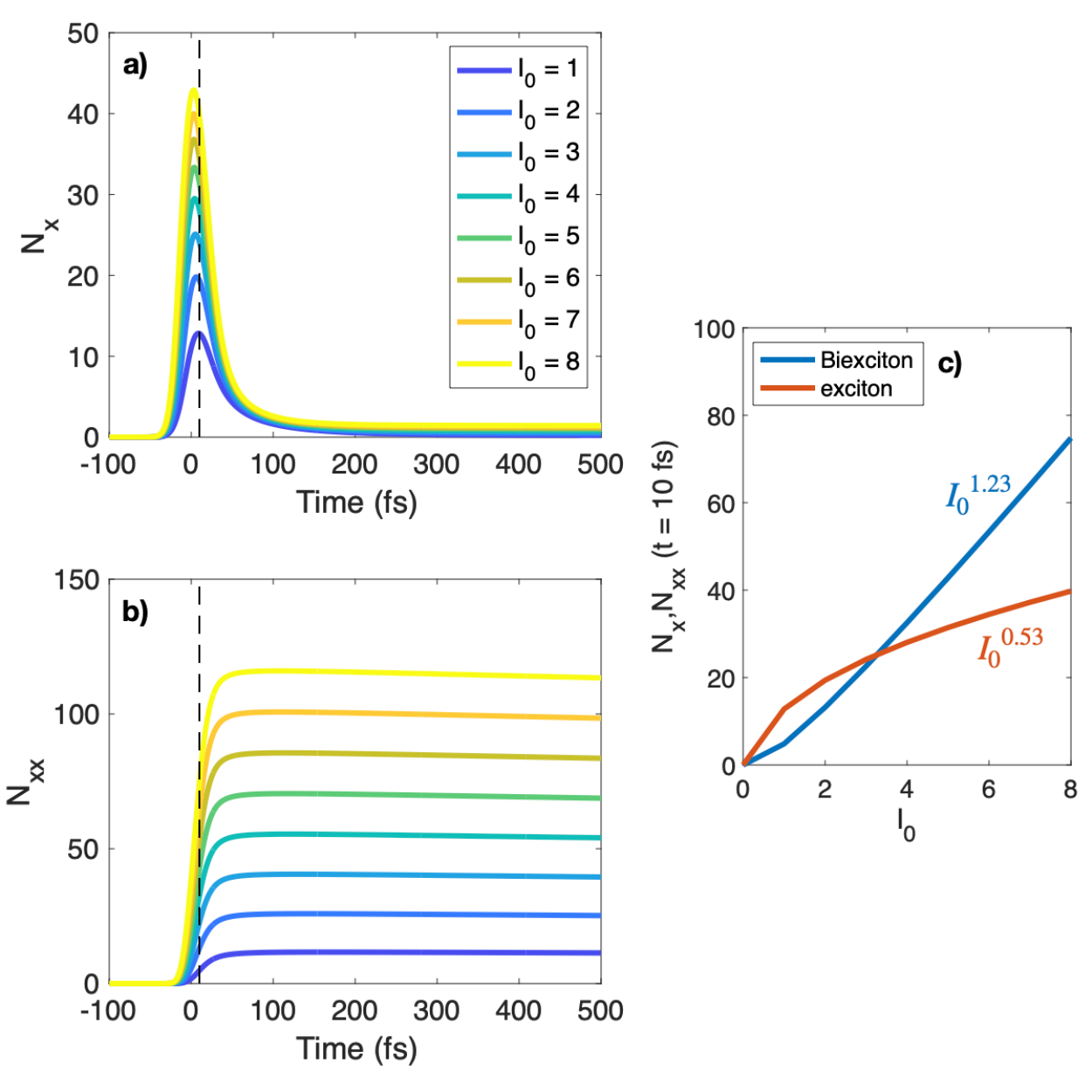}
\caption{a) and b) display the numerical solution of the two coupled differential equations \ref{SI eq: exciton rate eq} and \ref{SI eq: bi-exciton rate eq} for different values of the source term intensity $I_0$. c) Number of excitons and bi-excitons at $t = 10$ (dashed lines in a) and b)) as a function of excitation amplitude.}
\label{SI fig: biexciton}
\end{figure}

The dynamics of bi-exciton formation is modelled through the following rate equations \cite{yamamoto2001biexciton,you2015observation}
\begin{equation}
        \frac{d N_x}{dt} = I(t) -\gamma_x N_x + 2 \gamma_{xx} N_{xx} - 2 \beta N_{x}^2
        \label{SI eq: exciton rate eq}
\end{equation}

\begin{equation}
        \frac{d N_{xx}}{dt} = -\gamma_{xx} N_{xx} + \beta N_x^2
        \label{SI eq: bi-exciton rate eq}
\end{equation}
where $N_x$ and $N_{xx}$ represent the number of excitons and bi-excitons respectively and $\gamma_x$ and $\gamma_{xx}$ are the decay rate constants for exciton and bi-exciton respectively. \\
$I(t) = I_0 \exp\left[-4 \log(2) \left(\frac{t}{FWHM_t}\right)^2\right]$ is the source term that accounts for the ultra-short pump pulse generating excitons and it is described by a gaussian function of full width at half maximum $FWHM_t$ = 30 fs and centred at $t = 0$. The $\beta$ parameter describes the rate of bi-exciton formation, which originates from the interaction between two excitons. Numerical solutions of equations \ref{SI eq: exciton rate eq} and \ref{SI eq: bi-exciton rate eq} with $\gamma_x = 0.01$ fs$^{-1}$, $\gamma_{xx} = 0.0001$ fs$^{-1}$ and $\beta = 0.002$ fs$^{-1}$ return the dynamics displayed in Figure \ref{SI fig: biexciton}a and b. We observe that excitons are instantaneously excited and quickly decay with a relaxation rate dominated by $\gamma_x$. The population of bi-exciton states occurs on a delayed timescale thanks to the coupling term $\beta N_x^2$ that accounts for exciton-exciton interaction leading to the formation of a two-body bound state. 

In order to investigate the fluence-dependent behaviour, the simulation of the dynamics obtained from eq. \ref{SI eq: exciton rate eq} and \ref{SI eq: bi-exciton rate eq} is repeated for increasing amplitude $I_0$ of the source term $I(t)$. In order to compare with the experiment, we look at the number of excitons (red curve in Fig. \ref{SI fig: biexciton}c) and bi-excitons (blue curve in Fig. \ref{SI fig: biexciton}c) at a fixed time delay $t^*$ after pump excitation as a function of the amplitude $I_0$ of the Gaussian source term. We chose $t^* = 10$ fs which is close to the time of maximum $N_x$ and corresponds to the $t_2$ delay where fluence-dependent 2DECS are performed. Regardless of the choice of parameters $\gamma_x $, $\gamma_{xx} $ and $\beta $, we reveal an intensity-dependent trend that is always super-linear for the number of bi-excitons ($N_{xx} = c_{xx}{I_0}^{n_{xx}}$, $n_{xx} > 1$) and sub-linear trend for the excitons ($N_x = c_x {I_0}^{n_x}$, $n_x < 1$). The sublinearity of $N_x$ at short time delays is related to the excitons binding into bi-excitons already during the excitation pulse duration (30 fs), if the bi-exciton formation characteristic time $1/\beta N_x$ is comparable to $FWHM_t$. We find that there is a range of parameters $\gamma_x $, $\gamma_{xx} $ and $\beta $ whose power low scaling exponent is comparable to the experimental values ($n_{\mathrm{A}}$ = 0.59 $\pm$ 0.9 for peak A associated to the exciton, $n_{\mathrm{B}}$ = 1.19 $\pm$ 0.06 for peak B associated to the bi-exciton). For example, the experimental trend is compatible with $\gamma_x = 0.01$ fs$^{-1}$, $\gamma_{xx} = 0.0001$ fs$^{-1}$and $\beta = 0.002$ fs$^{-1}$, which give $n_x$ = 0.53 and $n_{xx}$ = 1.23 as shown in Figure \ref{SI fig: biexciton}. In this case, the dynamics of exciton decay and bi-exciton build-up are governed by the large exciton decay rate $\gamma_x$ that dominates over $\beta N_x$, resulting in characteristic decay and build-up times that are almost fluence-independent within the time resolution given by $FWHM_t$. We note that the dipole strength that determines the signal intensity of peak A and peak B measured in 2D spectroscopy will, in general, be different for the exciton and bi-exciton; this results in an exciton/bi-exciton intensity ratio ($I_A/I_B$) that can differ from the simulated $N_x/N_{xx}$.

\clearpage
\section{2DES anti-diagonal profiles  fitting}

\if
In order to extract a meaningful estimate of the broadening of each component from the temperature-dependent measurements, we fit the anti-diagonal profiles using a sum of two Lorentzian functions and a constant background, convoluted with a Gaussian that accounts for the spectral resolution of the 2DES setup. Each Lorentzian component has the form
\begin{equation}
    I_i \frac{1}{\pi} \frac{\Gamma_i/2}{(x-x_{0,i})^2+(\Gamma_i/2)^2}
\end{equation}
where $I_i$ is an amplitude parameter representing the area under the Lorentzian curve, $\Gamma_i$ is the full width at half maximum and $x_{0,i}$ is the centre of the Lorentzian peak. The spectral resolution is described by a Gaussian of FWHM 27 meV. This value is estimated as follows. The frequency resolution of our setup is determined by the maximum delay between the two pulse replicas \cite{preda2016broadband} generated by the GEMINI 2D and by the GEMINI for, respectively, the pump and probe photon energy axis. For the measurements presented here, the wedge scan ran between -0.8 mm and 1.6 mm for both GEMINI 2D and GEMINI, so that the resulting spectral resolutions is the same for the two frequency axis. A precise estimate of the energy resolution value was determined by recording the interferogram of the narrowband (4 meV) light emitted by a He-Ne laser and evaluating the spectral broadening observed in the FT spectrum due to the finite wedge scan in the [-0.8, 1.6] mm range. 
\fi

\begin{figure}[ph]
\includegraphics[width = 15 cm]{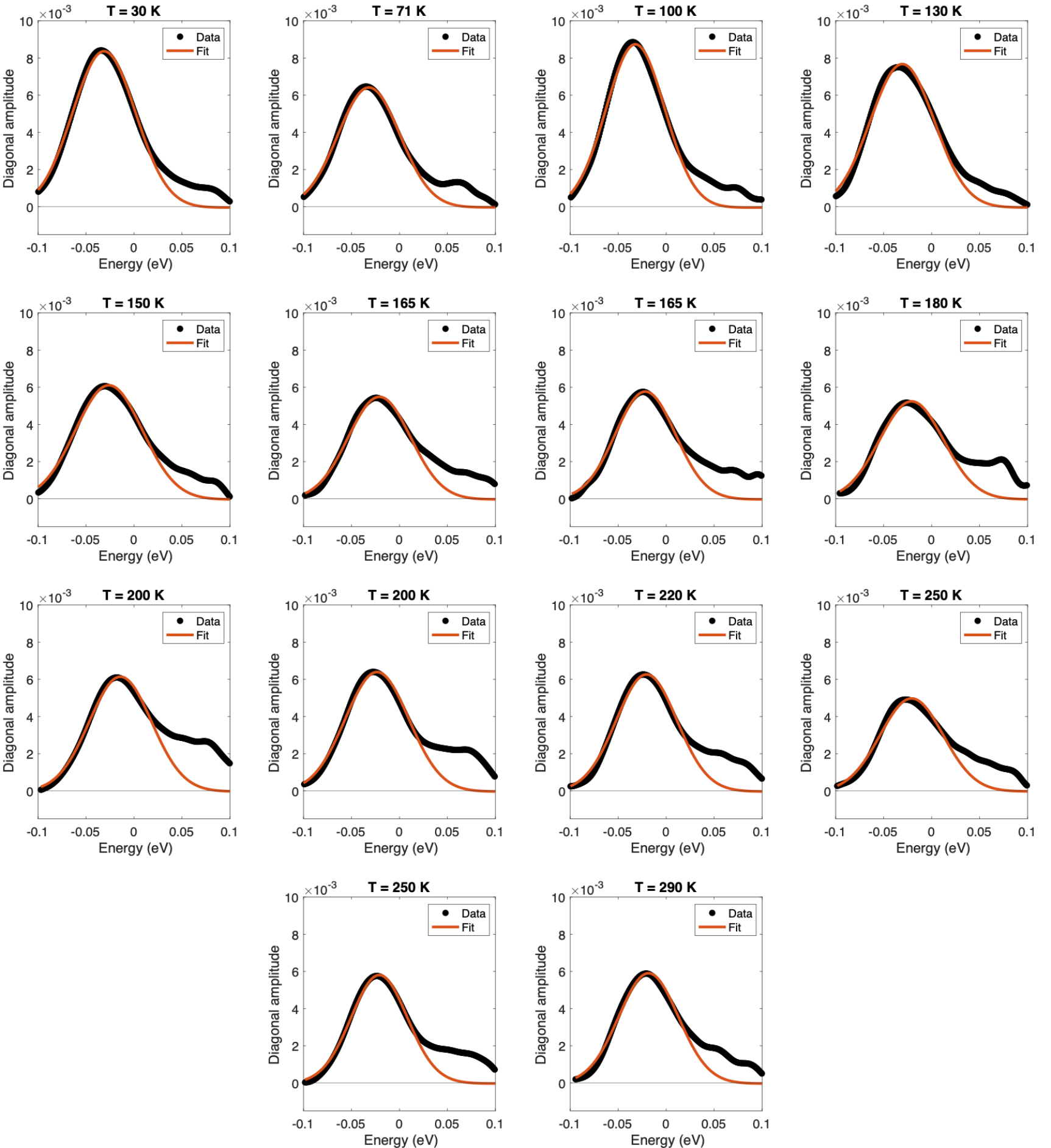}
\caption{a,b) Fit of the diagonal profiles of 2D spectra measured at short $t_2$ time delay ($t_2$ = 0) with $F$ = 175 \textmu J/cm$^2$ at different temperatures.  The curves are fitted with a function of the form $S^{R}_{\omega_0}(\omega') +S^{NR}_{\omega_0}(\omega')$ (sum of equations S6 and S8). }
\label{SI fig: diagonal fit}
\end{figure}

\begin{figure}[ph]
\includegraphics[width = 15 cm]{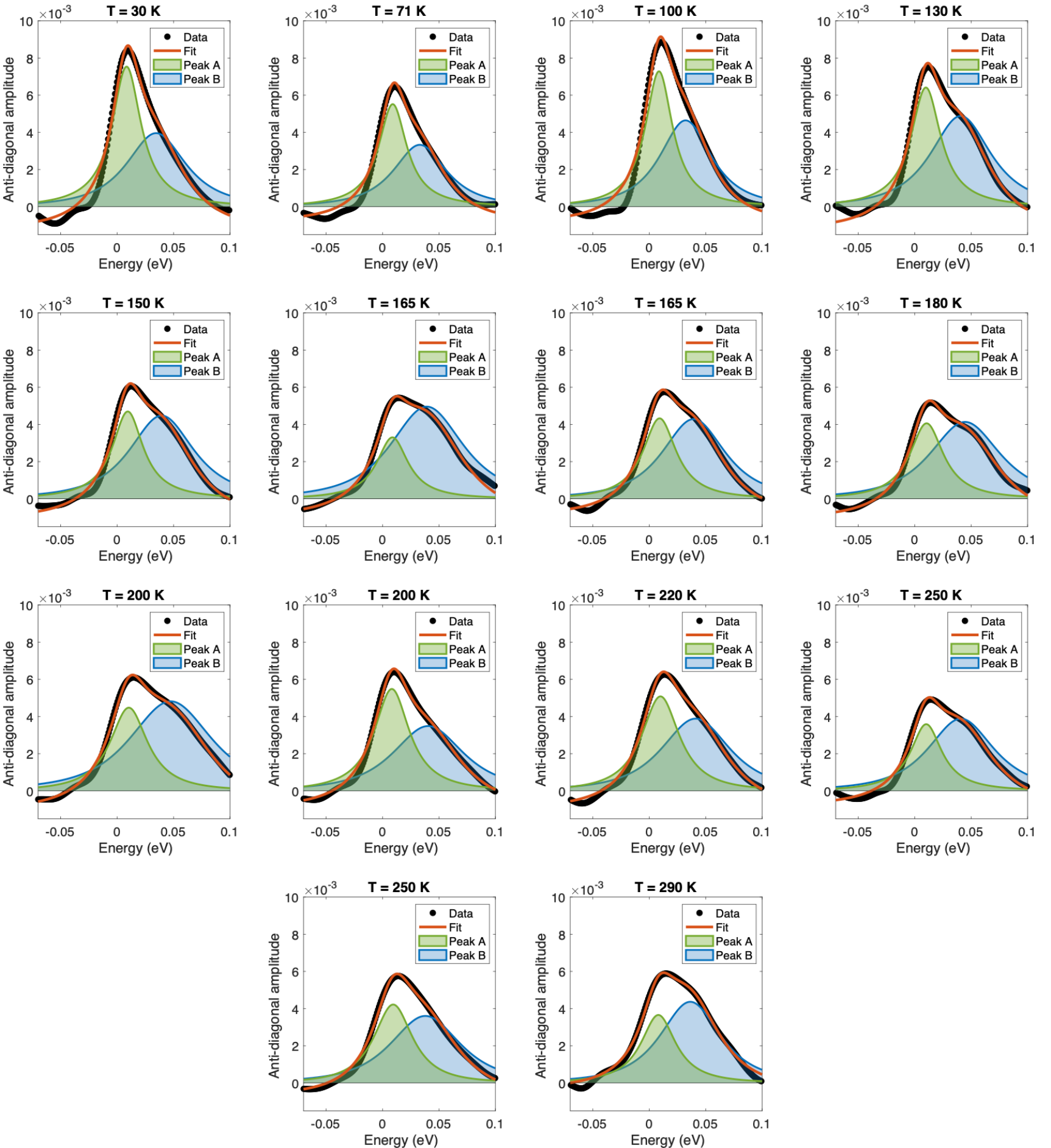}
\caption{Fit of the anti-diagonal profiles of 2D spectra measured at short $t_2$ time delay ($t_2$ = 0) with $F$ = 175 \textmu J/cm$^2$ as a function of sample temperature. The curves are fitted with the sum of two peaks, each given by $S^{R}_{\omega_0}(\omega) +S^{NR}_{\omega_0}(\omega)$ (sum of equations S5 and S7), and a constant background. }
\label{SI fig: temperature fit}
\end{figure}

\begin{figure}[]
\includegraphics[width = 17 cm]{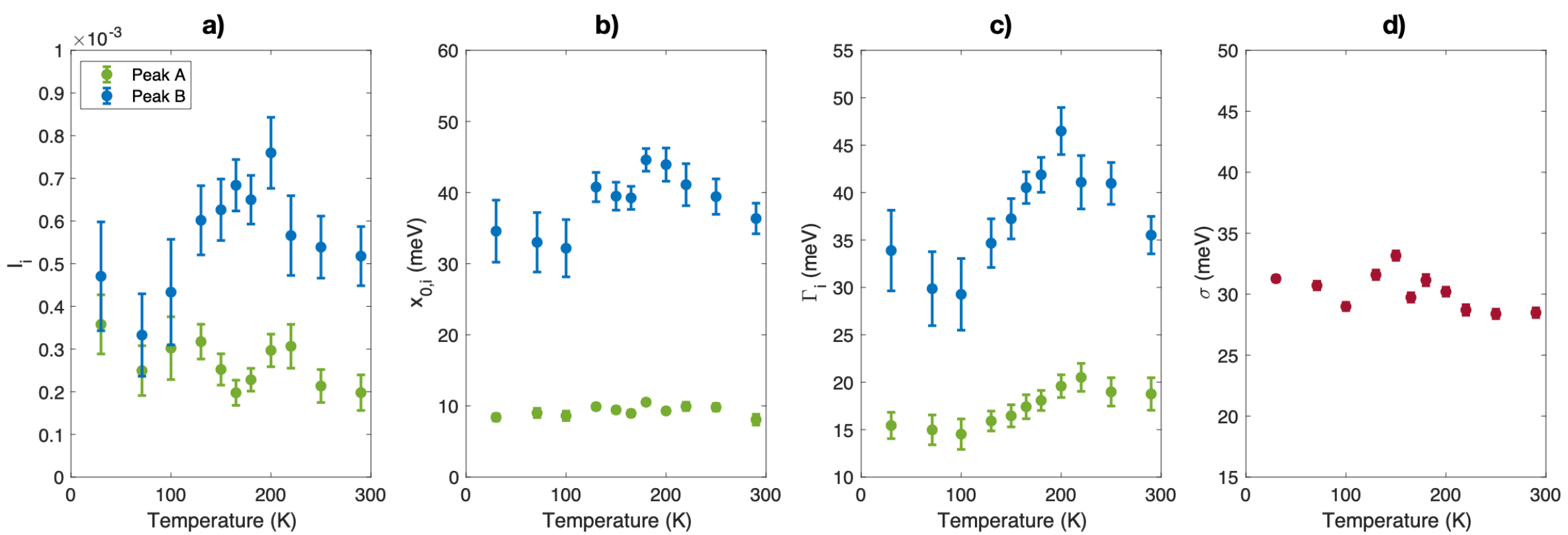}
\caption{Parameters obtained from the simultaneous fit of the diagonal and anti-diagonal profiles, as a function of sample temperature: a) amplitude parameters of peak A and peak B in the anti-diagonal slice, b) peak A and peak B central position along anti-diagonal direction, c) homogeneous widths of peak A and peak B and d) inhomogeneous broadening ($\sigma_A = \sigma_B$).}
\label{SI fig: temperature param}
\end{figure}

Line cuts of the excitonic response in the 2D spectra are analyzed with the aim of investigating homogeneous broadening. The elongation of the excitonic peak along the diagonal indicates the presence of inhomogeneity in the system. Since the linewidth along diagonal and antidiagonal directions are of the same order of magnitude, the inhomogeneous and homogeneous broadening are not completely decoupled along diagonal and anti-diagonal respectively. In this case, a global lineshape analysis is needed to extract quantitative information about the homogeneous linewidth. \\
We follow the procedure outlined in Ref. \citenum{siemens2010resonance}, where the fitting functions are derived starting from the solution $s(t_3,t_1)$ of the optical Bloch equations in 2D-time domain. For a two-level system with Gaussian inhomogeneous broadening, the signal in 2D-time domain at $t_2 = 0$ is
\begin{equation}
    s_R(t_3,t_1) = S_{0,0} e^{-\Gamma (t_3 + t_1) - i \omega_0 (t_3 - t_1) - \sigma^2 (t_3-t_1)^2/2} \Theta(t_3) \Theta(t_1)
\end{equation}
for the rephasing pulse sequence, and 
\begin{equation}
    s_{NR}(t_3,t_1) = S_{0,0} e^{-\Gamma (t_3 + t_1) - i \omega_0 (t_3 + t_1) - \sigma^2 (t_3+t_1)^2/2} \Theta(t_3) \Theta(t_1)
\end{equation}
for the non-rephasing pulse sequence \cite{mukamel1995principles, russo2022dephasing}, where $t_1$ is the coherence time, $t_3$ is the detection time, $\omega_0$ is the absorption peak central frequency, $\Gamma$ is the dephasing rate, $\sigma$ is the inhomogeneous width and $\Theta(x)$ is the Heavisede function. In 2D frequency domain, the lineshapes of the diagonal and anti-diagonal profiles can be obtained, for both the rephasing and non-rephasing parts, by applying the projection-slice theorem of 2D Fourier transforms. Indicating as $\omega$ and $\omega'$ the frequency axis along the anti-diagonal and diagonal directions of the 2D spectrum, respectively, the line-shapes for the rephasing signal are: 
\begin{equation}
    S^{R}_{\omega_0}(\omega) = \frac{1}{\sigma(\Gamma-i \omega)} \exp{\left[\frac{(\Gamma-i \omega)^2}{2 \sigma^2}\right]} \erfc{\left[\frac{\Gamma - i \omega}{\sqrt{2}\sigma}\right]}
\end{equation}
and 
\begin{equation}
    S^{R}_{\omega_0}(\omega') = \sqrt{\frac{2}{\pi \sigma^2}} e^{-\omega'^2/2 \sigma^2} \ast \frac{1}{\Gamma^2 + \omega'^2}
\end{equation}
where $\erfc$ indicates the complementary error function and $\ast$ the convolution. Analogously, the anti-diagonal and diagonal lineshapes for the non-repahsing signal are: 
\begin{equation}
    S^{NR}_{\omega_0}(\omega) = -\frac{ie^{\Gamma^2/2\sigma^2}}{\sqrt{2}\sigma}\frac{1}{\omega} e^{-\omega^2/2\sigma^2} \left[ e^{-i \omega\Gamma / \sigma^2} \erfc{\left( \frac{\Gamma - i \omega}{\sqrt{2}\sigma}\right)} - e^{i \omega\Gamma / \sigma^2} \erfc{\left( \frac{\Gamma + i \omega}{\sqrt{2}\sigma}\right)}\right]
\end{equation}
and 
\begin{equation}
    S^{NR}_{\omega_0}(\omega') = \frac{1}{ \sigma} e^{-\omega'^2/2\sigma^2} \ast \frac{1}{(\Gamma+i\omega')^2}
\end{equation}
In the partially collinear geometry employed in the present work, where rephasing and non-rephasing pulse sequences can't be distinguished, the measured signal is the sum of the two contributions. Therefore, in this configuration, the lineshape along the 2D spectrum diagonal can be described by $S_{\omega_0}(\omega') = S^{R}_{\omega_0}(\omega') +S^{NR}_{\omega_0}(\omega')$, while the anti-diagonal one is given by  $S_{\omega_0}(\omega) = S^{R}_{\omega_0}(\omega) +S^{NR}_{\omega_0}(\omega)$. \\
The measured 2DES maps are therefore sliced along the anti-diagonal and diagonal directions. The slices are centred on the maximum of the excitonic diagonal resonance in the 2D spectrum at short $t_2$ time delay. The asymmetry of the exciton peak in the 2D spectrum suggests the presence of two peaks, one centred on the diagonal (peak A) and one a few tens of meV above the diagonal (peak B). The presence of overlapping contributions hinders the performance of the global fitting, required by the approach presented above, for peak A and peak B separately.  We therefore assume that the inhomogeneous broadening is the same for both features (peak A and peak B), i.e. $\sigma_A = \sigma_B $. Diagonal and anti-diagonal profiles are then fitted simultaneously: the diagonal one is fitted with $S_{A,\omega_0}(\omega')$, describing the linshape of feature A along the diagonal direction; for the anti-diagonal direction, where two peaks are needed to fit the data, the linecuts are fitted with $I_A S_{A,\omega_0}(\omega) + I_B S_{B,\omega_0}(\omega)$, with $I_i$ amplitude parameters and the subscript $i = A,B$. \\
The fit is performed as described above for the short time delays t$_2$ 2DES data, collected at temperatures between 30 K and 290 K. The fitted diagonal and anti-diagonal profiles are plotted in Figures \ref{SI fig: diagonal fit} and \ref{SI fig: temperature fit}, respectively. For the diagonal slices, the tail around 70 meV that is not reproduced by the fitting functions may suggest the presence of a weak response on the diagonal around 1.61 eV, as further discussed in the main text. 
Figure \ref{SI fig: temperature param} displays the obtained fit parameters parameters. For some temperatures the experiment was repeated twice; in these cases, the reported output parameters are the average values obtained from the two measurements. 

\clearpage

\begin{figure}[h!]
\includegraphics[width = 15 cm]{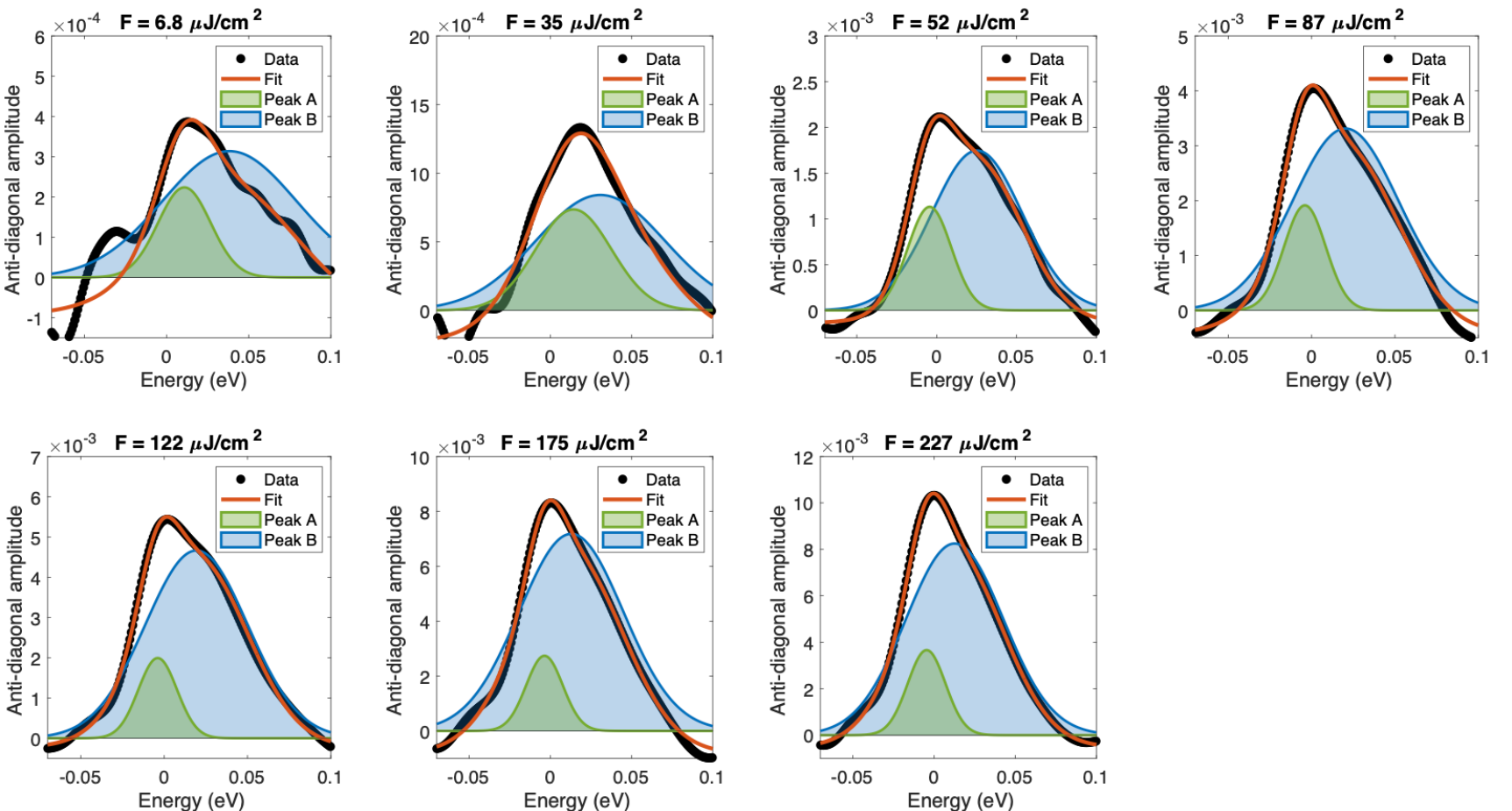}
\caption{Fit of the anti-diagonal profiles of 2D spectra measured at short $t_2$ time delay ($t_2$ = 0) and $T$ = 200 K, as a function of pump fluence. The curves are fitted with the sum of two Gaussian functions (green and blue filled areas) and a constant background.}
\label{SI fig: fluence fit}
\end{figure}

\begin{figure}[h]
\includegraphics[width = 16 cm]{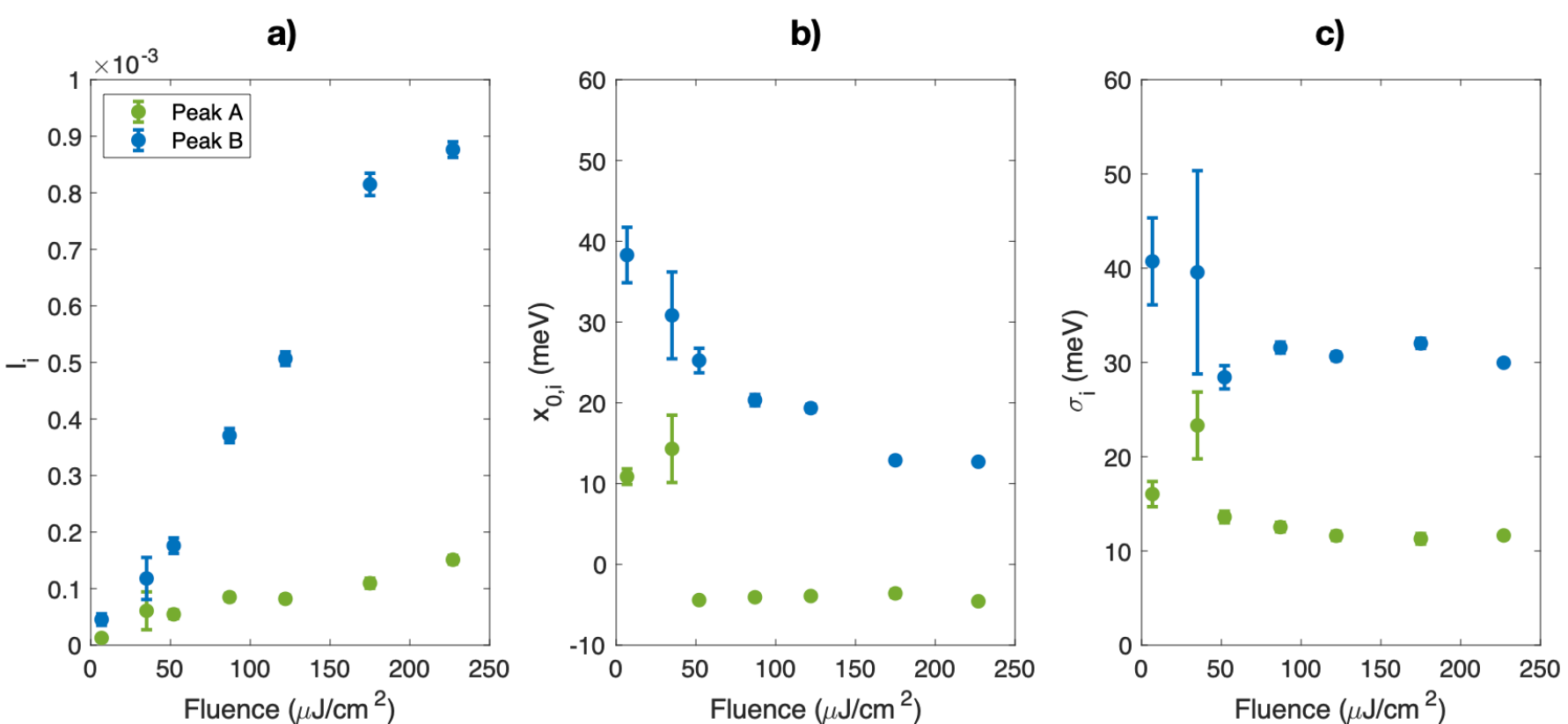}
\caption{Parameters of the two Gaussian peaks obtained from the fit of the anti-diagonal profiles as a function of sample temperature: a) amplitude, b) peak central position, c) width. }
\label{SI fig: fluence param}
\end{figure}

The fluence-dependent data are analyzed with a similar approach. Since in this case we are mostly interested in estimating the amplitude of the two components rather than extracting the purely homogeneous linewidth, we fit only the anti-diagonal profiles by employing a sum of two Gaussians and a constant background as fit function. This choice results in a more robust fit for the fluence scan, due to the larger noise of the data collected at very low excitation intensity. Each component is described by a curve define as 
\begin{equation}
    \frac{I_i}{\sqrt{2 \pi} \sigma_i} \exp{\left[-\frac{(x-x_{0,i})^2}{2\sigma_i^2}\right]}
\end{equation}
with $I_i$, $x_{0,i}$ and $\sigma_i$ ($i = 1,2$) the fit free parameters, representing respectively the area, central energy and width (standard deviation) of each peak.  The fit of fluence-dependent 2D spectra anti-diagonal profiles are plotted in Figure \ref{SI fig: fluence fit} and the output parameters are displayed in Fig. \ref{SI fig: fluence param}.

\begin{figure}[h]
\includegraphics[width = 15 cm]{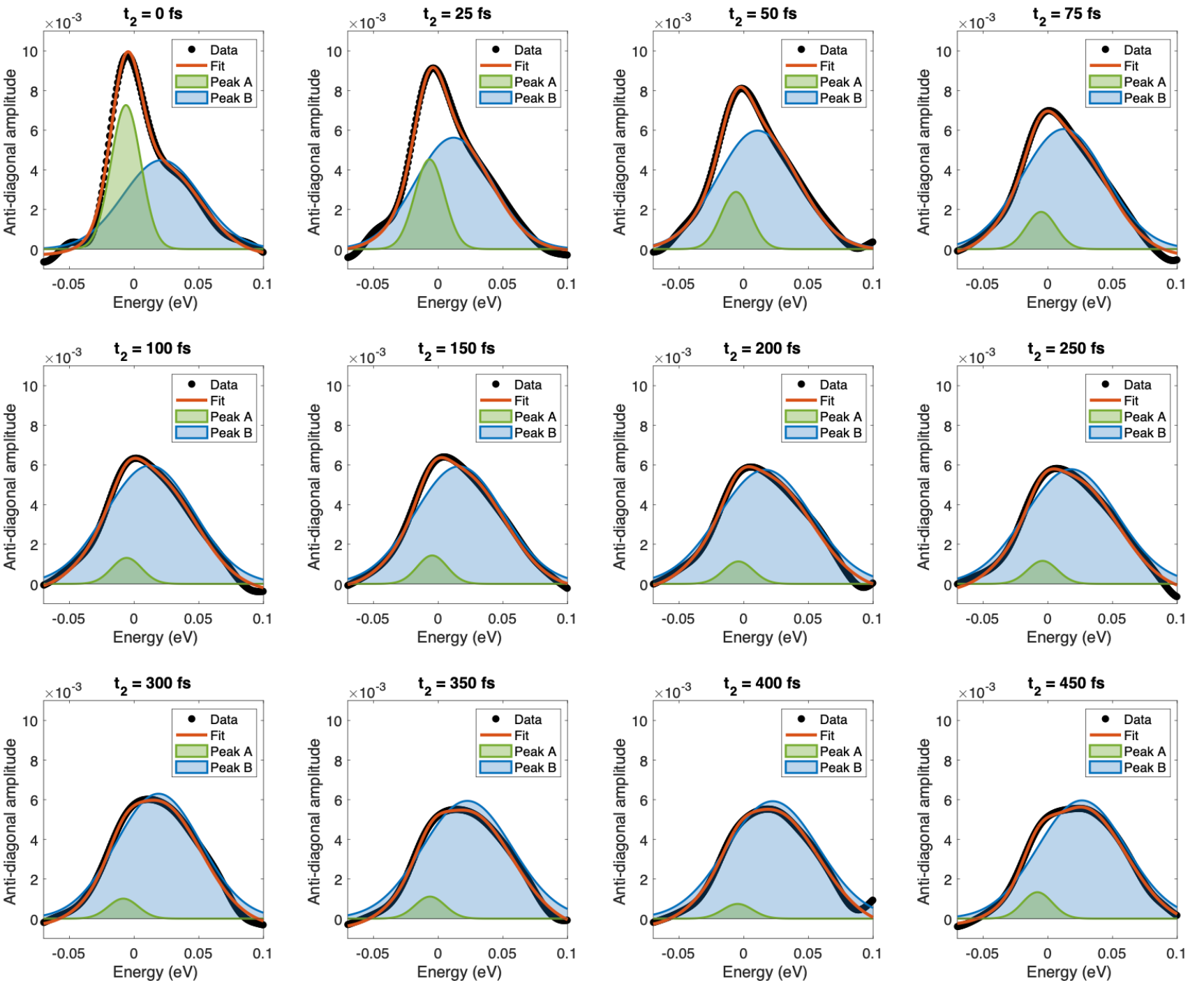}
\caption{Fit of the anti-diagonal profiles of 2D spectra measured as a function of $t_2$ time delay at $T$ = 200 K with $F$ = 175 \textmu J/cm$^2$. The curves are fitted with the sum of two Gaussian functions (green and blue filled areas) and a constant background.}
\label{SI fig: t2 scan fit}
\end{figure}

\begin{figure}[h]
\includegraphics[width = 16 cm]{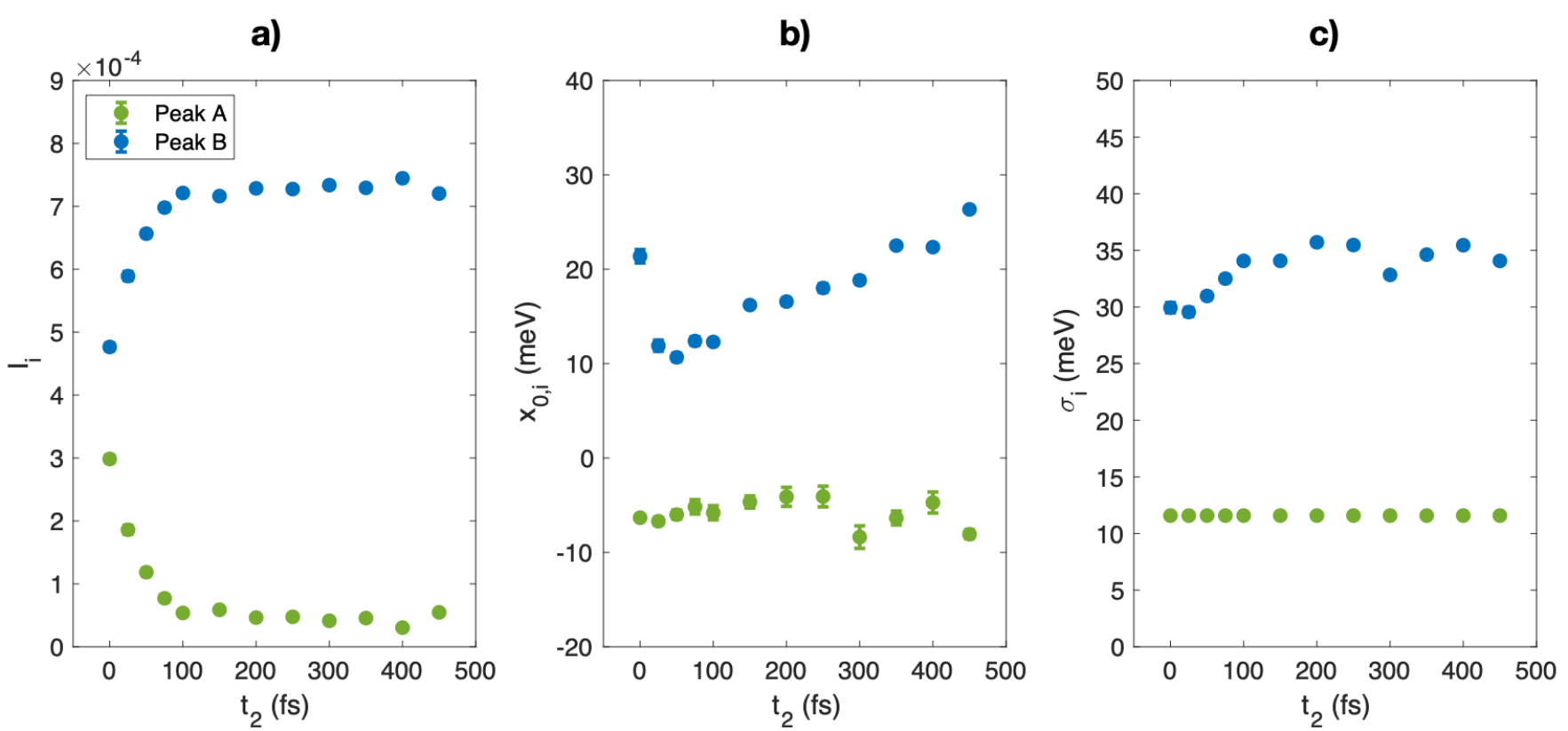}
\caption{Parameters of the two Gaussian peaks obtained from the fit of the anti-diagonal profiles as a function of $t_2$ time delay: a) amplitude, b) peak central position, c) width. }
\label{SI fig: t2 scan param}
\end{figure}

Similarly, the time-resolved 2D spectra anti-diagonal profiles are fitted with the sum of two Gaussian peaks and a constant background. From a first fit, we observe that the total spectral weight of the two components, given by $A_1 + A_2$, is constant over the scanned $t_2$ range. We, therefore, fix $A_1 + A_2$ to a constant value in order to improve fit quality and robustness. The results of the fitting procedure are displayed in figures \ref{SI fig: t2 scan fit} and \ref{SI fig: t2 scan param}.

\clearpage
\section{Fluence dependent 2DES at  \lowercase{t}$_2$ = 3 \lowercase{ps}}
\begin{figure}[h]
\includegraphics[width = 13 cm]{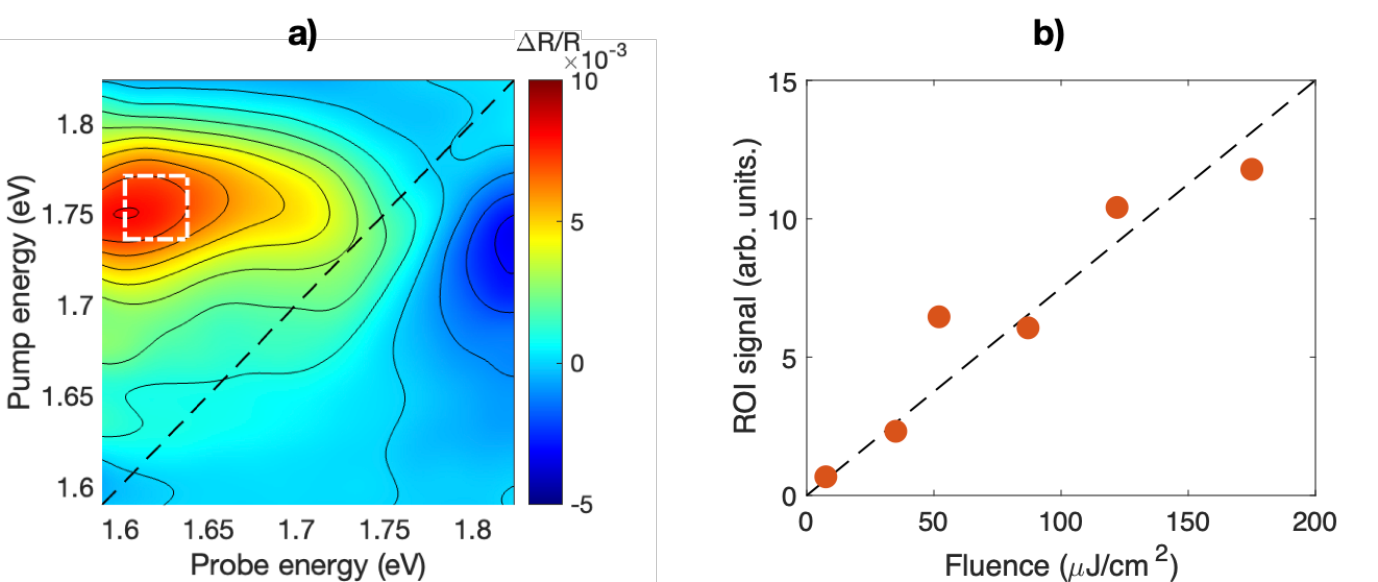}
\caption{a) 2D spectrum collected at $T$ = 21.3 K, $t_2$ = 3 ps and $F$ = 175 \textmu J/cm$^2$. The white dashed rectangle indicates the region of interest where the signal is integrated for all spectra acquired at the same temperature and $t_2$ delay, as a function of excitation intensity. b) The red dots display the fluence dependence of the 2DES signal integrated in the ROI highlighted in a), the black dashed line indicates the observed linear trend.}
\label{SI fig: fluence 3ps}
\end{figure}

The fluence dependence of the off-diagonal structure in the 2D spectra is investigated from 2DES measurements performed at 21.3 K and $t_2$ = 3 ps. Integration of the signal in the squared region of interest (ROI) highlighted in Figure \ref{SI fig: fluence 3ps}a shows a linear behaviour of the intensity of the off-diagonal spectral feature with excitation intensity, as displayed in Figure \ref{SI fig: fluence 3ps}b.

\section{2DES in transmission geometry}

2DES experiments in transmission geometry can be performed by employing samples deposited on a transparent glass substrate. We however find that, in the transmission configuration, there is stronger scattering of the pump beam, which affects the measured signal along the diagonal of the 2D spectrum. For this reason, the detailed experiments analysed in the main text are performed in reflection configuration.

\begin{figure}[h]
\includegraphics[width = 17 cm]{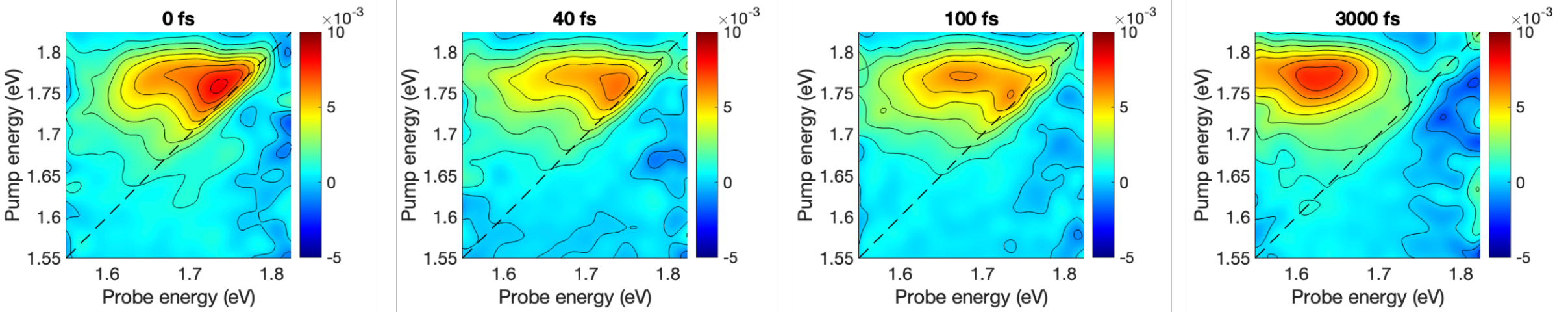}
\caption{2D spectra measured in transmission geometry for a FAPI NC superlattice sample on a transparent substrate. The measurements are performed at T = 23 K and fluence 175 $\textmu$J/cm$^2$ excitation fluence for four different time delays t$_2$.}
\label{SI fig: transmission}
\end{figure}

For comparison, one 2DES experiment in transmission geometry is reported in figure \ref{SI fig: transmission}. In this experimental configuration, the key features revealed from 2D spectroscopy are the same, although more noisy, as those observed in reflection geometry and discussed in the main text. Specifically, it is possible to observe an excitonic response centred on the spectrum diagonal and an off-diagonal feature emerging on a longer timescale.

\newpage
\section{2DES on FAPI disordered nanocube film}

\begin{figure}[h]
\includegraphics[width = 12 cm]{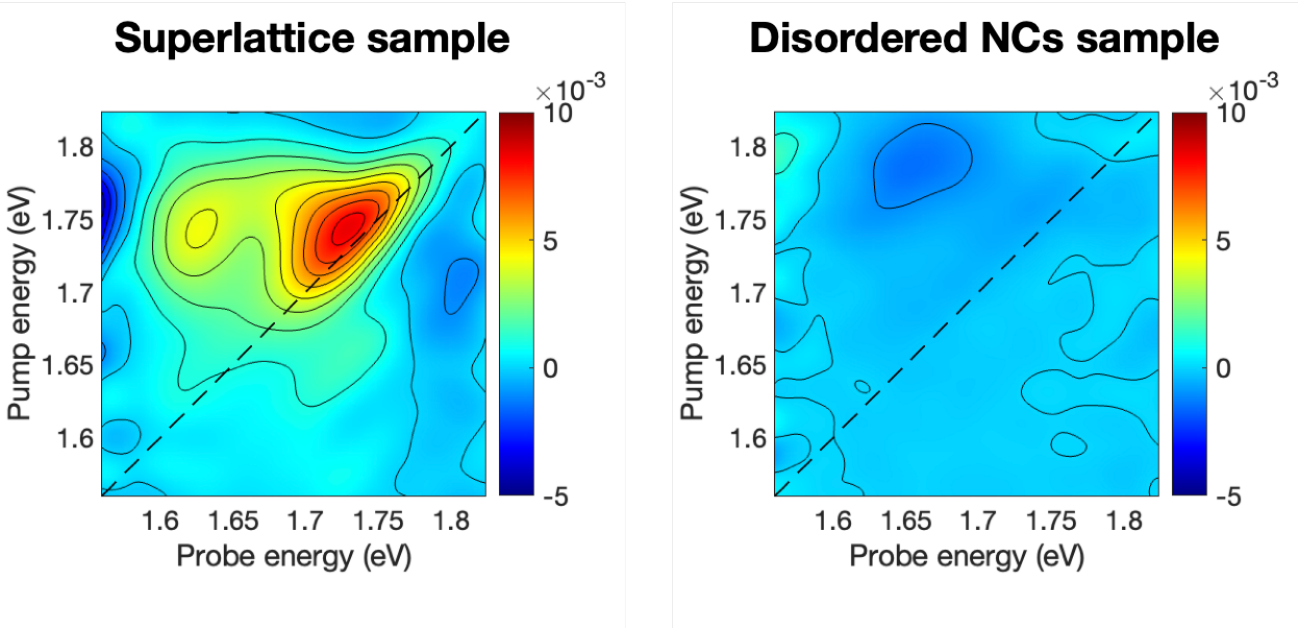}
\caption{2D spectra collected for the nanocube superlattice sample (left panel) and disordered nanocube film (right panel) in the same experimental conditions, namely cryogenic temperature (T = 30 K for the superlattice sample and T = 23 K for the disordered sample), $t_2$ = 0 ps and $F$ = 175 \textmu J/cm$^2$.}
\label{SI fig: disordered sample}
\end{figure}

Figure \ref{SI fig: disordered sample} reports the 2DES spectrum measured for a sample composed of randomly oriented nanocrystals. In this case, the transient reflectivity signal was largely suppressed. The left panel shows the 2D spectrum of Figure 3a of the main text, which was obtained for the superlattice sample (T = 30 K, t$_2 = $ 0 fs, fluence $F = $175 $\textmu$J/cm$^2$), whereas the right panel displays the 2D spectrum acquired for randomly oriented nanocubes in equivalent experimental conditions (T = 23 K, t$_2 = $ 0 fs, fluence  $F = $175 $\textmu$ J/cm$^2$). The extremely low signal of the disordered sample prevents any further analysis and investigation of the response in this system. The almost complete suppression of the signal in disordered samples however suggests some cooperative effect and calls for further investigation that is beyond the scope of the present work.

\section{Exciton density}
The number of excited excitons per nanocube can be estimated based on the experimental absorbance of the sample under investigation, the incoming energy per laser pulse, and the number of nanocubes present in the excited volume. \\
The measured absorbance $A$, reported in Figures S6 and S7, is obtained from the sample transmission $T$ according to $A = -\log_{10}{T}$. It therefore includes contributions from sample absorption, reflection and scattering. We neglect reflection and scattering effects and compute the sample absorption as $\alpha = 1 - T = 1 - 10^{-A}$, which results in an overestimate of the absorption. In this case, we obtain $\alpha \simeq$ 0.5 in the photon energy range of the experiment. 
\\
The number of excited nanocubes is limited either by the penetration depth $l_p$ of the material or by the sample thickness $h$. $l_p$ can be estimated using the refractive index of bulk FAPI reported in literature \cite{alonso2019spectroscopic}, which returns $l_p \simeq$ 800 nm in the 1.55-1.8 eV photon energy range of interest. The thickness of the sample is expected to be around 300-500 nm and thus turns out to be the limiting factor. Therefore, the number of nanocubes excited in an excitation area $L^2$, where $L$ is nanocube edge size, will be $h/L$. \\
The number of absorbed photons per nanocube is thus given by 
\begin{equation}
\langle N \rangle = \frac{F \alpha}{\hbar \omega_1} \frac{L^2}{h/L}
\end{equation}
where $F$ is the incoming pump fluence, $\hbar \omega_1$ is the pump photon energy. Based on the fitting of the sample absorbance, we expect that a large fraction of $\langle N \rangle$ will go into excitons rather than free carriers. Using $L$ = 10 nm and $\alpha = 0.5$, we obtain $\langle N \rangle \sim 8-14$ (depending on the sample thickness) at the highest excitation fluence of 230 \textmu J/cm$^2$. While this value represents only a rough estimate, it provides a useful indication of the order of magnitude of $\langle N \rangle$ and shows that the number of excitons per nanocube varies between 1 and a few tens in the fluence range spanned in the 2D spectroscpy experiments.

\newpage
\bibliography{Refs}